 \documentclass{emulateapj}

  \usepackage{amsmath}

\newcommand{\gsim}{\; ^>_\sim \;}
\newcommand{\lsim}{ \; ^<_\sim \;}

\newcommand {\Lya}    {Ly$\alpha$}   
\newcommand {\Lyb}    {Ly$\beta$}    
\newcommand {\Lyg}    {Ly$\gamma$}
\newcommand {\Lyd}    {Ly$\delta$}
\newcommand {\Lye}    {Ly$\epsilon$}
\newcommand {\Lyz}    {Ly$\zeta$}
\newcommand {\Lyeta}  {Ly$\eta$}
\newcommand {\Lytheta}  {Ly$\theta$}

\newcommand {\HI}        {\ion{H}{1}}   

\newcommand {\OI}      {\ion{O}{1}}   
\newcommand {\OII}     {\ion{O}{2}}   
\newcommand {\OIII}    {\ion{O}{3}}   
\newcommand {\OIV}    {\ion{O}{4}}   
\newcommand {\OV}     {\ion{O}{5}}    
\newcommand {\OVI}    {\ion{O}{6}}   

\newcommand {\CII}     {\ion{C}{2}}   
\newcommand {\CIII}    {\ion{C}{3}}   
\newcommand {\CIV}    {\ion{C}{4}}

\newcommand {\NI}       {\ion{N}{1}}
\newcommand {\NIII}     {\ion{N}{3}}

\newcommand {\MgII}     {\ion{Mg}{2}}

\newcommand {\SiIII}  {\ion{Si}{3}}
\newcommand {\SiII}   {\ion{Si}{2}}

\newcommand {\NeV}  {\ion{Ne}{5}}      
\newcommand {\NeVIII}  {\ion{Ne}{8}}   

\newcommand {\FeII}   {\ion{Fe}{2}}

\newcommand {\kms}    {km~s$^{-1}$}

\newcommand {\FUSE}   {{\it FUSE}} 
\newcommand {\HST}    {{\it HST}}
\newcommand {\IUE}    {{\it IUE}}
\newcommand {\GALEX}    {{\it GALEX}}

\newcommand {\etal}   {et~al.}

\begin{document}

\title{An Ultraviolet Survey of Low-Redshift Partial Lyman-Limit Systems \\
with the HST Cosmic Origins Spectrograph } 

\author{J. Michael Shull,  Charles W. Danforth, Evan M. Tilton,  Joshua Moloney}
\affil{CASA, Department of Astrophysical \& Planetary Sciences, \\
University of Colorado, Boulder, CO 80309}

\author{and Matthew L. Stevans}
\affil{Dept.\ of Astronomy, University of Texas at Austin, Austin, TX 78712 } 

\email{michael.shull@colorado.edu, danforth@colorado.edu} 


\begin{abstract} 

We present an ultraviolet spectroscopic survey of strong \HI\ absorbers in the intergalactic medium, probing
their evolution over the last 6-7~Gyr at redshifts $0.24 \leq z \leq 0.84$.  We measure column 
densities $N_{\rm HI} \,( {\rm cm}^{-2})$ from the pattern of Lyman-series absorption lines and flux decrement 
at the Lyman limit (LL) when available.   We analyzed 220 \HI\ absorbers in ultraviolet spectra of 102 active 
galactic nuclei (AGN) taken by the Cosmic Origins Spectrograph aboard the {\it Hubble Space Telescope} 
with G130M/G160M gratings (1134-1795~\AA).  For 158 absorbers with $\log N_{\rm HI} \geq 15$, the mean
frequency is $d {\cal N}/dz = 4.95 \pm 0.39$ over pathlength $\Delta z = 31.94$ ($0.24 \leq z \leq 0.84)$.  
We identify 8 Lyman Limit Systems (LLS, $\log N_{\rm HI} \geq 17.2$) and 54 partial systems (pLLS)
with $16.0 \leq \log N_{\rm HI} < 17.2$.   Toward 159 AGN between $0.01 < z_{\rm abs} < 0.84$ with
$\Delta z \approx 48$, we find four damped \Lya\ absorbers (DLA) with 
$(d {\cal N}/dz)_{\rm DLA} = 0.083^{+0.066}_{-0.040}$ at $\langle z \rangle = 0.18$. 
The mean LLS frequency between $z = 0.24-0.48$ is $(d {\cal N}/dz)_{\rm LLS} = 0.36^{+0.20}_{-0.13}$ 
fitted to $N(z) = (0.25^{+0.13}_{-0.09})(1+z)^{1.14}$.  
For 54 pLLS we find $(d {\cal N}/dz)_{\rm pLLS} = 1.69\pm0.23$ at $\langle z \rangle = 0.39$, a frequency
consistent with gaseous halo sizes $R \approx 100 h^{-1}~{\rm kpc}$ for ($0.3-3L^*$) galaxies. 
A maximum-likelihood analysis yields a distribution $f(N,z) = C_0 N^{-\beta} (1+z)^{\gamma}$ with 
$\beta = 1.48 \pm 0.05$ and $\gamma = 1.14^{+0.88}_{-0.89}$ for $15 \leq \log N_{\rm HI} \leq 17.5$. The 
far-UV opacity gradient is $d \tau_{\rm eff} / dz \approx (0.444)(1+z)^{1.14}$ over the range  
$15 \leq \log N_{\rm HI} \leq 17$, implying mean LyC optical depth $\tau_{\rm eff} \approx 0.3-0.5$ toward
sources at $z = 1-2$.  

\vspace{0.1cm}

\end{abstract} 



\section{INTRODUCTION}

Over the past decade, astronomers have uncovered large reservoirs of gas in the outer portions of galaxy halos 
(Tumlinson \etal\ 2011a, 2013;  Stocke \etal\ 2013, 2014) and in the intergalactic medium  or IGM (Penton \etal\ 
2004; Shull \etal\ 2012a), far from the gravitational influence of individual galaxies.  These gaseous structures 
are detected by absorption-line spectra of quasars and other active galactic nuclei (AGN) using resonance lines 
of neutral hydrogen (\HI) and metal ions (e.g., \CIV, \OVI, \SiIII, \CII) in the rest-frame far-ultraviolet.  Shortward 
of the \Lya\ emission line at 1215.67~\AA, numerous weak \HI\ absorption lines blanket the AGN continuum
in the ``Lyman-alpha forest" with column densities that we have been able to measure reliably through \HST/COS 
absorption-line spectra over the range $12.5 \leq \log N_{\rm HI} ({\rm cm}^{-2}) \leq 15.0$ (Danforth \etal\ 2016). 
Over the range $15 < \log N_{\rm HI} \lsim 17$, the Lyman lines are highly saturated, and column densities are
difficult to measure from Lyman series absorption alone.  
When $\log N_{\rm HI} \geq 17.2$, photoelectric absorption in the Lyman continuum (LyC) produces optical depth 
$\tau_{\rm LL} \geq 1$ at the Lyman limit (LL) at $\lambda_{\rm LL} = 911.753$~\AA.  These redshifted absorbers 
are called Lyman-limit systems (LLS), while those with slightly lower column densities are termed  ``partial 
Lyman-limit systems" (pLLS).  In this paper, we use the term Lyman Limit (LL) in reference to the redshifted 
wavelength (912~\AA\ rest frame) and the types of absorbers (pLLS or LLS).  The term Lyman decrement refers 
to the drop in transmitted flux at the LL, which is used to define continuum optical depth ($\tau_{\rm LL}$).

Spanning a loosely defined range ($16 < \log N_{\rm HI} < 17.2$), the pLLS  are commonly associated with galaxies 
(Sargent \etal\ 1989; Steidel 1990; Simcoe \etal\ 2006) and their gaseous halos, now more fashionably called the 
circumgalactic medium (CGM).   Although previous \HST\ surveys (Storrie-Lombardi \etal\ 1994;  Stengler-Larrea
 \etal\ 1995;  Ribaudo \etal\ 2011a) focused primarily on true LLS, we note that two recent surveys of ``Lyman limit 
 systems" (Ribaudo \etal\ 2011a; Lehner \etal\ 2013) also included many partial LLS in their tables.  
The LLS and pLLS dominate the continuum opacity of the IGM (Shull \etal\ 1999; Haardt \& Madau 2012; 
O'Meara \etal\ 2013) and probe the metallicities in the environment around galaxies (Ribaudo \etal\ 2011b; 
Tripp \etal\ 2011; Lehner \etal\ 2013; Fox \etal\ 2013).   Even rarer are the Damped Lyman-alpha (DLA) 
absorbers (Wolfe \etal\ 2005;  Meiring \etal\ 2011; Turnshek \etal\ 2015) whose column densities, 
$N_{\rm HI} \geq 2 \times 10^{20}$ cm$^{-2}$, are sufficient to produce strong Lorentzian wings in their 
absorption profiles.  These \HI\ lines provide the dominant technique for measuring the baryon content of the
IGM (Shull \etal\ 2012a), and the LLS/pLLS are a benchmark for determining the metallicity of the CGM 
(Tripp \etal\ 2011).  Strong \HI\ absorbers have been linked (Simcoe \etal\ 2006; Lehner \etal\ 2009; 
Ribaudo \etal\ 2011b) to the extended regions of galaxies.  The semantic question of ``where galaxies end" 
(Shull 2014) depends on their gravitational influence as well as dynamical effects of gaseous outflow and
infall from the cosmic web (Tripp \etal\ 2011).  
  
In our recent survey (Danforth \etal\ 2016) of low-redshift \Lya\ absorbers with the Cosmic Origins Spectrograph 
(COS) on the {\it Hubble Space Telescope} (\HST), we fitted the column densities to a power-law differential 
distribution, $f(N_{\rm HI})  \propto N_{\rm HI}^{-\beta}$, with $\beta = 1.65 \pm 0.02$ over the range
$12.5 \leq \log N_{\rm HI} \leq 15.0$.   Column densities determined from strong \Lya\ lines are uncertain owing 
to line saturation at $\log N_{\rm HI} \geq 14$ for typical Doppler parameters $b \approx 20-35$~\kms.  The 
distribution is also poorly constrained at $\log N_{\rm HI} > 15$ because strong \Lya\ absorbers are rare.  Some 
progress in defining their column densities has been made with access to higher Lyman lines (\Lyb, \Lyg, \Lyd)
from the \FUSE\ satellite (Shull \etal\ 2000) and \HST\ (Danforth \& Shull 2008; Danforth \etal\ 2016).  The higher 
Lyman lines become available to COS at modest redshifts ($z \geq 0.107$ for \Lyb, $z \geq 0.167$ for \Lyg) and 
they yield more accurate measurements of the curve of growth (CoG).   The LL shifts into the COS/G130M
window at $z_{\rm abs} \geq 0.24$.  At $\log N_{\rm HI} \gsim 16.2$, in data with good signal-to-noise 
(S/N $\gsim10$) we can use the flux decrement at the Lyman edge to confirm and supplement the CoG solutions.

In this paper, we explore the pLLS/LLS distribution in $N_{\rm HI}$ and redshift, employing a ``Lyman-comb" 
technique to find strong \HI\ absorbers between $z_{\rm abs} = 0.24-0.95$ from their pattern of Lyman-series 
absorption lines and Lyman edge.  An accurate determination of the Ly decrement depends on reliable continuum 
placement longward and shortward of the LL at observed wavelength 
$\lambda_{\rm obs} = (911.753~{\rm \AA})(1+z_{\rm abs})$. We use high-S/N spectra of 102 AGN at 
$z_{\rm AGN} \geq 0.24$ with the COS G130M/160M gratings.  Figure 1 shows the redshift coverage of our 
survey, plotting the number of AGN sight lines sensitive to the LL (at 912~\AA\ rest-frame).  Our survey has much 
higher spectral resolution ($R \approx 17,000)$ than earlier surveys with the low-resolution ($R \approx 1000-1300$) 
gratings on \HST/FOS and \HST/STIS.  In well-exposed spectra with S/N $\gsim 10$, we are able to resolve velocity 
components $\Delta v = 40-400$~\kms\ within absorbers and construct multi-component CoGs  when needed. For 
absorbers with $\log N_{\rm HI} > 16.2$, the continuum optical depth at the LL is usually detectable by HST/COS 
with optical depth $\tau_{\rm LL} =  (6.304 \times 10^{-18}~{\rm cm}^2) \, N_{\rm HI} > 0.1$.  By combining the 
Lyman decrement with CoG fitting, we can confirm the \HI\ column density and its range of uncertainty.


\begin{figure}
\includegraphics[angle=90,scale=0.29]{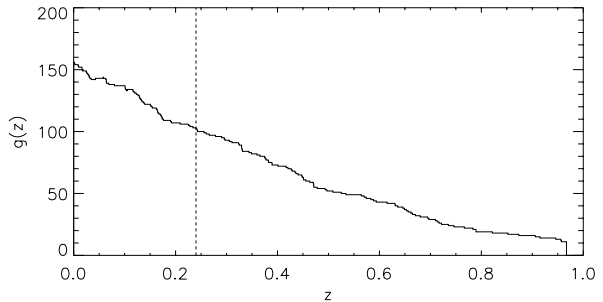}
\caption{\small{  Distribution in redshift, $g(z)$, showing the number of AGN that contribute pathlength 
capable of detecting the redshifted LL of \HI\ between $z = 0.003 - 0.96$.  We consider only absorbers 
at $z \geq 0.24$ (dashed line) whose LL is redshifted into the COS/G130M band ($\lambda>1130$~\AA).   
These 102 AGN sight lines allow us to use the LL flux-decrement, when available, and curve of growth for
Lyman absorption lines to determine \HI\ column densities and assess systematic errors. }
}
\end{figure}



\begin{figure*}
 \includegraphics[angle=0,scale=0.68] {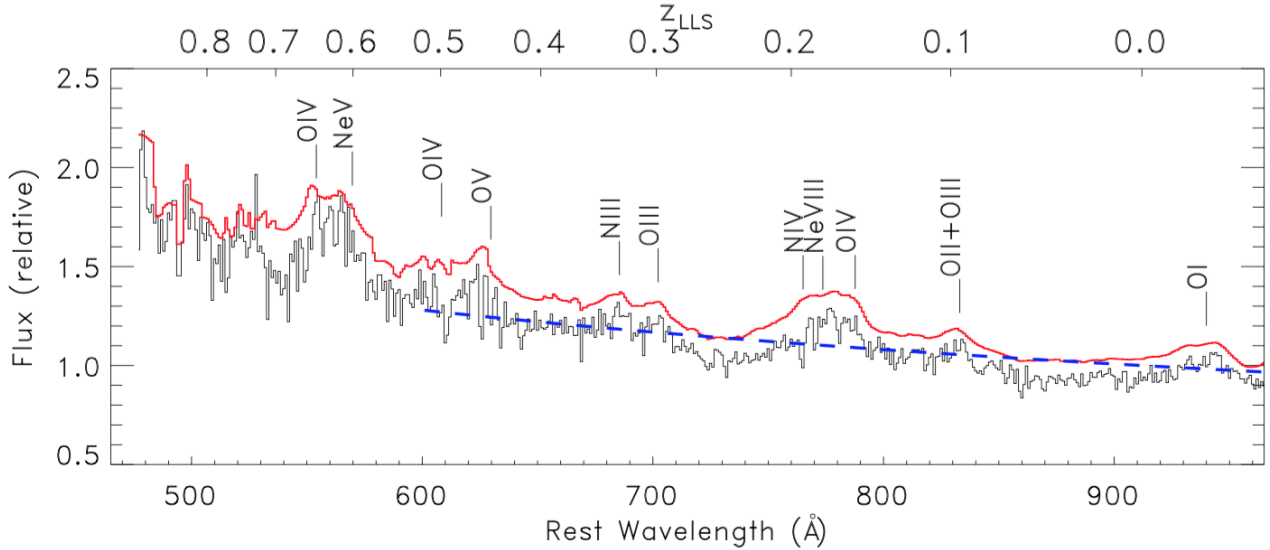}  
 \caption{\small{ AGN composite spectrum based on Stevans \etal\ (2014) and annotated with prominent
 EUV broad emission lines of metal ions (O, N, Ne).  Red spectrum shows spline fit to AGN flux passing 
 above \Lya\ absorption lines (black).  The AGN emission lines must be identified and fitted before 
 placing the underlying EUV continuum (blue dotted line).  Measuring the LL optical depth of absorbers
 requires careful attention to the the continuum on either side of the LL.  The true continuum can be 
 contaminated by broad emission lines such as the \NeVIII\ $\lambda 770,780$ doublet, \OIV\ $\lambda 788$, 
 and the 685~\AA\ and 702~\AA\ lines of \NIII\ and \OIII.  Shorter wavelength EUV lines of \OIV, \OV, and 
 \NeV\ can affect  the continuum around absorbers at $z_{\rm LLS} = 0.45-0.70$.  }
  } 
\end{figure*}


The continuum can be influenced by AGN emission lines in the UV and EUV, many of which are broad and blended
features that produce bumps and undulations in the underlying power-law continuum.  Fortunately, we have a good 
template for the location of these emission features (Figure 2) obtained from the composite rest-frame UV/EUV 
spectrum of AGN (Shull \etal\ 2012b; Stevans \etal\ 2014).  We refer the reader to our papers on AGN composite
spectra (Shull \etal\ 2012; Stevans \etal\ 2014) which describe our choice of line-free windows.  Typical errors in
continuum choice typically result in $\pm 0.02$ errors in $\log N_{\rm HI}$.  
In the rest-frame far-UV, the most prominent emission lines are the 
\OVI\ doublet (1032 and 1038~\AA), \CIII\ (977~\AA), and a blend of \OI\ features (930-950~\AA).  The dominant 
emission lines in the rest-frame EUV (550~\AA\ to 912~\AA) are lines of \OII\ and \OIII\ (833~\AA\ and 834~\AA), a 
broad complex (760-800~\AA) consisting of the \NeVIII\ doublet (770 \AA\ and 780~\AA) and \OIV\ 788~\AA, the 
700~\AA\ blend of \OIII\ (702~\AA) and \NIII\ (686~\AA), and strong emission lines of \OIV\ (608~\AA, 554~\AA) and
\OV\ (630~\AA).  A full list of EUV lines in the AGN composite spectrum is provided in Table~4 of Shull \etal\ (2012b).    

The sample used in this survey began with 159 AGN targets chosen because their UV brightness allowed them to 
be observed by \HST/COS with both G130M and G160M gratings.  Of these 159 AGN, 102 had redshifts 
$z_{\rm AGN} \geq 0.24$, sufficient to shift the LL and higher Lyman-series lines into the G130M window.   
This sample includes many AGN previously observed by the {\it International Ultraviolet Explorer} (IUE) and \HST\ 
spectrographs.  Many were used by the COS-GTO team (Danforth \etal\ 2016) and COS-Halos project (Tumlinson 
\etal\ 2013) for the purpose of studying the low-redshift IGM and galactic halo gas.  Stevans \etal\ (2014) used 159 
AGN with redshifts $0.001 < z_{\rm AGN} \leq 1.476$ to produce a COS composite spectrum of AGN in their
rest-frame UV and EUV.  The AGN in the COS-Halos program were selected to avoid strong \MgII\ absorbers at 
$z > 0.4$, which would bias the survey against LLS.  Because our survey of LLS and pLLS used only AGN with 
$z_{\rm abs} \geq 0.24$, it excludes nearby Seyfert galaxies whose sight lines might be biased against LLS.   
Further discussion of potential sample biases for LLS and DLA is given in Section 3.3 and in Ribaudo \etal\ (2011a) 
and Neeleman \etal\ (2016).  

Spectra taken with the COS G130M/G160M gratings (Green \etal\ 2012) provide moderate spectral resolution 
($R = \lambda / \Delta \lambda \approx 17,000$) allowing us to resolve individual Lyman-series absorbers and 
fit the underlying AGN continuum. With the combined coverage from the G130M grating 
($1134~{\rm \AA} - 1460~{\rm \AA}$) and G160M grating ($1390~{\rm \AA} - 1795~{\rm \AA})$ we can 
identify LLS and pLLS out to $z_{\rm abs} \approx 0.95$, using a template of Lyman-series absorption lines
at the same redshift.  Even at $\log N_{\rm HI} < 16.2$, when the LyC optical depth is weak ($\tau_{\rm LL} < 0.1$), 
we are able to identify the Lyman-line pattern down to $\log N_{\rm HI} \approx 14.5$ in well-exposed spectra
($S/N \gsim 10$).  Figure 3 shows three AGN sight lines with \HI\ absorbers between $\log N_{\rm HI} = 15.0-15.25$, 
found through higher Lyman-series absorption lines (\Lyg\ through Ly8).  In  Figure 4, we demonstrate the 
effectiveness of using the Lyman-line pattern, compared to the injection of weak ``mock absorbers" 
300~\kms\ to the red of the actual absorber.  This confirms our our ability to detect systems below 50~m\AA\ 
equivalent width in data with S/N $\gsim 10$.  

Over the redshift range of \HI\ absorbers in our full sample, $0.237 \leq z_{\rm abs} \leq 0.928$, we surveyed 
total absorption pathlength $\Delta z_{\rm eff} = 31.94$ and identified 211 absorbers:  
8 were LLS with $\log N_{\rm HI} \geq 17.2$, 54 were pLLS between $16.0 < \log N_{\rm HI} < 17.2$, and 
the remainder lay between $14.0 < \log_{\rm HI} < 16.0$.  Our COS survey contains the largest number of 
low-$z$ LLS and pLLS to date, a distribution that we compare to absorbers in the HST/FOS Key-Project 
surveys of LLS at $0.4 < z < 1.4$ (Storrie-Lombardi \etal\ 1994;  Stengler-Larrea \etal\ 1995) and the 
FOS/STIS survey of LLS at $z < 2.6$ (Ribaudo \etal\ 2011a).

In Sections 2 and 3, we describe the data set of LLS and pLLS absorbers and our determination of their 
distribution, $f(N, z)$, in column density and redshift.   We find a line frequency $d {\cal N}/dz = 4.95 \pm 0.39$ 
for $15.0 < \log N_{\rm HI} < 18.5$ averaged over $0.24 < z < 0.84$.  We quantify the pLLS distribution in \HI\
column density and redshift through maximum-likelihood fitting to the form, 
$f(N,z) = C_0 \, N^{-\beta} \,(1+z)^{\gamma}$, with best-fit parameters $\beta = 1.48\pm0.05$ and 
$\gamma = 1.14^{+0.88}_{-0.89}$.  Although the evolutionary index $\gamma \approx 1.1$ is uncertain,
owing to the limited redshift coverage and statistics, it is consistent with cosmological expectations for
a population of non-evolving pLLS absorbers with constant space density and absorption cross section.
By integrating the distribution of absorbers,  $f(N_{\rm HI}, z)$, over column density, we compute the redshift
gradient in LyC opacity, $d \tau_{\rm eff} / dz$, for absorbers between $12.5 \leq \log N_{\rm HI} \leq 17.2$.  
We estimate a far-UV (1130~\AA) continuum opacity, $\tau_{\rm eff} \approx 0.4-0.5$, toward AGN at 
$z \approx 1.5-2.0$ and discuss the possible effects on AGN source selection and their rest-frame EUV 
spectra. In Section~4, we summarize our results and their implications for IGM photoelectric opacity in
the UV,  QSO UV-survey selection, and the spatial association of pLLS with galaxy halos.   Appendix~A 
gives narratives of our analysis of 73 strong \HI\ absorbers, 23 of which were studied in previous surveys.  
Appendix~B describes our statistical analysis and the maximum-likelihood approach to obtaining 
distribution parameters.


\section{SURVEY TECHNIQUES}

Our survey of strong \HI\ absorbers comes as a natural by-product of the UV composite spectra of AGN 
constructed from moderate-resolution HST/COS data (Stevans \etal\ 2014).  To find the underlying AGN 
continuum, we used G130M/G160M spectra to identify the numerous \Lya\ forest lines, as well as the less 
frequent but stronger (LLS and pLLS) absorbers.  We did {\it not} consider absorbers associated with the host 
galaxy of the AGN.   Owing to the high resolution of COS, we are able to distinguish individual absorption
lines and resolve the true continuum level between them (Figures 3 and 4).  As discussed in our AGN
composite paper (Stevans \etal\ 2014), we corrected the AGN continuum for \HI\ photoelectric absorption 
by pLLS and LLS absorbers.  These corrections are important for establishing the underlying continuum 
longward and shortward of the \HI\ absorption features.  The LyC optical depth is related to \HI\ column density
by $\tau_{\lambda} \approx (6.304 \times 10^{-18}~{\rm cm}^2)  (\lambda / \lambda_{\rm LL})^3 \, N_{\rm HI}$
for $\lambda \leq \lambda_{\rm LL}$.  After determining $N_{\rm HI}$ from the flux decrement, we multiply 
the observed flux shortward of the LL by $\exp(\tau_{\lambda})$ to restore the true AGN continuum.


\begin{figure} 
\includegraphics[angle=0,scale=0.45]{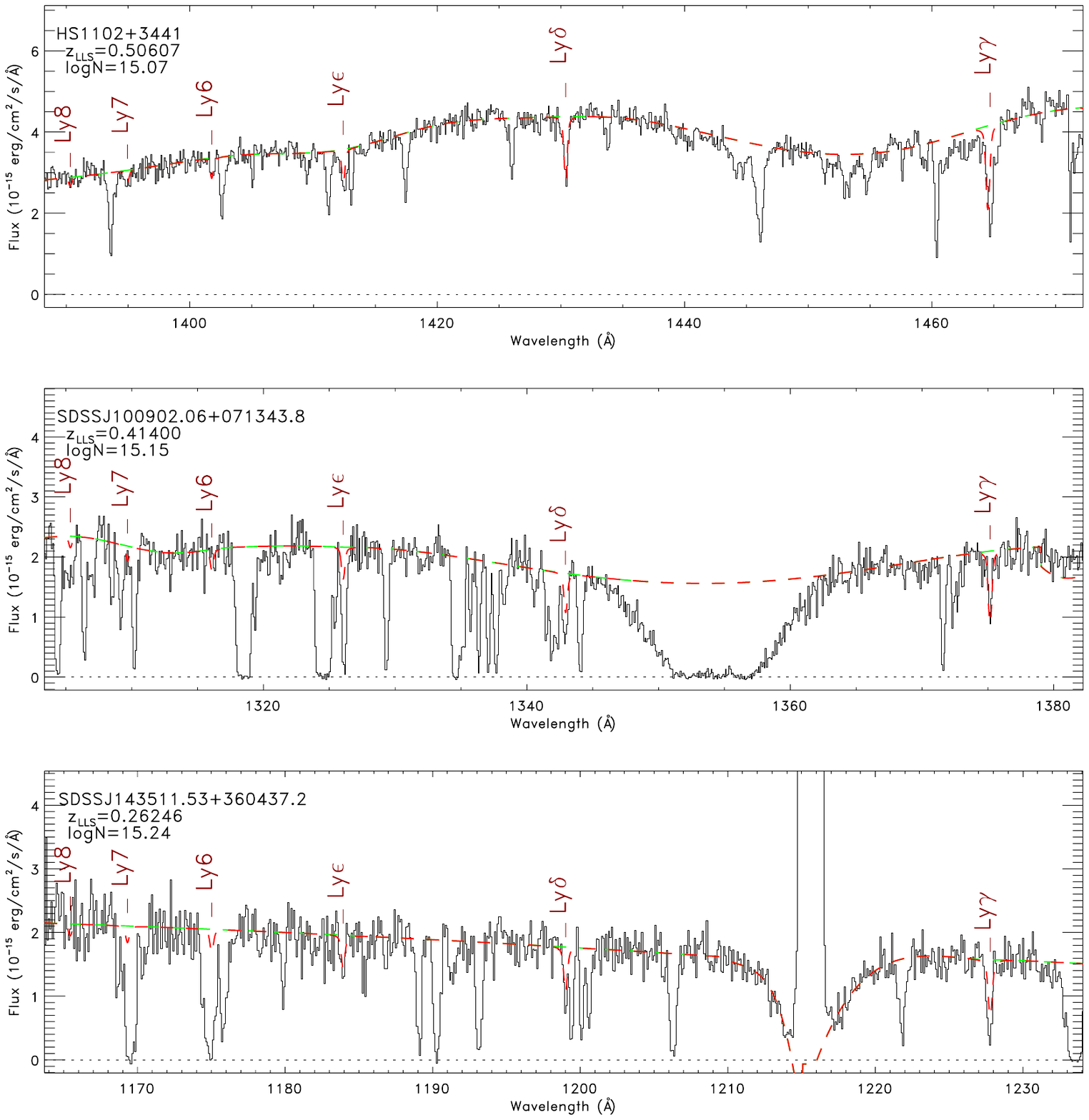}
 \caption{\small{ Three examples of COS/G130M detections of \HI\ absorbers with 
 $\log N_{\rm HI} = 15.07\pm0.02$ (at $z = 0.5061$ toward HS\,1102+3441), 
 $\log N_{\rm HI} = 15.15\pm0.02$ (at $z = 0.4140$ toward J1009+0713), and
 $\log N_{\rm HI} = 15.24\pm0.02$ (at $z  = 0.2625$ toward J1435+3604).
 Not shown are detections in \Lyb\ in all three systems and in \Lya\ for the bottom two systems;
 \Lya\ shifts beyond the G160M window at $z > 0.47$.   The panels show detections in the higher 
 Lyman-series lines (\Lyg, \Lyd, \Lye, etc).  Middle panel includes a strong DLA at $z_{\rm a} = 0.114$ 
 ($\log N_{\rm HI} = 20.68 \pm 0.10$) toward J1009+0713.  } 
  } 
\end{figure}



\begin{figure} 
\includegraphics[angle=0,scale=0.74]{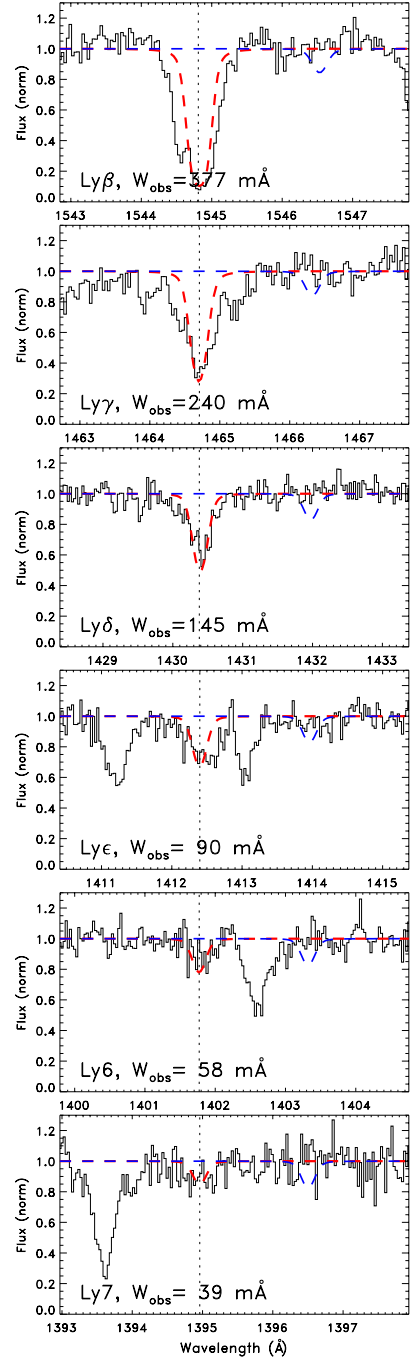}
 \caption{\small{ Six Lyman-series absorption lines toward HS\,1102+3441 detected at the same redshift 
 ($z_{\rm abs} = 0.50607$) and aligned in all panels.  Observed-frame equivalent widths (EW) range 
 from 377~m\AA\ (\Lyb) to 39~m\AA\ (Ly7) and are fitted (red-dashed lines) to a CoG with 
 $\log N_{\rm HI} = 15.07\pm0.02$ and $b = 25$~\kms.  Mock absorbers with observed-frame EWs of 
 50~m\AA\ (blue-dashed lines) are inserted 300~\kms\ redward of each Lyman line.  This simple example 
 illustrates our ability to find such absorbers in data with S/N $\gsim 10$.   } 
  } 
\end{figure}


The LLS and pLLS absorption systems are identified by a ``Lyman comb" technique (Stevans \etal\ 2014)
in which we search for a pattern of lines in the \HI\ Lyman series together with the corresponding Lyman
decrement when detectable. Table 1 lists the wavelengths, redshift bands, and column densities for which the 
first eight Lyman lines and decrement are easily detectable (greater than 50~m\AA\ equivalent width).  
To implement the method, we inspect the spectra for flux decrements at the LL, employing a computer script
that scans  for correlated down-pixels at the locations of higher-order Lyman lines of strong absorbers.   
When a system is confirmed, we measure the equivalent widths of up to the first 12 Lyman lines and fit them 
to a CoG to determine the column density and Doppler parameter ($b$ in \kms).  
Our technique depends primarily on identifying the pattern of Lyman lines and less on detecting the Lyman 
decrement.  For the standard wavelength coverage in the COS/G130M grating (1134--1459~\AA), the LL 
becomes detectable\footnote{In some cases, the G130M spectra extend down to 1130~\AA, allowing detection 
of the LL at $z \geq 0.24$.   Because our Lyman-comb method relies on finding pLLS through a pattern of
Lyman lines, we could also detect Lyman lines from pLLS at lower redshifts: $z > 0.107$ for \Lyb,  $z > 0.167$ 
for \Lyg, and $z > 0.195$ for \Lyd.   
In this survey we only report on systems at $z \geq 0.24$.}  at $z_{\rm LL} > 0.244$ and shifts out of the 
G130M band at $z_{\rm LL}  > 0.60$.  By including wavelength coverage with the G160M grating 
(1400--1795~\AA) we can observe the far-UV range with access to the LL out to $z_{\rm LL} \approx 0.95$ 
and to various Lyman lines (\Lya\ - \Lyz) over the redshift ranges shown in Table~1.  

Figure~3 illustrates our method for the higher Lyman lines of three \HI\ absorbers with 
$\log N_{\rm HI} = 15.07 - 15.24$ at redshifts $z_{\rm abs} = 0.2625$, 0.4140, and 0.5061.  All three systems 
were easily detected in \Lyb, and two of them have \Lya\ (for the absorber at $z = 0.5061$,  \Lya\ has 
shifted beyond the G160M window).  From the expected detection limits (Table~1) in COS spectra like those
shown in Figure 3, our survey is able to identify strong absorbers down to $\log N_{\rm HI} \approx 15$.  
In some AGN sight lines with low S/N ($\sim5$), particularly those at redshifts ($z > 0.75$) where we 
lose \Lyb, we may miss some absorbers at $\log N_{\rm HI} = 15.0-15.5$.

We began with the 221 strong absorption systems identified in the 159 AGN sight lines studied by Stevans \etal\ 
(2014).  Because the LL shifts into the G130M band at $z \geq 0.24$, we only searched for strong absorbers 
toward those 102 quasars with redshifts $z_{\rm AGN} \geq 0.24$.  For our statistical sample, we dropped a weak 
absorber at low column density ($\log N_{\rm HI} = 13.5$).  One LLS had a redshift ($z = 0.2374$) just below
our cutoff at $z = 0.24$.  We carefully re-analyzed all systems with $\log N_{\rm HI} \geq 15.75$, combining a 
multi-component CoG with measurements of the Ly decrement, when detectable.  To derive $\tau_{\rm LL}$ we 
examined the AGN spectrum for broad emission-line contamination of the continuum on either side of the LL.  
In our re-analysis, we found an additional strong absorber omitted in Stevans \etal\ (2014), a DLA at 
$z_{\rm abs} = 0.3221$.  Over redshifts $0.237 \leq z _{\rm abs} \leq 0.928$ we found eight true LLS 
($\log N_{\rm HI} \geq 17.2$), one DLA ($\log N_{\rm HI} = 20.34 \pm 0.12$), and 54 pLLS 
($16.0 \leq \log N_{\rm HI} < 17.2$).  Including the lower redshift portions of the spectrum ($z < 0.24$) we found
a total of four DLAs towards all 159 surveyed AGN, at redshifts $z_{\rm DLA} = 0.0963$, 0.1140, 0.185, and 0.3211.  
Appendix A presents a narrative discussion of 73 systems with $\log N_{\rm HI} \geq 15.75$, many of which 
exhibit multiple velocity components separated by $\Delta v \approx 40-400$~\kms.

\subsection{Measuring $N_{\rm HI}$ from Lyman Lines and Lyman-Edge} 

In a HST/COS survey of \HI\ column densities in the low-redshift IGM along 82  AGN sight lines,  Danforth \etal\ 
(2016) found 2577 distinct \HI\ absorption systems, some of them single-line (\Lya) systems.  For statistical analysis, 
they defined a ``uniform sample" of 2256 systems in which $N_{\rm HI}$ was found by multi-line CoG analysis, using 
either (\Lya\ + \Lyb)  or (\Lyb\ + \Lyg) at a minimum for $\log N_{\rm HI} \geq 13.5$, and \Lya\ alone for well-measured 
weaker lines.  Of these 2256 absorbers, 65  had column densities $N_{\rm HI} \geq 10^{15}$~cm$^{-2}$.  These column 
densities were determined by a traditional CoG, which works best for absorption lines that are unsaturated or mildly 
saturated.  At $N_{\rm HI} < 10^{13.5}$~cm$^{-2}$, IGM surveys typically rely on \Lya\ absorbers, since \Lyb\ is too weak 
to detect at typical COS sensitivity (15-20~m\AA\ equivalent widths).   When higher Lyman lines become available at 
higher $z$ and greater $N_{\rm HI}$, the CoG yields reliable parameters ($N, b$).  \\

Once the Lyman lines become strongly 
saturated, with equivalent widths on the ``flat portion" of the CoG, the inferred column densities are more uncertain.  
The onset of saturation is gauged by the line-center optical depth, $\tau_0 = (\pi e^2 / m_e c) (N f \lambda / \pi^{1/2} b)$, 
where $b = (25~{\rm km~s}^{-1}) b_{25}$ is a typical doppler parameter.  For the first four Lyman lines, these optical 
depths are
\begin{eqnarray}
   \tau_0({\rm Ly}\alpha)     &  = &  (30.3) N_{15}\,  b_{25}^{-1}  \; \; \; \; \; \; 
   \tau_0({\rm Ly}\beta)        =    (4.86) N_{15} \, b_{25}^{-1}  \nonumber  \\
   \tau_0({\rm Ly}\gamma)  & =  &  (1.69)   N_{15}\,  b_{25}^{-1}  \; \; \; \; \; \; 
   \tau_0({\rm Ly}\delta)        =   (0.793)  N_{15} \, b_{25}^{-1}  \; , 
 \end{eqnarray}
for $N_{\rm HI}$  scaled to $(10^{15}$ cm$^{-2})N_{15}$.   As long as some Lyman lines remain near the 
linear portion of the CoG, this method works well.  For $b \approx 25$~\kms, the \Lya\ line begins to saturate 
($\tau_0 > 1$ at $\log N_{\rm HI} >13.5$) producing large uncertainties at $14.5 < \log N_{\rm HI} < 16.5$.  \\
 
In this paper, we focus on the stronger \HI\ absorbers, using their Lyman decrements to derive accurate column 
densities over the range $16.2 \leq \log N_{\rm HI} \leq 17.85$.  With our Lyman-comb technique and $S/N \geq 10$, 
we can detect the Lyman decrement at optical depths $\tau_{\rm LL} \geq 0.1$ corresponding to 
$\log N_{\rm HI} \geq 16.2$.  Once $\log N_{\rm HI} \geq 17.85$ ($\tau_{\rm LL} > 4.46$) it becomes difficult to detect 
transmitted flux in the Lyman continuum.  In a few cases with high S/N ($\gsim 20$) we detected or limited the
residual flux transmission equivalent to $\tau_{\rm LL} \geq 5.0$ ($\log N_{\rm HI} \geq 17.9$).  For high column 
density systems, we can constrain $N_{\rm HI}$ from damping wings in the \Lya\ line, when present (\Lya\ shifts 
out of the COS/G160M band at $z \gsim 0.47$.)  By combining the Lyman decrement (for $\log N_{\rm HI}  > 16.2$) 
with CoG methods on higher Lyman lines, typically up to Ly12 and occasionally to Ly15, we obtain more accurate 
column densities than with CoG alone.   \\

The CoG fitting uses a series of Voigt profiles convolved with an appropriate 
COS line spread function\footnote{www.stsci.edu/hst/cos/performance/spectral$\_$resolution}.  
Widely separated velocity components are identified semi-automatically (see Danforth \etal\ 2016), but closely 
blended components require interactive identification and fitting.  We compare the models to the observed spectrum 
in normalized flux space via a $\chi^2$ minimization package {\sc mpfit} (Markwardt 2009) with equivalent widths fitted 
to line profiles of each component (not to the observed flux).   Moderately saturated lines with simple component 
structure are well constrained by this method, and the CoG gives much better $(N, b)$ solutions for \HI\ than a 
single-line profile fit.   \\

Figures 5--8 illustrate our technique for Lyman-series and Lyman decrement absorption for two LLS with 
$\log N = 17.85 \pm 0.02$ and $\log N = 17.67 \pm 0.10$ and two pLLS absorbers with
$\log N_{\rm HI} = 16.41 \pm 0.03$ and $17.01 \pm 0.05$.  The presence of a Lyman decrement typically yields 
$\log N_{\rm HI}$ accurate to $\pm 0.05$ or better.  Figure 9 shows spectra of the other eight strong \HI\ absorbers
with $\log N_{\rm HI} > 17.0$.  
Strong \HI\ absorbers  are often composed of multiple, blended components.  Lower-order Lyman lines are typically 
too strong to see blended components, but higher-order lines can reveal their presence.   
Absorbers where the minimized $\chi^2$ solution fails to match the data may harbor unresolved component structure. 
For example, in the strong absorber toward SBS\,1108$+$560 (Figure~5), a weaker component is seen in the blue 
wing of \Lye\ and higher lines.  When blended components are present, we fit a CoG to each component, using only 
lines in which they are unambiguously separable, eliminating lines contaminated with airglow emission or unrelated 
absorption.  Line profiles and total column densities for the combined solution (e.g., $N_1, b_1, z_1$ and $N_2, b_2, z_2$) 
are then calculated and compared qualitatively to the stronger, lower-order Lyman lines.  In several cases, the CoG solution 
does not reproduce the observed line profiles, or it differs from the Lyman decrement.  The CoG is determined from the 
measured equivalent widths of the lines.   Sometimes a solution with a smaller $b$ and larger $N$ (or vice versa) is 
required to match the observed line profiles or decrement.   

 \subsection{Lyman line overlap and velocity components}  
 
The CoG techniques generally give accurate results with the availability of higher Lyman lines.  However, line overlap sets 
in at Ly15 ($\lambda = 915.329$~\AA) or Ly16 ($\lambda = 914.919$~\AA) as the higher Lyman series converges on the 
Lyman limit at $\lambda_{\rm LL} = 911.753$~\AA.  Line crowding and uncertain continuum placement makemeasurements
of equivalent widths difficult when the wavelength separation, $\Delta \lambda_{n,n+1}$, between sequential Lyman lines 
is comparable to their line width.  Table~2 shows line separations and line-center optical depths for Ly12 - Ly24,  scaled to 
the ratio, $N_{17}/b_{25}$, for column densities $N_{\rm HI} = (10^{17}~{\rm cm}^{-2}) N_{17}$.  The lines are distinguishable 
up to Ly15, where $\Delta \lambda_{n,n+1} \approx 0.5$~\AA.  For Gaussian line profiles, the full width at half maximum is 
$\Delta \lambda_{\rm FWHM} = 2(\ln 2)^{1/2}(\Delta \lambda_D) \approx (0.127~{\rm \AA})\, b_{25} \, \lambda_{914}$, 
where $\lambda_{914}$ is a typical (L15 - Ly20) wavelength in units of 914~\AA\  and $\Delta \lambda_D = \lambda_0 (b/c)$ is
the doppler width.   Severe overlap sets in above Ly20, where separations become less than 0.2~\AA.  At this point, higher 
Lyman lines overlap in their wings, 10\% below the continuum, defined by width 
$\Delta \lambda_{\rm 10\%} = 2 (\ln 10)^{1/2} (\Delta \lambda_D) \approx (0.231~{\rm \AA})\, b_{25} \, \lambda_{914}$.   
A few absorbers have  $b = 40 - 50$~\kms, with wing overlap affecting Ly 14 - Ly17 at line separations of 0.4 - 0.5~\AA. 

Overlap creates difficulties in measuring equivalent widths, with offsetting effects of shared line absorption and continuum 
placement.  Multiple velocity components complicate the problem further., and a proper treatment requires multi-line radiative
transfer.   Therefore, we do not include lines above Ly15 in our analysis, as illustrated in Figure~10 for the absorber toward 
PKS\,0552-640.  The standard CoG up to Ly15 gives  $\log N_{\rm HI} = 17.01\pm0.05$, whereas including additional 
(overlapping) lines from Ly16 to Ly24 gives an erroneous fit with $\log N_{\rm HI} = 16.81\pm0.02$.  In this case, the 
observed Lyman decrement provides an accurate column density, $\log N_{\rm HI} = 17.01\pm0.03$, verifying the CoG 
solution up to Ly15.  

Appendix A provides narrative discussion for our analysis of 73 strong \HI\ absorbers with $\log N_{\rm HI} > 15.75$.   
For several absorbers with $\log N_{\rm HI} > 16.25$, our combination of CoG fits and Ly decrement measurements 
uncovered a few discrepancies with previous values in the literature (Lehner \etal\ 2013; Fox \etal\ 2013; Stevans \etal\ 
2014).  For four systems with well-resolved velocity components separated by 150-200 \kms\ or greater, we treated the
components as separate absorbers:   Systems \#24abc, \#47abc, \#59ab, and \#60ab.



\begin{figure*}
\includegraphics[angle=0,scale=0.70]{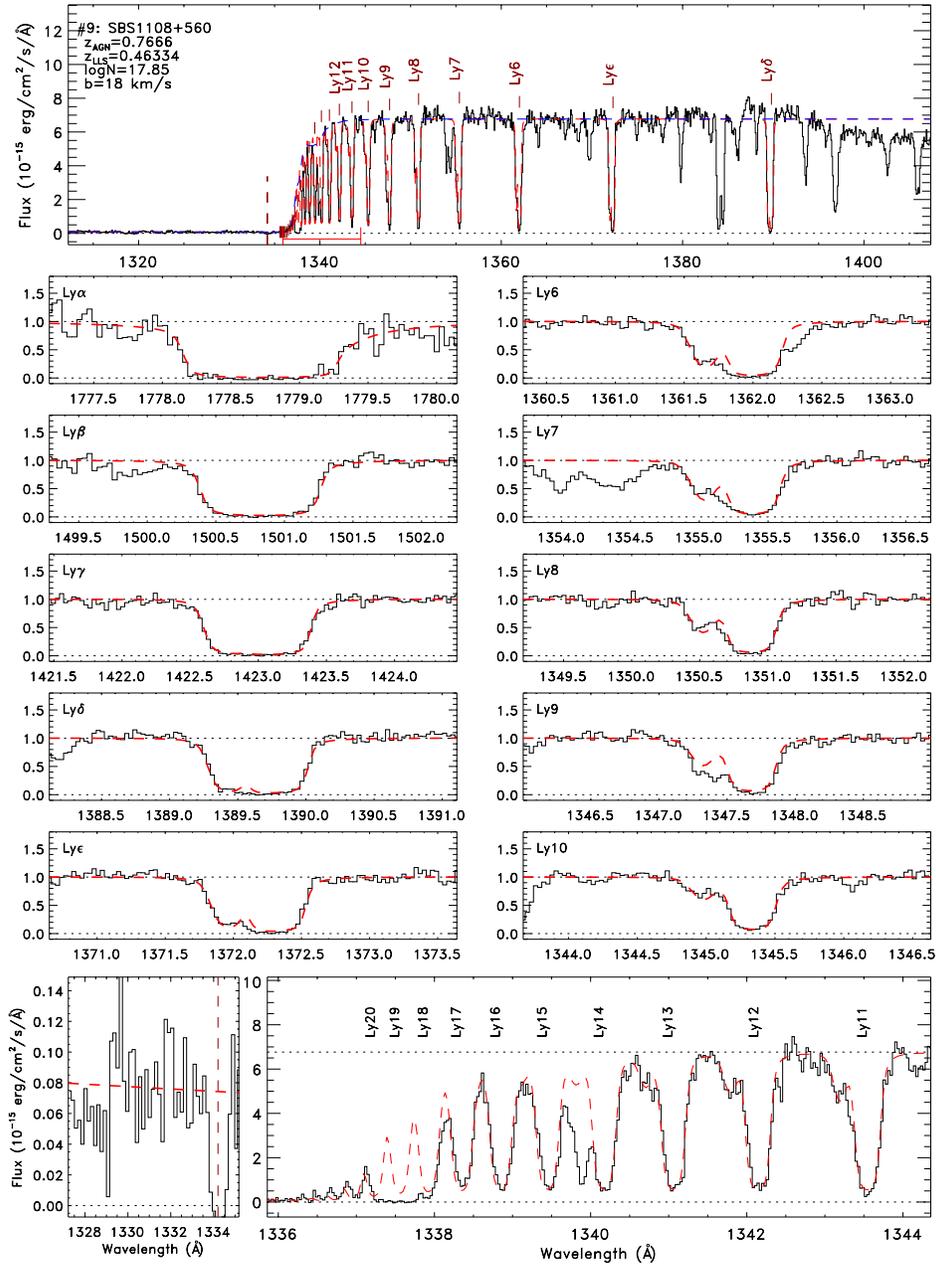}
\caption{\small{ (Top) Normalized COS/G130M spectrum of SBS\,1108+560 showing lines of System \#9 at 
$z_{\rm abs} = 0.46334$, a LLS with Lyman lines (Ly6 - Ly12) converging on LL at 1334.2~\AA\ (vertical red 
dashed line).  Red dashed lines in spectrum show model with column density $\log N_{\rm HI} = 17.85\pm0.02$ 
and doppler parameter $b = 18\pm2$~\kms, not fits to individual lines.  (Middle panels)  Line profiles of Lyman 
lines (\Lya\ - Ly10).  (Bottom) Higher Lyman lines (Ly11 - Ly20) and transmitted flux in LyC.
Note \CII\  $\lambda1334.53$ interstellar absorption in the LyC (bottom left panel).  }
}
\end{figure*}



\begin{figure*}
\includegraphics[angle=0,scale=0.70]{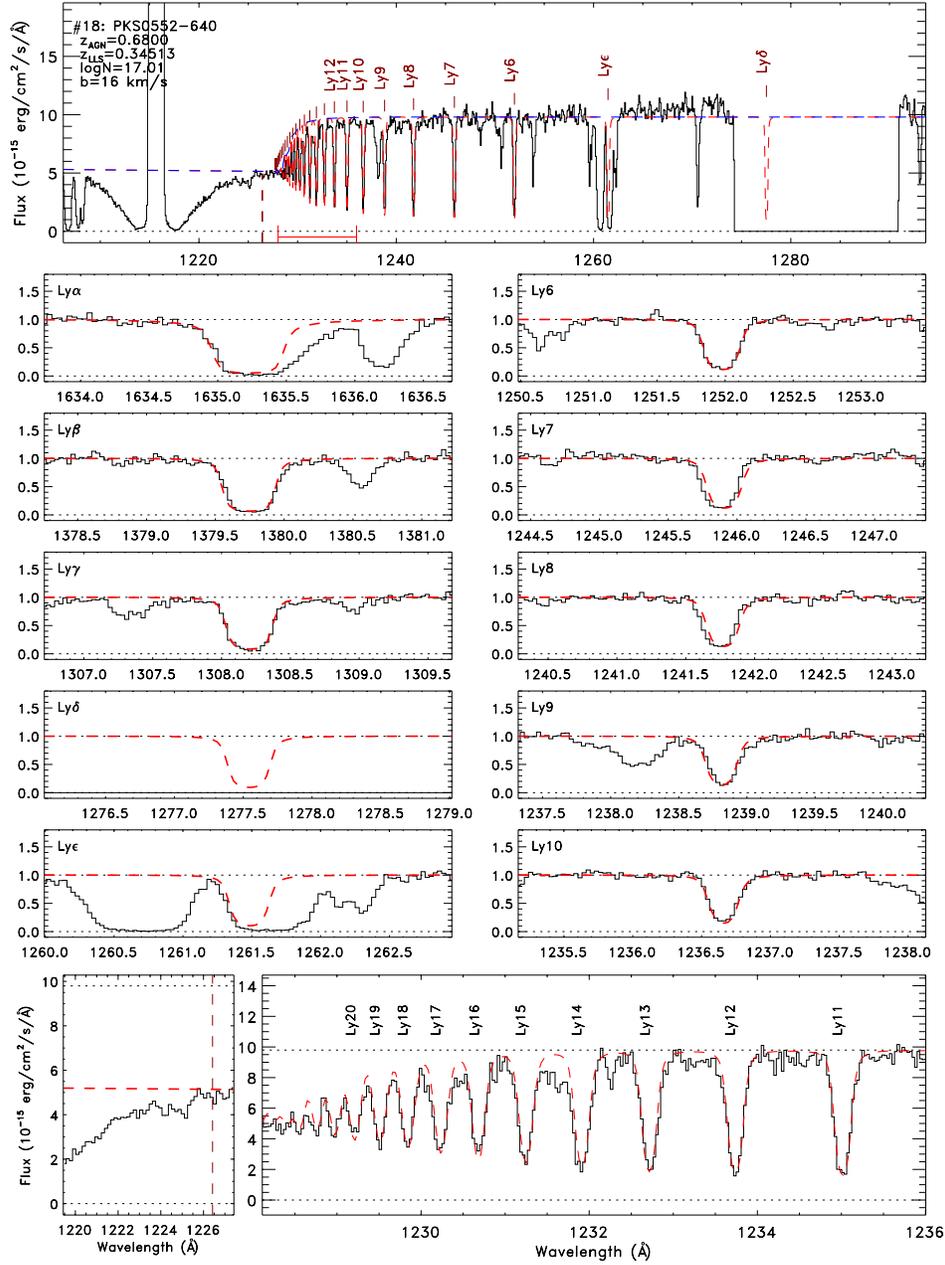}
\caption{\small{ Same as Figure 5 for pLLS toward PKS\,0552-640 (System \#18 at $z_{\rm abs} = 0.34513$)
with LL at 1226.4~\AA.   Redward wing of Galactic \Lya\ absorption extends from 1216-1226~\AA.   Red dashed
lines in spectrum show profiles of Lyman lines (\Lya\ - Ly10) for model with $\log N_{\rm HI} = 17.01\pm0.05$ 
and $b = 16\pm2$~\kms.   (Bottom) Higher Lyman-series lines (Ly11 - Ly20) and transmitted flux in LyC above
the Galactic DLA.  }
}
\end{figure*}



\begin{figure*}
\includegraphics[angle=0,scale=0.70]{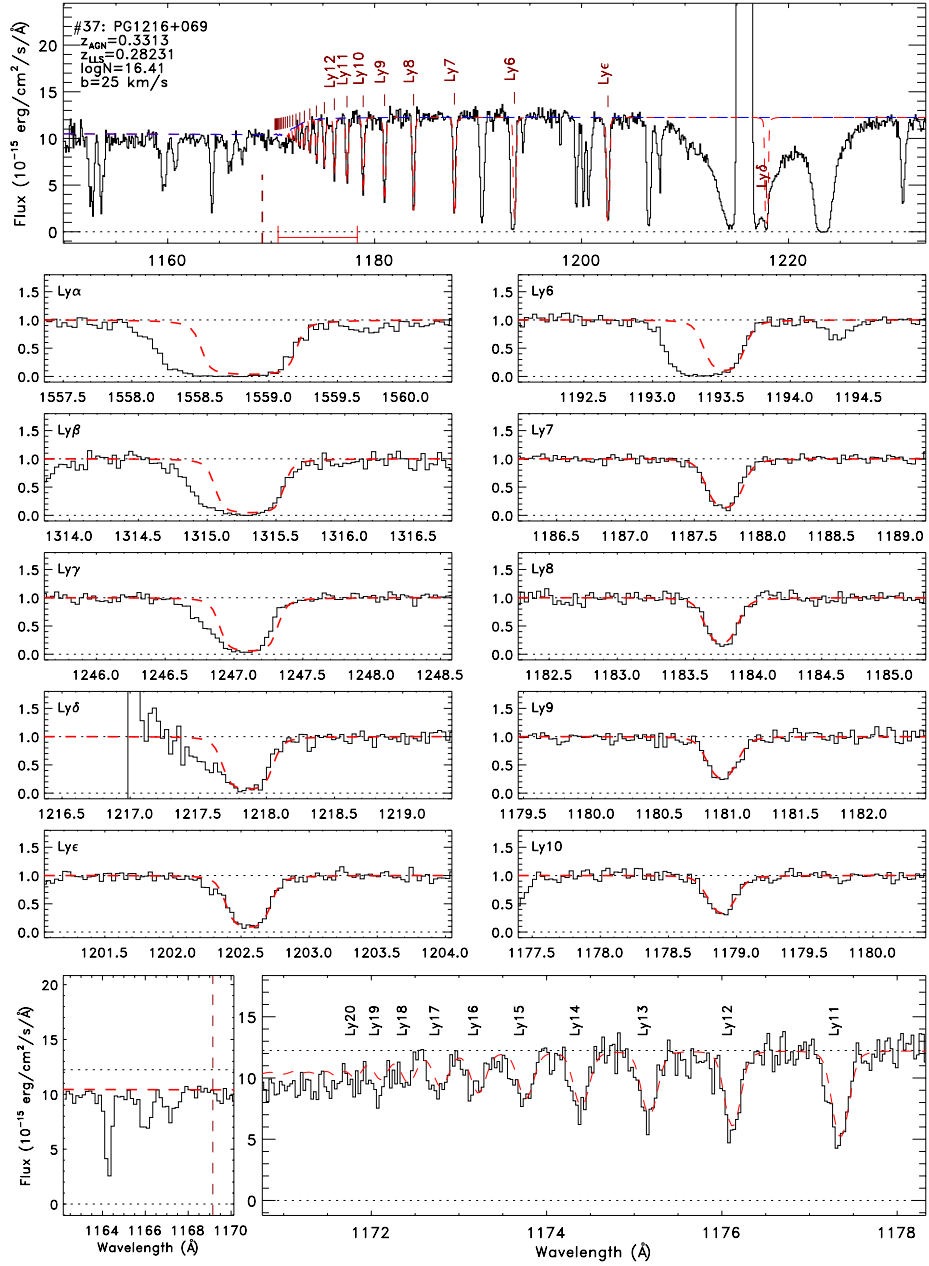}
\caption{\small{ Same as Figure 5 for pLLS toward PG\,1216+069 (System \#37 at $z_{\rm abs} = 0.28231$).
Absorption lines converge on LL at 1161.15~\AA.  Red dashed lines (\Lye\ - Ly10) show strongest component 
fitted to CoG with $\log N_{\rm HI} = 16.41 \pm 0.03$ and $b = 25\pm5$ km~s$^{-1}$.  Lyman decrement gives 
$\tau_{\rm LL} = 0.16\pm0.02$ or $\log N_{\rm HI} = 16.40 \pm 0.06$.  
(Bottom) Higher Lyman lines  (Ly11 - Ly20) and transmitted flux in LyC.  }
 }
\end{figure*}



\begin{figure*}
\includegraphics[angle=0,scale=0.70]{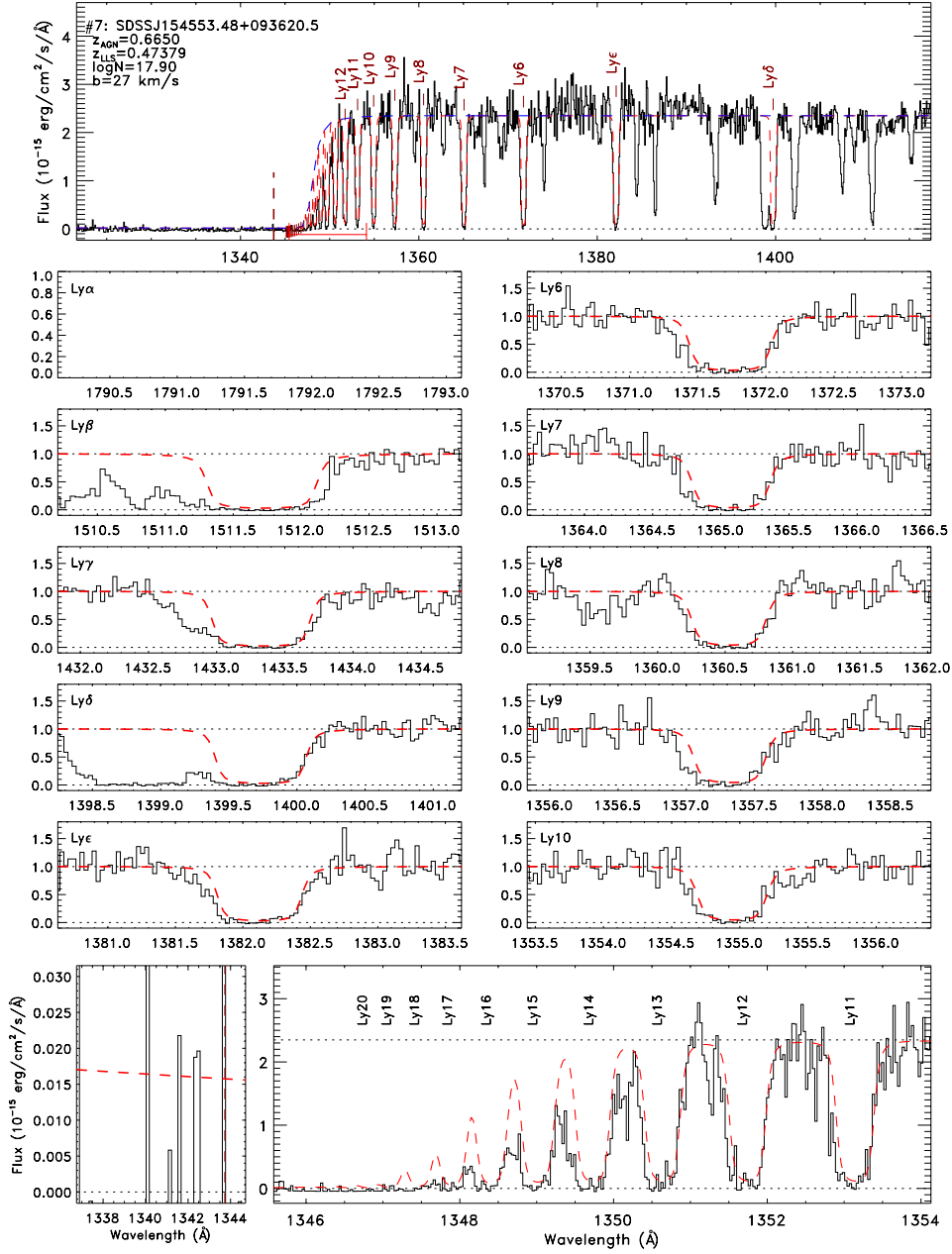}
\caption{\small { Same as Figure 5 for LLS toward SDSS~J154553.48+093620.5 (System \#7 at 
$z_{\rm abs} = 0.47379$).  Absorption lines of Ly6 - Ly12 converge on LL at 1343.7~\AA.   
Red dashed lines (\Lyb\ - Ly10) show strongest component fitted to CoG with
$\log N_{\rm HI} = 17.67 \pm 0.15$ and $b = 35 \pm 1$ km~s$^{-1}$.  This column density would 
give $\tau_{\rm LL} = 2.95$ and transmitted flux of 5.25\%, inconsistent with observations of flux 
below the LL (bottom panels) which imply $\log N_{\rm HI} \geq 17.9$. } 
 }

\end{figure*}



\begin{figure*}
\includegraphics[angle=90,scale=0.70]{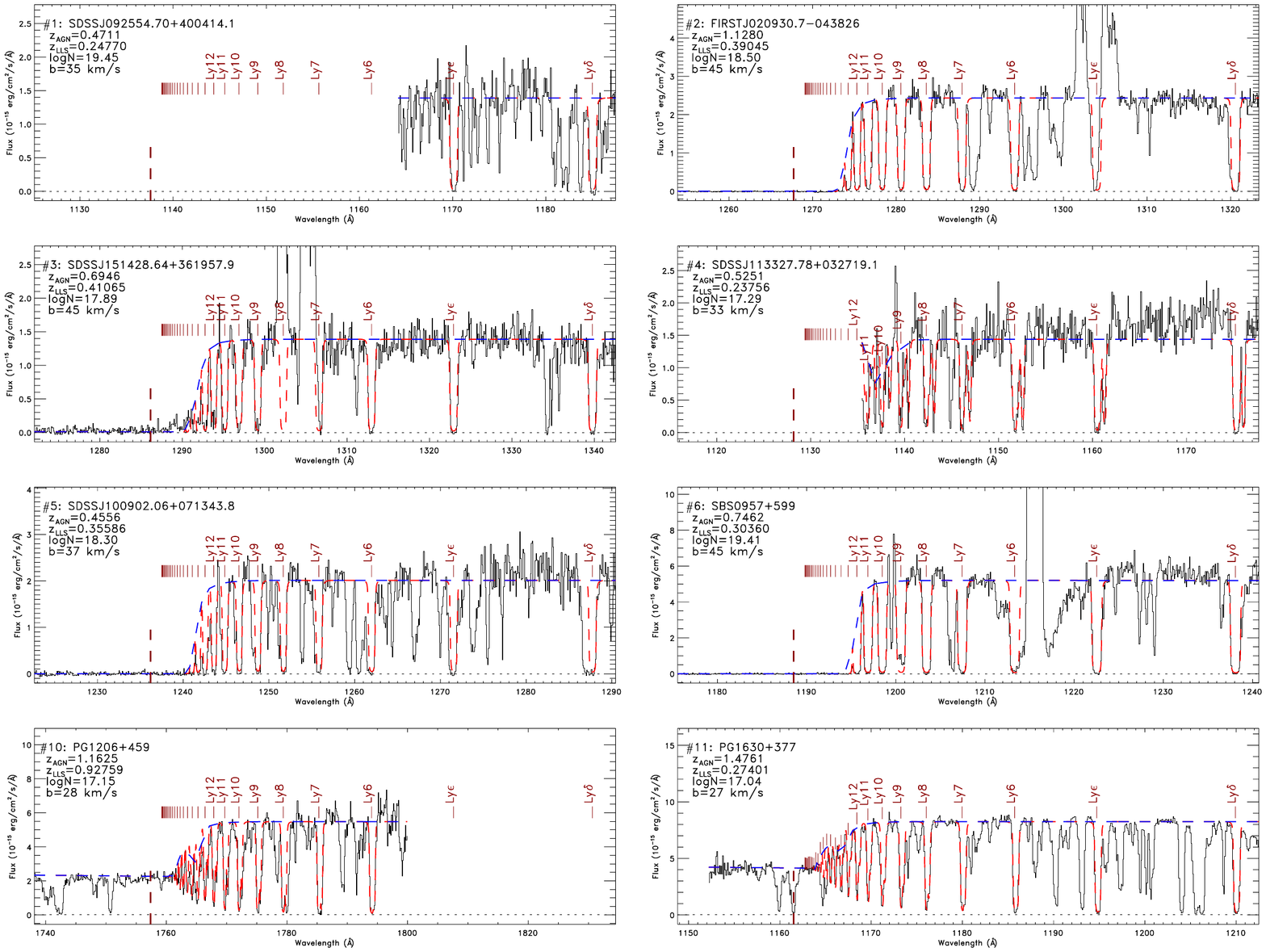}
\caption{\small { COS Spectra of eight strong \HI\ absorbers ($\log N_{\rm HI} > 17.0$), listed in Appendix~A
as systems \#1--6, \#10,  \#11 (see labels in top-left corner).  These include six LLS and two pLLS 
($\log N_{\rm HI} = 17.15$ and 17.04).  
Other strong absorbers were shown in Figures 4, 5, and 7.  Spectra for Systems \#1 and \#10 are limited by 
cutoffs of the settings of the G130M and G160M gratings.  
Additional Lyman lines beyond those shown were available in many cases. } 
 }

\end{figure*}




\begin{figure}
\includegraphics[angle=0,scale=0.70] {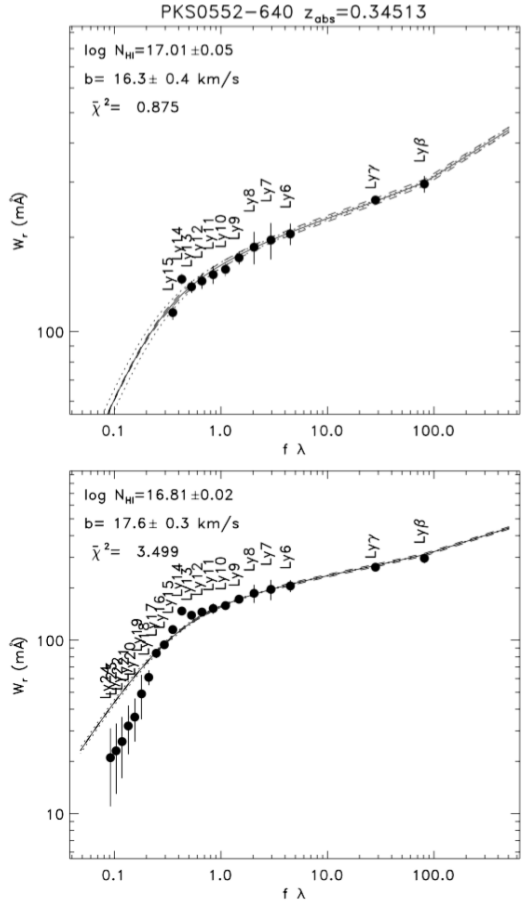}
\caption{ \small { (Top) Curve of growth of the Lyman series (\Lyb\ through Ly15) for pLLS toward 
PKS\,0552-640 at $z_{\rm abs} = 0.34513$ (see Figure~6). Equivalent widths ($W_{\lambda}$ in m\AA), 
are plotted vs.\ line strengths, $f \lambda$, with oscillator strengths $f$ and wavelengths $\lambda$(\AA)
from Morton (2003).  The best fit gives $\log N_{\rm HI} = 17.01 \pm 0.05$ with $b = 16.3\pm0.4$~\kms\
(reduced $\chi^2$ noted in header.)  
(Bottom) CoG including lines of Ly16 - Ly24.  Line overlap and crowding result in reduced equivalent widths 
and an erroneous fit with $\log N_{\rm HI} = 16.81\pm0.02$ and $b = 17.6\pm0.3$~\kms. }
 }
\end{figure}


\section{RESULTS }
 
To analyze the bivariate distribution, $f(N,z)$, of \HI\ absorbers, we group them into a binned array, $F(i,j)$, 
shown in Table~3, with redshift indices $(i = 1-15)$ and column density indices $(j = 1-15)$. The 15 redshift  
bins have equal width $\Delta z = 0.04$ from $z = 0.24-0.84$.  The first 12 column-density bins have width 
$\Delta (\log N_{\rm HI}) = 0.25$ spanning $14 < \log N_{\rm HI} \leq 17$.  The last three bins 
($j = 13, 14, 15$) are wider and cover the 10 strongest absorbers with $\log N_{\rm HI} = 17.0-20.5$.  
As shown in Table~1 (column 4), the higher Lyman lines (\Lyg, \Lyd, \Lye) are easily detectable, at 50~m\AA\ 
equivalent width, for column densities  $\log N_{\rm HI} \geq 14.3-14.9$.  Because the best COS data have 
S/N $>20$, we have regularly detected the first three Lyman lines (\Lya, \Lyb, \Lyd) and often even higher 
Lyman lines (Figure~3).  This allows us to identify absorbers with $\log N_{\rm HI} = 14-15$, even though the 
Lyman edge  is undetectable at these column densities.  The LLS and stronger pLLS absorbers with 
$\log N_{\rm HI} > 16.5$ are typically observed in higher Lyman lines up to Ly12 and often beyond 
(see Figures 5--9), and they yield precise \HI\ column densities via CoG.  At higher redshifts, the two strongest 
Lyman lines move out of the G160M window (\Lya\ at $z > 0.47$ and \Lyb\ at $z > 0.75$).   Because our higher
redshift \HI\ detections rely on \Lyg\ and higher lines, we could miss a few absorbers with 
$\log N_{\rm HI} = 15.0-15.5$.  
Column densities for strong absorbers can sometimes be influenced by velocity components.  In well-exposed 
COS spectra, we can identify components with $\Delta v = 40-400$~\kms, all of which contribute to the Lyman 
decrement.  We derive individual column densities with multi-component CoGs, which could affect absorber 
counts for the bins between $\log N_{\rm HI} = 14.75 - 16.00$.  We have taken a conservative approach, 
only splitting the velocity components in four well-separated systems with $\Delta v \geq 150-200$~\kms.  

The observed distribution in column density (Table~3) exhibits the expected falloff in numbers at high 
column densities owing to their scarcity.  We believe the decrease in absorber numbers in bins $j = 1-4$ 
($\log N_{\rm HI} = 14-15$) arises primarily from the lower detection sensitivity of weak Lyman lines at higher 
redshift.  Because the LL is not detectable at $\log N_{\rm HI} < 16.2$, we rely on finding a pattern of higher
Lyman lines (\Lyg, \Lyd, \Lye) whose detection in the G160M window requires $\log N_{\rm HI} \geq 14.3-14.9$ 
(Table 1).  These higher Lyman lines are easily detected in G130M (see Figure 3), but at $z > 0.5$ we rely on
G160M and higher (weaker) lines in the Lyman series.
For this reason, we focus our attention on pLLS and LLS absorbers with $\log N_{\rm HI} \geq 16.0$.  Our 
statistical analysis in Appendix B restricts the \HI\ absorber sample to the range $0.24 \leq z_{\rm abs} \leq 0.84$ and 
$15 \leq \log N_{\rm HI} \leq 20$, where we feel confident in detecting most systems in AGN sight lines with 
well-exposed COS spectra (S/N $ > 10$).  Our LLS sample is restricted to $0.24 \leq z \leq 0.48$, as described
in Section 3.3.   This technique differs from that in our recent IGM survey (Danforth \etal\ 2016), which detected 
primarily weak \HI\ absorbers through \Lya\ lines.  
 
 \subsection{Redshift Coverage per Bin}  
 
  The total (effective) redshift pathlength, $\Delta z_{\rm eff}^{\rm (tot)}$, is found from the spectral coverage of the 
  AGN in our survey.  We chose redshift bins of width $\Delta z = 0.04$, starting at bin~1 ($0.24 < z < 0.28$) where 
  the Lyman edge at wavelength $\lambda_{\rm LL} = (911.753~{\rm \AA})(1+z)$ first falls within the range of most
 COS/G130M data (1134--1459~\AA).   Table~4 shows the number of AGN with sensitivity to detecting a LL in 
 each redshift bin.  Each AGN with redshift above the bin contributes pathlength $\Delta z = 0.04$, plus partial 
 redshift coverage for a few AGN whose redshifts fall within the bin.  The redshift-bin pathlengths, 
 $\Delta z_{\rm eff}^{(i)}$, are shown in column~5 of Table~4.  To determine this redshift path, we subtracted a few 
 spectral regions blocked by strong absorbers ($\log N_{\rm HI} > 17$) along ten AGN sight lines shown in Table~5.  
 In practice, only a few of the strongest LLS absorbers, with $\log N_{\rm HI} > 17.5$, produce significant blockage.  
 Strong foreground absorption by the Galactic interstellar \HI\ (\Lya) does not impact $\Delta z_{\rm eff}$ or the 
 Lyman comb, since it usually blocks only one of the Lyman lines.  
 
 With the far-UV spectral coverage between 1134 \AA\ and 1795~\AA\ and availability of many lines in the Lyman 
 series, we were able to detect strong \HI\ systems between $z_{\rm abs} = 0.237-0.928$.   Because of the reduced 
 spectral coverage of higher-redshift AGN, we limited our analysis to 15 redshift bins between $z = 0.24-0.84$.  
 Sensitivity to higher Lyman lines declines at $z > 0.846$ as \Lyg\ shifts out of the COS/G160M band.   The total 
 pathlength in this sample (Table 4) is $\Delta z_{\rm eff}^{\rm (tot)} = 31.94$.  For individual bins, the pathlength 
 decreases from $\Delta z_{\rm eff}^{(i)} = 3.83$ for bin~1 ($\bar{z} = 0.26$ and $\lambda_{\rm LL} = 1149$~\AA) 
 to $\Delta z_{\rm eff} ^{(i)}= 2.80$ for bin 5 ($\bar{z} = 0.42$ and $\lambda_{\rm LL} = 1295$~\AA).  Although bin~5 
 covers the redshift ($z = 0.423$) at which \Lyb\ shifts out of G130M, spectral overlap with G160M (1400--1459~\AA) 
 allows \Lyb\ to be continuously observed to $z = 0.75$.  The Lyman edge shifts out of G130M beyond bin~9 
 ($0.56 < z < 0.60$), and the AGN pathlength drops to $\Delta z_{\rm eff}^{(i)} \approx 0.74-0.96$ in bins~13--15.   
 With G160M, we have detected pLLS out to $z = 0.928$.  Wavelength overlap between G130M and G160M gratings 
 (1400--1459~\AA) slightly enhances our ability for discovering \Lya\ absorbers at $z = 0.15-0.20$,  \Lyb\ absorbers at 
 $z = 0.36-0.42$, and \Lyg\ absorbers at $z = 0.44-0.50$.  However, this effect is minor.  Because pLLS identification
 requires finding a pattern of several Lyman lines, our survey is most sensitive between $z \approx 0.24-0.84$.  

Selection biases arise toward several high-$z$ AGN that are less likely to contain strong LLSs and DLAs (see
Sections 3.3 and 3.4). Our sample of 159 AGN (Stevans \etal\ 2014) contains 29 targets at $z > 0.70$ and 43 at 
$z > 0.60$, many of them observed previously by \IUE,  \HST, and \FUSE.  To be detected in the G130M/G160M 
gratings, these AGN were ``UV-qualified" for sufficient far-UV flux, usually by \GALEX\ or \IUE.  A number of these 
AGN came from the COS-Halos project (Tumlinson \etal\ 2013) whose QSOs were selected to have high fluxes in 
the \GALEX\ far-UV band and avoid strong \MgII\ absorbers at $z_{\rm abs} > 0.4$.   

 \subsection{Bivariate Distribution in $\log N$ and $z$}
 
Table~6 lists 41 strong \HI\ absorbers with $\log N_{\rm HI} \geq 16.0$, classified as LLS or pLLS (systems \#1-40 and 
\#73 from Appendix~A).  The last columns compare our column densities to previous measurements from Lehner \etal\ 
(2013) and estimates from Stevans \etal\ (2014).  The Lehner values are the same as those reported by Fox \etal\ (2013).   
Those from Stevans \etal\ (2014) were estimates, many of which were revised in this paper after careful analysis of CoGs 
and Lyman decrements.  For 158 absorbers above our expected detection limit ($\log N_{\rm HI} \geq 15.0$) the absorber 
frequency per redshift is $d {\cal N} / dz  \approx 4.95 \pm 0.39$ averaged over $0.24 < z < 0.84$.  We see some indication 
of an increase in frequency with redshift.  Figure~11 shows the distribution in column density for our pLLS data, together with 
the \Lya\ forest distribution from Danforth \etal\ (2016).  Some offset is expected, as the COS \Lya\ forest survey covers 
$z < 0.47$ with median redshift $z_{\rm abs} = 0.14$, whereas our pLLS survey spans $0.24 < z < 0.93$ with median 
$z_{\rm abs} = 0.43$.  The eight detected LLS range from $z = 0.237$ to 0.474, with median redshift $z_{\rm abs} = 0.39$.  
The absence of LLS at $z > 0.48$ is surprising statistically, as we expected to detect $3.6^{+3.1}_{-1.8}$ LLS between 
$0.48 < z < 0.84$ based on the derived redshift evolution (Section 3.3).  This may be evidence for a bias toward AGN with 
high far-UV fluxes, unblocked by LLS at $z > 0.5$.  We do not see this effect for the stronger pLLS with 
$\log N_{\rm HI} = 16.5-17.0$.  For pLLS statistics, we use the full redshift range ($0.24 \leq z \leq 0.84$) but restrict the 
LLS sample to bins 1--6 ($0.24 \leq z \leq 0.48$).


\begin{figure}
 \includegraphics[angle=0,scale=0.3] {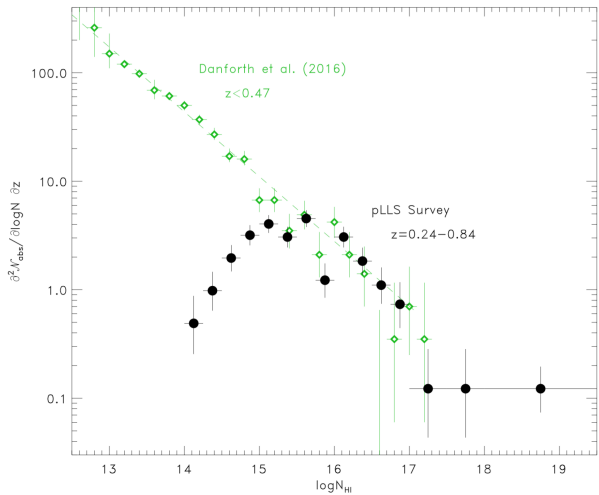} 
\caption{\small{ Distribution of \HI\ absorbers in column density (per $\log N_{\rm HI}$) over total redshift
pathlength $\Delta z_{\rm eff} = 31.94$ between $0.24 \leq z \leq 0.84$.  Solid diamonds (green) are from 
the HST/COS survey (Danforth \etal\ 2016) of \Lya\ forest at $0 < z < 0.47$, and filled circles (black) show 
the current survey of pLLS and LLS at $0.24 < z < 0.84$.   Green dotted line is a least-squares fit to the 
differential distribution, $f(N,z) \propto N^{-\beta}$, with $\beta = 1.60\pm0.02$. } 
 }
\end{figure}


The \Lya\ forest and pLLS distributions match fairly well at $\log N_{\rm HI} \geq16$.   A possible turnover in the pLLS 
distribution appears at $\log N_{\rm HI} < 15.5$, which could mark the onset of survey incompleteness.  Because of the 
small number of absorbers, one expects fluctuations in the range $\log N_{\rm HI} = 15.0 - 16.0$ (bins $j = 5-8$ in Table~3).  
The two surveys agree for bin~7 (15.50--15.75) and bin 9 (16.00--16.25), whereas bin 8 (15.75--16.00) has a low number of 
absorbers that may arise from the finite spacings in our survey grid, $\Delta z = 0.04$, $\Delta (\log N_{\rm HI}) = 0.25$.   
As noted above and in Table~1, we should be able to identify most absorbers between $15.0 < \log N_{\rm HI} < 15.5$.   
However, a small decrease in detection efficiency of weak absorbers could occur at $z > 0.47$, when \Lya\ shifts out of 
the G160M window.   The absorber numbers in bins 6--9 could also be affected by velocity components within strong 
systems, which separate into distinct absorbers.  The distinction between ``systems" and ``components" is a subtle one  
that we have not pursued beyond the obvious cases in Systems \#24abc, \#47abc, \#59ab, and \#60ab (see Appendix~A).

To analyze the bivariate differential distribution of absorbers in \HI\ column density and frequency, we have applied a 
maximum-likelihood analysis (Appendix B) to the parameterized form 
$f(N,z) \equiv d^2 {\cal N} / dz \, dN_{\rm HI} = C_0 N_{\rm HI}^{-\beta} (1+z)^{\gamma}$.  We fitted all absorbers in the 
ranges $0.24 \leq z \leq 0.84$ and $15 \leq \log N_{\rm HI} \leq 20$, using Monte-Carlo Markov-Chain (MCMC) simulations 
to derive values $\beta = 1.48 \pm 0.05$ and $\gamma = 1.14^{+0.88}_{-0.89}$.   Figure~12 shows the $2D$ joint probability 
distributions in power-law parameters and their $1\sigma$ error bars.   Both distributions are close to Gaussian, although the 
column-density distribution is much better constrained (3.4\% error in $\beta$) than the redshift evolution index, $\gamma$.  
The uncertainty in $\gamma$ is illustrated in Figure~13 with a histogram of the redshift distribution from $z = 0.24-0.84$.  
This evolution index is similar to those determined in our \Lya\ survey (Danforth \etal\ 2016).  Over redshifts $0 < z < 0.47$,
we found $\gamma_{\rm all} = 1.24\pm0.04$ for all absorbers with $\log N_{\rm HI} > 13$.  Re-fitting the distribution of
65 absorbers at $\log N_{\rm HI} \geq 15.0$, we find  $\gamma = 1.43\pm0.34$.

The normalization ($C_0)$ of our (MCMC) distribution follows from integrating over observed (and fitted) ranges in redshift 
($z_1 < z < z_2$) and column density ($N_1 < N_{\rm HI} < N_2$),
\begin{eqnarray}
       N_{\rm tot} &=&  C_0  \int_{z_1}^{z_2} (1+z)^{\gamma} \, \Delta z_{\rm eff} (z) \, dz 
                 \int_{N_1}^{N_2} N_{\rm HI}^{-\beta} \, dN_{\rm HI}  \nonumber \\
              &=& (4.11\times10^{-6}) \, C_0   \; .  
 \end{eqnarray}  
In evaluating the integral over column density, we take $N_1 = 10^{15}~{\rm cm}^{-2}$, $N_2 = 10^{20}~{\rm cm}^{-2}$, 
and $\beta = 1.48$ to give $1.31 \times 10^{-7}$.  The redshift integral gives 0.983, for $z_1 = 0.24$, $z_2 = 0.84$, 
$\gamma = 1.14$, and assuming $\Delta z_{\rm eff} (z)$ to be constant across the full redshift range with pathlength
$\Delta z_{\rm eff}^{\rm (tot)} = 31.94$.  Thus, we find $C_0 = 3.87 \times 10^7$ for $N_{\rm tot} = 159$ absorbers.   
If we weight $f(N,z)$ by the effective pathlengths, $\Delta z_{\rm eff} (z_i)$, of the 15 individual redshift bins (Table 4), we 
find a slightly larger normalization, $C_0 = 4.12 \times 10^7$.  We adopt $C_0 = 4 \times 10^7$ for $N$(cm$^{-2}$).



\begin{figure}
\includegraphics[angle=0,scale=0.4]{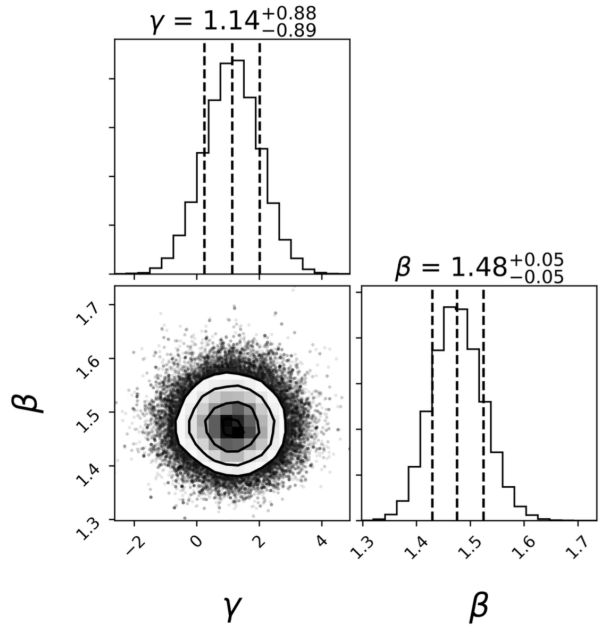}
\caption{\small{ Parameterized evolution of the distribution of \HI\ absorbers in column density and redshift,
$f(N,z) = C_0 N^{-\beta} \, (1+z)^{\gamma}$, from Monte-Carlo Markov Chain (MCMC) samples for power-law 
parameters.  Lower-left panel plots the $2D$ joint-probability of $\gamma$ and $\beta$, and other panels show 
the marginalized distributions with median and $1\sigma$ bounds marked by dashed lines. }
}
\end{figure}



\begin{figure}
\includegraphics[angle=0,scale=0.4]{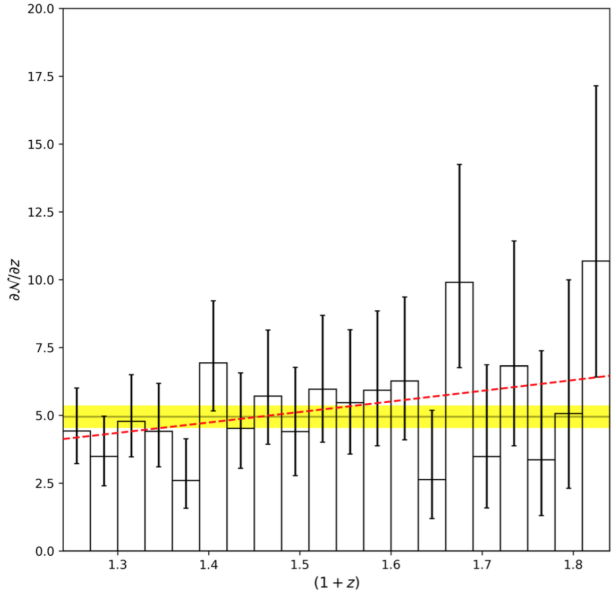}
\caption{\small{ Parameterized evolution of the number of \HI\ absorbers per unit redshift, $d{\cal N}/dz$, for
$z = 0.24-0.84$, integrating  $f(N,z) = C_0 N^{-\beta} \, (1+z)^{\gamma}$ over $15 \leq \log N_{\rm HI} \leq 20$. 
The MCMC realizations yield a fit with considerable uncertainty in index $\gamma = 1.14^{+0.88}_{-0.89}$.   
Horizontal (yellow) band shows the average line frequency, $\langle d {\cal N}/dz \rangle = 4.95 \pm 0.39$, and
red-dashed line is the MCMC fit with $\gamma = 1.14$.  Histogram shows measured number of absorbers per 
unit redshift in 20 equal bins with error bars determined according to Poisson statistics.  A non-integer number of 
absorbers may contribute to bin values because uncertainty in $N_{\rm HI}$ measurements can place a portion
of an absorber's probability density outside the range used in fit. }
}

\end{figure}


\subsection{True Lyman-Limit Systems}  
 
By convention, true LLS are defined as absorbers with $\tau_{\rm LL} \geq 1$ or $\log N_{\rm HI} \geq 17.2$.  Although 
most surveys follow this definition, several recent papers included pLLS absorbers with lower column densities in their lists.  
Because of the scarcity of low-redshift LLS, extending the definition of LLS into the pLLS range creates samples with better
statistics.   In our survey, we retain standard definitions of LLS and pLLS ($16.0 < \log N_{\rm HI} < 17.2$) and analyze a 
total of 158 \HI\ absorbers with $\log N_{\rm HI} \geq 15.0$ between $z = 0.237-0.928$.  Of these, eight are true LLS, one
is a DLA ($\log N_{\rm HI} = 20.34 \pm 0.12$) and 54 are pLLS.  This \HST/COS survey is one of the largest samples of 
strong \HI\ absorbers at low redshift (full range from $0.237 \leq z \leq 0.928$).  The eight LLS are detected through their
Lyman decrements and in multiple Lyman lines using the COS G130M/160M gratings with resolution $R \approx 17,000$. 
The eight LLS in our COS survey are distinct from those in the FOS study, and they range from $z_{\rm abs} = 0.2374-0.4738$ 
with median redshift $z_{\rm abs} = 0.39$. Extrapolating the pLLS distribution (Section 3.2) into the LLS regime, we expect 
a frequency,
\begin{equation}
   \left( \frac {d {\cal N}} {dz} \right)_{\rm LLS}  = C_0 (1+z)^{\gamma}  \int_{N_1}^{N_2}  N^{-\beta} \, dN 
    \approx 0.44 \, (1+z)^{\gamma}  \;  , 
\end{equation}
for $C_0 = 4 \times 10^7$ and $\beta = 1.48$.  For $\gamma = 1.14$, the predicted number of LLS over the full survey,
with $\Delta z_{\rm eff} = 31.94$ for $0.24 \leq z \leq 0.84$, would be $N_{\rm LLS} = 13.8^{+4.8}_{-3.8}$, larger than the 
observed 7 (or 8) LLS.  We found no LLS in redshift bins 7--15 ($0.48 \leq z \leq 0.84$), whereas we would expect 
statistically to observe $N_{\rm LLS} = 3.6^{+3.1}_{-1.8}$.  For the restricted LLS range ($0.24 \leq z \leq 0.64$) in which 
the Lyman edge falls at $\lambda < 1495$~\AA, the observed pathlength $\Delta z_{\rm eff} = 26.82$ over bins 1-10.  
We would then expect to observe $N_{\rm LLS} = 7.2^{+3.8}_{-2.8}$, in agreement with our survey numbers.   This deficit 
suggests that some of the ``high-redshift" ($z > 0.5$) AGN sight lines in our sample are biased {\it against} finding strong
LLS and DLAs that block their far-UV flux in the most sensitive portion (1420--1650~\AA) of the \GALEX\ far-UV band often used in 
QSO target selection.  Many of these AGN have been studied with previous UV spectrographs (\IUE, \FUSE, \HST) based 
on their far-UV brightness.  Although we found no LLS between $0.48 \leq z \leq 0.84$, these AGN sight lines do contain
strong pLLS absorbers ($16.5 \leq \log N_{\rm HI} \leq 17.2$) in the expected numbers. Thus, the AGN bias may only affect 
the statistics for stronger LLS with $\tau_{\rm LL} > 2$.  

To guard against potential LLS bias at $z > 0.5$, we restrict our LLS sample to $z \leq 0.48$ and analyze the seven LLS in
bins 1-6 over surveyed pathlength $\Delta z_{\rm eff} = 19.24$.  This leads to a LLS frequency, 
$d {\cal N}/dz = 0.36^{+0.20}_{-0.13}$, at the median $z_{\rm LLS} = 0.39$. Translating to standard form, 
$d {\cal N}/dz = N_0 (1+z)^{\gamma}$, this frequency corresponds to $N_0 = 0.25^{+0.13}_{-0.09}$, after dividing by 
$(1+z)^{\gamma} \approx 1.46$ at the median redshift with $\gamma = 1.14$.   This LLS frequency is in agreement with 
previous studies (see Table~7 for a summary).  We begin with the \HST/FOS Key Project survey of QSO absorption lines at 
$0.4 < z < 1.4$ (Bahcall \etal\ 1993).  The LLS in that sample were analyzed in two papers that parameterized the redshift 
evolution.  Storrie-Lombardi \etal\ (1994) found $d {\cal N}/dz = (0.27^{+0.20}_{-0.13} )(1+z)^{1.55\pm0.45}$ based 
on seven LLS observed by FOS at $0.456 < z < 1.036$ with median redshift $z_{\rm abs} = 0.649$.  Using the same seven 
LLS, Stengler-Larrea \etal\ (1995) found $d {\cal N}/dz = (0.25^{+0.17}_{-0.10})(1+z)^{1.50 \pm 0.39}$.  In a recent survey 
of strong absorbers at $z < 2.6$, Ribaudo \etal\ (2011a) analyzed 206 LLS and pLLS absorbers with low-resolution 
gratings ($R \approx 1000$) on STIS (G140L, G230L) and FOS (G140L).  Most of the \HI\ absorbers in that survey were at 
$z > 1$; Table 4 in their paper lists five absorbers at $z < 0.5$ (3 true LLS) and 22 absorbers at $z < 0.84$ (17 LLS).  From 
their LLS sample with $\tau_{\rm LLS} \geq 1$, they fitted the evolution to $d {\cal N}/dz = (0.28) (1+z)^{1.19\pm0.56}$ over 
$0.25 < z < 2.59$.  In general, the number of low-redshift ($z < 1$) LLS reported in the literature is small, with just seven LLS 
in the \HST/FOS Key Project and eight in our \HST/COS survey (these 8 are distinct from those seen with FOS).  We are also
aware of three LLS at $z < 0.24$, found by the \FUSE\ satellite in the far-UV:
 
\medskip

\indent PHL\,1811   $(z_{\rm abs} = 0.080923)$  $\log N_{\rm HI} = 17.98\pm0.05$ (Jenkins \etal\ 2005) \\
\indent PKS\,1302-102  $(z_{\rm abs} = 0.09487)$  $\log N_{\rm HI} = 17.2\pm0.2$ (Cooksey \etal\ 2008) \\
\indent  PKS\,0312-77   $(z_{\rm abs} = 0.2028)$  $\log N_{\rm HI} = 18.22^{+0.19}_{-0.25}$ (Lehner \etal\ 2009) 

\medskip
 
We return to the observed deficit of LLS at $z > 0.48$, below the expected numbers given the fitted line frequency.  
The COS-Halos survey (Tumlinson \etal\ 2013) selected 39 UV-bright QSOs (\GALEX\ far-UV magnitudes $<18.5$) 
with median redshift $z_{\rm AGN} = 0.525$.  They avoided QSOs with strong (equivalent width $> 1$~\AA) \MgII\ 
absorbers at $z > 0.4$ and advise that their QSOs at $z > 0.4$ are not expected to provide an unbiased sample 
of LLS.  Our sample of 102 AGN with $z_{\rm AGN} \geq 0.24$ includes 37 targets from the COS-Halos sample:
20 at $z_{\rm AGN} > 0.5$, 14 at $z_{\rm AGN} > 0.6$, and 9 with $z_{\rm AGN} = 0.713-0.887$.  We also have 
seven UV-bright (Palomar-Green) quasars observed with \IUE\ (Tripp \etal\ 1994):  
PG~1407+265 ($z = 0.946$), PG~1148+549  ($z = 0.975$), PG~1206+459 ($z = 1.1625$),  PG~1338+416 
($z = 1.21422$), PG~1522+101 ($z = 1.32785$), Q\,0232-042 ($z = 1.437$), PG~1630+377 ($z = 1.476$).  
Our sample includes several UV-bright AGN (e.g., SBS~1108+560 at $z = 0.766619$, SBS~1122+594 at $z = 0.8514$) 
from the \HST\ GTO Program and Guest Investigator Programs 11248, 11264, 11585, 11598, 11741, with diverse 
scientific goals including intergalactic absorbers (Danforth \etal\ 2016), high-redshift absorbers (Tripp \etal\ 2011;
Ribaudo \etal\ 2011b), galaxy-quasar pairs (Keeney \etal\ 2006;  Crighton \etal\ 2010; Meiring \etal\ 2011; 
Tumlinson \etal\ 2013; Stocke \etal\ 2013; Bordoloi \etal\ 2014), and interstellar high-velocity clouds (Shull \etal\ 2009).
Several of these high-$z$ AGN were selected to study intervening \NeVIII\ absorbers and the hot phase of the IGM
(Narayanan \etal\ 2011; Tripp \etal\ 2011; Savage \etal\ 2011; Meiring \etal\ 2013; Hussain \etal\ 2015).    

Our derived LLS coefficient, $N_0 = 0.25^{+0.13}_{-0.09}$, is consistent with previous values (Table~7) from
low-resolution \HST\ surveys with FOS and STIS.  Our fit to pLLS redshift evolution over $0.24 < z < 0.84$ gives 
$\gamma_{\rm pLLS} = 1.14^{+0.88}_{-0.89}$, consistent with estimates noted above (Storrie-Lombardi \etal\ 1994; 
Stengler-Larrea \etal\ 1995; Ribaudo \etal\ 2011a).  However, the \HST\ surveys are all based on small numbers of LLS 
with different redshift coverages.  The median LLS redshifts are $z = 0.65$ for the seven FOS Key Project absorbers
and $z =  0.39$ for the eight LLS studied by COS.  The Ribaudo \etal\ (2011a) survey was dominated by systems at 
$z > 1$, with only three LLS at $z \leq 0.5$.  Thus, the redshift evolution of LLS and pLLS remains uncertain.  Our best 
fit, $\gamma_{\rm pLLS}= 1.14$, is similar to the value $\gamma_{\rm LF} = 1.24\pm0.06$ (Danforth \etal\ 2016)
for weak \Lya\ forest absorbers and also to that, $\gamma_{\rm LLS} = 1.19 \pm 0.56$ (Ribaudo \etal\ 2011a) fitted 
to the LLS ($\tau_{\rm LLS} \geq 1$) absorbers at $z < 2.6$ in low-resolution FOS and STIS spectra.
Differences in $N_0$ inferred from extrapolating $f(N,z)$ could be used to detect a turnover in the power-law slope 
($\beta$) at $\log N > 17$.  However, the number of low-$z$ LLS absorbers is currently too small to provide reliable 
statistics.  Obviously, larger UV surveys of LLS and pLLS that cover a wider range of redshifts would have greater 
leverage for determining the evolution index $\gamma$.  

Accurate values of pLLS/LLS evolution out to $z \approx 2$ are also important for assessing the far-UV opacity to 
intermediate-redshift AGN, as we discuss in Section 3.5.   The scarcity of strong absorbers ($\log N_{\rm HI} > 16.5$) 
results in large fluctuations about these mean optical depths.  However, the accumulated far-UV absorption from pLLS 
could decrease $F_{\lambda}$ and flatten the spectral energy distribution in the far-UV.  There may also be bias in the 
selection of the intermediate-redshift quasars ($1.0 < z < 2.2$) used to construct rest-frame LyC composite spectra
(Telfer \etal\ 2002; Shull \etal\ 2012b; Stevans \etal\ 2014; Tilton \etal\ 2016). To be selected, these AGN needed to have 
detectable fluxes in the \GALEX\ far-UV channel covering 1344--1786~\AA\ (Morrissey \etal\ 2005).  Thus, the observed 
AGN sight lines generally avoided encountering LLS at $z_{\rm LLS} \approx 0.5-1.0$ with $\log N_{\rm HI} > 17.5$ that 
block the far-UV absorption.   How strongly this selection bias affects the intrinsic AGN composite spectrum is still unknown.

 \subsection{Damped \Lya\ Absorbers}
 
Although we have focused primarily on pLLS and LLS absorbers, we also include a discussion of the four DLAs
detected along the sight lines to all 159 AGN in our sample  We provide these low-redshift statistics with only
moderate astronomical interpretation, because of the small numbers and the accompanying uncertainties in 
deriving effective pathlengths  We compare our values with several previous DLA surveys that used UV data at 
$z < 1.65$ (Rao \etal\ 2006; Meiring \etal\ 2011; Battisti \etal\ 2012; Turnshek \etal\ 2015; Neeleman \etal\ 2016).  
Table~8 lists these four DLAs, together with six sub-DLAs ($\log N_{\rm HI} = 19.0-20.3$) and two strong absorbers 
($\log N_{\rm HI} \approx 18.5\pm0.5$) whose large error bars place them near the sub-DLA range.  As with our 
LLS statistics, we only consider redshifts $z_{\rm abs} < 0.48$. Indeed, all four DLAs in our sample are at low 
redshifts ($z < 0.4$).  Three were found serendipitously toward AGN in the COS-Halos survey, which excluded 
QSOs with strong \MgII\ absorbers at $z > 0.4$.  The fourth DLA ($z_{\rm abs} = 0.185$) is toward B0120-28, a 
UV-bright QSO used to probe the Magellanic Stream.  Because of the small numbers of low-redshift DLAs and 
the range of QSO targeting strategies, the statistics that follow are uncertain, as are those in other low-redshift surveys.  

In our COS survey of 102 QSOs at $z \geq 0.24$, we had an effective pathlength $\Delta z = 19.24$ sensitive to 
detecting LLS and DLA between $0.24 < z < 0.48$.  We include the extra pathlength ($\Delta z =  29.26$) at the
redshifts between $0.01 < z < 0.24$ available to all 159 AGN, and subtract a small amount ($\Delta z = 0.282$) 
of LLS-blocked spectra.  Thus, we have total redshift coverage $\Delta z \approx 48$ sensitive to DLAs.   Our 
statistics for DLA frequency are therefore based on four low-redshift DLAs:
J1619+3342 ($z_a = 0.0963$) with $\log N_{\rm HI} = 20.55 \pm 0.10$; 
J1009-0713 ($z_a = 0.1140$) with $\log N_{\rm HI}  = 20.68 \pm 0.10$; 
B0120-28 ($z_a = 0.185$) with $\log N_{\rm HI}  = 20.50 \pm 0.10$; and
J1616+4153 ($z_a = 0.3211$) with $\log N _{\rm HI} = 20.60 \pm 0.20$.
Over pathlength $\Delta z = 48$, these four DLAs correspond to a line frequency,
\begin{equation}
    \left( \frac { d {\cal N} } {dz }\right)_{\rm DLA}  =  0.083^{+0.066}_{-0.040}   \; \; ,
 \end{equation}
at mean redshift $\langle z \rangle = 0.18$.  We now compare this DLA frequency to those inferred from other 
surveys at $z < 1.65$.  Owing to the small numbers of low-$z$ DLAs, these surveys also have large uncertainties, 
ranging from $d {\cal N}/dz = 0.033^{+0.026}_{-0.015}$ (Neeleman \etal\ 2016) to $0.25^{+0.24}_{-0.14}$ 
(Meiring \etal\ 2011).   Most of the DLAs in those surveys are at higher redshifts ($z > 1$) compared to the 
4 DLAs in our COS survey ($z = 0.096$, 0.114, 0.185,and  0.321).  As we now discuss, our value is generally 
consistent with these prior estimates.  

Rao, Turnshek, \& Nestor (2006) used UV spectra from \HST/STIS, identifying DLAs from a \MgII\ and \FeII\ 
selected sample. They found $d {\cal N} / dz = 0.079 \pm 0.019$  for DLAs at $ z < 1.65$ based on 18 DLAs found in 
108 \MgII\ systems at ($0.11 < z < 0.9$) with median redshift $z = 0.609$.  Their fit to DLA evolution over all redshifts 
gave $n(z) = n_0 (1+z)^{\gamma}$ with $n_0 = 0.044 \pm 0.005$ and $\gamma = 1.27 \pm 0.11$.  
Meiring \etal\ (2011) found $d {\cal N} /dz = 0.25^{+0.24}_{-0.14}$ from three DLAs along a COS-surveyed path 
with $\Delta z = 11.94$.  These three DLAs were also identified in our survey: 
J1616+4153 ($z_a = 0.3211$), J1619+3342 ($z_a = 0.0963$), J1009-0713 ($z_a = 0.1140$).  Their much higher
LLS absorber frequency was based on a smaller redshift pathlength, only 22\% of that in our survey.   
Neeleman \etal\ (2016) conducted a large UV survey of 463 quasars at $z < 1.6$ over significant pathlength 
($\Delta z = 123.3$).  They found a line frequency $d {\cal N} /dz = 0.033^{+0.026}_{-0.015}$ at median 
redshift $z = 0.623$.  However, these statistics were based on just 4 DLAs with data from FOS, STIS, and COS.  
They originally had a total of 47 DLAs in their sample, but they excluded most of them in their statistics because 
they were found through \MgII\ targeting or because of a galaxy close to the QSO sightline.   
 
In the most recent survey, Rao \etal\ (2017) examined the statistical properties of DLAs at $0.11 < z < 1.65$ using 
UV measurements (\HST/ACS, \GALEX\ and \HST/COS).  Over a wide range of redshifts ($z = 0-5$) they fit 
$d {\cal N} /dz = (0.027\pm0.007)(1+z)^{1.682\pm0.200}$.  This fit corresponds to $d {\cal N} /dz = 0.036$
at the mean redshift ($z = 0.18$) of our survey.  They find no bias with \MgII-selected samples of DLAs.  Their 
COS sample turned up no true DLAs (all had $\log N_{\rm HI} \leq 20.0$) but two of them were sub-DLAs, also 
found in our survey (Systems \#1 and \#6 in Appendix A).

 \subsection{Opacity of the Low-$z$ IGM } 
 
After the epoch of reionization of neutral hydrogen at $z \approx 7$, the IGM becomes mostly ionized.  However, 
the UV continuum of intermediate-redshift AGN is still blanketed by \Lya\ absorbers that produce photoelectric
opacity in their Lyman continua.  This LyC opacity determines the mean-free-path of ionizing photons in the IGM,
attenuates the ionizing photons from galaxies and quasars, and determines the metagalactic EUV background 
radiation (Fardal \etal\ 1998; Miralda-Escud\'e 2003; Haardt \& Madau 2012; O'Meara \etal\ 2013).  
This continuum opacity is also relevant to measuring the composite spectrum of quasars in their rest-frame LyC.  
Our recent studies of AGN at intermediate redshifts ($z \approx 1.5-2.2$) exhibit frequent pLLS absorption 
(Shull \etal\ 2012b; Stevans \etal\ 2014; Tilton \etal\ 2016) whose LyC optical depth and recovery at shorter 
wavelengths is used to restore the underlying AGN continuum.

The primary observable of our survey is the bivariate distribution of \HI\ absorbers in redshift and column density.
For a Poisson-distributed ensemble of \HI\ absorbers (Paresce \etal\ 1980), one can compute the {\it average} 
photoelectric continuum opacity in the low-redshift IGM by integrating over the relevant range of column densities. 
We find the gradient of optical depth with redshift,
\begin{equation}
   \frac {d \tau_{\rm eff}} {dz} = \int  \left(  \frac {d^2 {\cal N}_{\rm abs} } {dz \; d N}  \right) \;
    \left[ 1 - \exp \{-N \sigma (z) \}  \right]  \; dN    \; \; .
\end{equation}
Here,  $N$ denotes $N_{\rm HI}$ and we define the \HI\ photoelectric optical depth as 
$\tau(\lambda_{\rm o}) = N \sigma (\lambda_0)$ at a typical {\it observed} far-UV wavelength, 
$\lambda_{\rm o}  \equiv \lambda_{\rm LL} (1+z_{\rm o})$.  The Lyman continuum cross section is approximated as
$\sigma(\lambda_{\rm o}) \approx \sigma_0 (\lambda_{\rm o} / \lambda_{\rm LL})^{3}$ where 
$\sigma_0 = 6.30 \times 10^{-18}~{\rm cm}^2$.  We adopt a fiducial far-UV wavelength
$\lambda_{\rm o} = 1130$~\AA\ where the COS/G130M coverage begins.  By construction, 
$\lambda_{\rm o} < \lambda_{\rm LL} (1+z)$ and $z_{\rm o} <  z$ in order for $\lambda_{\rm o}$ to lie in the 
LyC of the absorber. 

We now compute the cumulative optical depth, $\tau_{\rm eff} (z_{\rm o}, z)$, due to a population of \HI\ absorbers 
at redshift $z$.  For absorbers with $\tau_{\rm LL} < 1$ ($N < 1.59 \times 10^{17}$ cm$^{-2}$) we approximate 
$[1 - \exp(-\tau)] \approx \tau$.  We express the \HI\ column density distribution as separable power laws 
in column density and redshift,
\begin{equation}
   f(N,z) \equiv  \frac {d^2 {\cal N}_{\rm abs} } {dz \; d N}  = C_0 N^{-\beta} \, (1+z)^{\gamma} \;  ,
\end{equation} 
for column densities $N$ measured in cm$^{-2}$.   In our pLLS survey, a maximum-likelihood fit finds 
$\beta_{\rm pLLS} = 1.48 \pm 0.05$, $\gamma_{\rm pLLS} = 1.14^{+0.88}_{-0.89}$, and 
$C_0 = 4  \times 10^7$ over the higher range in column density ($15 \leq \log N \leq 20$).  
The redshift-gradient of LyC optical depth at observed wavelength $\lambda_{\rm o}$ depends on an integral 
over the column-density distribution,
\begin{eqnarray}
   \frac {d \tau_{\rm eff} (z_{\rm o}, z)} {dz} &=&  C_0  \, \sigma_0 \left[ \frac {1+z_0} {1+z} \right]^3  (1+z)^{\gamma}
         \int_{N_1}^{N_2} N^{(-\beta + 1)} \;  dN     \nonumber \\
              &=&    \frac {C_0 \, \sigma_0 \, (1+z)^{\gamma} } {(2 - \beta)}  \left[ \frac {1+z_0} {1+z} \right]^3  
              \left[ N_2^{(2-\beta)} - N_1^{(2-\beta)} \right] \; .
\end{eqnarray}
For $\beta_{\rm pLLS} = 1.48 \pm 0.05$, the \HI\ opacity is weakly dominated by the higher-column density absorbers 
with $\tau_{\rm eff} \propto N_2^{0.52}$.  Here, $\sigma(\lambda_{\rm o}) \approx \sigma_0 [(1+z_0)/(1+z)]^3$ is the 
cross section at wavelength $\lambda_{\rm o}$.   From the observed distribution parameters we derive an opacity 
gradient, $d \tau_{\rm eff} / dz = (0.408)(1+z)^{\gamma_{\rm pLLS}}  [ (1+z_0)/(1+z) ]^3$ for the strong absorbers, 
$15 \leq \log N_{\rm HI} \leq 17.2$.  We then integrate $\tau_{\rm eff} (z_{\rm o}, z)$ from redshift $z_{\rm o}$ out to 
higher redshifts,
\begin{eqnarray}
   \tau_{\rm eff}^{\rm (pLLS)} (z_{\rm o}, z) &=& (0.396) \int_{z_{\rm o} }^{z} \left[  \frac {1+z_{\rm o}} {1+z} \right]^3 
                (1+z)^ {\gamma} \; dz                                      \nonumber \\
    &=& \frac { (0.396) (1+z_{\rm o})^{\gamma+1} }  {(2 - \gamma) } 
   \left[ 1 -  \left( \frac { 1+z_{\rm o} } { 1+z } \right) ^ { (2 - \gamma) }  \right]  \; . 
\end{eqnarray} 
For our best-fitting index, $\gamma_{\rm pLLS} = 1.14$, the optical depths at $\lambda_{\rm o} = 1130$~\AA\
($z_{\rm o} = 0.24$) are $\tau_{\rm eff} (z_{\rm o}, z) = (0.25, 0.33, 0.39)$ for sources at $z = (1.0, 1.5, 2.0$).
These opacities depend somewhat on the index $\gamma_{\rm pLLS}$.  
For $\gamma_{\rm pLLS} = 2$, the integral has a logarithmic dependence,
\begin{eqnarray}
    \tau_{\rm eff}^{\rm (pLLS)} (z_{\rm o}, z) = (0.396) (1+z_{\rm o})^3 \ln \left[ \frac { 1+z } { 1+z_{\rm o} } \right]  \; .
\end{eqnarray}
At $z = (1.0, 1.5, 2.0)$, the corresponding optical depths are $\tau_{\rm eff} (z_{\rm o}, z) = (0.37, 0.55, 0.69)$.  
Because several LLS surveys have coverage to higher redshifts, it is possible that $\gamma_{\rm pLLS}$ could 
be higher than our assumed value of $\gamma = 1.14$.  However, this best-fit index is similar to previous studies.
For example, Ribaudo \etal\ (2011a) found $\gamma = 1.19 \pm 0.56$ for LLS with $\tau_{\rm LL} \geq 1$ over
the range $0.25< z < 2.59$.   However, their survey only had 3 LLS at $z \leq 0.5$ and 17 at $z \leq 0.84$.  \\

Weaker lines in the \Lya\ forest, with $\log N < 15$, contribute less optical depth, 
$d \tau_{\rm eff} / dz = (0.0674)(1+z)^{\gamma_{\rm LF}}$, from which we find
\begin{equation}
   \tau_{\rm eff}^{\rm (LF)} (z_{\rm o}, z) = \frac { (0.0674) (1+z_{\rm o})^{\gamma+1} }  {(2 - \gamma) } 
   \left[ 1 -  \left( \frac { 1+z_{\rm o} } { 1+z } \right) ^ { (2 - \gamma) }  \right]   \; . 
\end{equation} 
For the index, $\gamma_{\rm LF} = 1.24\pm0.06$, that fits all absorbers at $\log N_{\rm HI} > 13.0$
(Danforth \etal\ 2016), the \Lya-forest optical depths at $\lambda_{\rm o} = 1130$~\AA\ ($z_{\rm o} = 0.24$) 
are $\tau_{\rm eff} (z_{\rm o}, z) = 0.0453$, 0.0614, and 0.0727 for sources at $z = 1.0$, 1.5,  and 2.0.  
Combining the pLLS and \Lya\ forest, we find average optical depths of 
$\tau_{\rm eff} (z_{\rm o}, z) = 0.29$, 0.39, 0.46 in the far UV (1130~\AA) corresponding to the 
Lyman edge redshifted to $z_{\rm o} = 0.24$ and sources at $z = 1.0$, 1.5, 2.0.  \\


 \section{SUMMARY AND DISCUSSION}
 
This \HST/COS survey is one of the largest sets of 220 strong \HI\ absorbers at low redshift.  Our survey spans column 
densities ($14.0 \leq \log N_{\rm HI} \leq 20.4$) and redshifts ($0.237 \leq z \leq 0.928$) sampled in bins of width 
$\Delta z = 0.04$ from $z = 0.24-0.84$ and $\Delta (\log N_{\rm HI}) = 0.25$.  With sensitivity to multiple lines in the 
\HI\ Lyman series, our pLLS survey should be nearly complete for the 158 absorbers with $\log N_{\rm HI} > 15$. 
The column densities of a few absorbers in two bins ($\log N_{\rm HI} = 15.0-15.5$) may have shifted to adjoining bins
owing to CoG effects of velocity components.  We may also have missed some weak absorbers at $z > 0.5$, detected 
only in higher Lyman-series lines, since \Lya\ and \Lyb\ shift out of the COS/G160M window at $z > 0.47$ and $z > 0.75$ 
respectively.  Our survey includes 8 true LLS ($\log N_{\rm HI} \geq 17.2$), one DLA ($\log N_{\rm HI} = 20.34 \pm 0.12$), 
and 54 pLLS ($16.0 \leq \log N_{\rm HI} < 17.2$).   All absorbers were detected with the G130M and G160M gratings at 
spectral resolution ($R \approx 17,000$ or $\Delta v \approx 18$~\kms) sufficient to measure multiple lines in the \HI\ 
Lyman series, along with the Lyman decrement at $\log N_{\rm HI} \geq 16.2$.  Our Lyman-comb technique is superior 
for column-density determination compared to methods based on low resolution ($R \approx 1000$) data. By combining 
the Lyman-series CoG with the Lyman decrement, we obtain reliable values of $N_{\rm HI}$  and its bivariate distribution, 
$f(N_{\rm HI}, z)$.  

The COS distribution is in good agreement with previous \HST\ surveys of LLS at $0.4 < z < 1.4$, with an absorber 
frequency per unit redshift parameterized as $d {\cal N}/dz  = N_0 (1+z)^{\gamma}$ and summarized in Table~7.  
Storrie-Lombardi \etal\ (1994) found $d{\cal N}/dz = (0.27)(1+z)^{1.55\pm0.45}$ and Stengler-Larrea \etal\ (1995) found  
$d {\cal N}/dz = (0.25)(1+z)^{1.50 \pm 0.39}$.   Each of these FOS studies analyzed the same seven LLS.  
From FOS and STIS data, Ribaudo \etal\ (2011a) fitted an index $\gamma = 1.19\pm0.56$ for LLS absorbers between
$0.25 < z < 2.59$. The eight LLS systems in our COS survey are distinct from those in the FOS study and range 
from $z_{\rm abs} = 0.2374$ to $z = 0.47379$ with median redshift $z_{\rm abs} = 0.39$.  Over total pathlength 
$\Delta z = 19.24$ between $0.24 < z < 0.48$, we find a mean frequency, 
$\langle d {\cal N}/dz \rangle_{\rm LLS} = 0.36^{+0.20}_{-0.13}$.  With median redshift $z_{\rm LLS} = 0.39$, this 
translates to $N(z) = N_0 (1+z)^{\gamma}$ where $N_0 = 0.25^{+0.13}_{-0.09}$ at $z = 0$.  An index
$\gamma \approx 1.1$ is what one would expect for non-evolving pLLS absorbers at $z < 1$.   For a $\Lambda$CDM 
cosmology with standard (Planck-2016) parameters, the expected redshift evolution for absorbers with constant space 
density $\phi_0$ and cross section $\sigma_0$ is:
\begin{equation}
   \frac {d {\cal N}} {dz} = \frac { \phi_0 \, \sigma_0  \, c}{H_0} 
         \frac {(1+z)^2} {[ \Omega_m (1+z)^3 + \Omega_{\Lambda}]^{1/2} }  
         \equiv \left( \frac { \phi_0 \, \sigma_0  \, c}{H_0} \right)  S(z)    \;  \; .
\end{equation}  
Over our sampled range in redshift, $0.24 \leq z \leq 0.84$, with $\Omega_m \approx 0.3$ and 
$\Omega_{\Lambda} \approx 0.7$, the cosmological factor is well-fitted by $S(z) = (1.08)(1+z)^{1.1}$ to 3\% 
accuracy.  Thus, our derived maximum-likelihood value $\gamma = 1.14^{+0.88}_{-0.89}$, is consistent with non-evolving
absorbers, although the allowed range in $\gamma$ provides poor constraint on their low-$z$ evolution.  Ultraviolet
spectroscopic surveys of pLLS at $z = 0.7 - 1.5$, with better statistics and redshift leverage, would narrow the range 
of $\gamma_{\rm pLLS}$ and help to characterize their redshift evolution.

In the Introduction, we alluded to the relationship between LLS absorbers and the extended regions of galaxies. 
Using our statistical sample of pLLS, with observed frequency $d {\cal N} / dz = 1.69 \pm 0.33$, equation (11) implies
a cross section, $\sigma_0 = \pi R_0^2$, with effective radius $R_0$ if we associate the pLLS with an appropriate
space density of galaxies.  We use the luminosity function of low-redshift ($z = 0.1$) galaxies from the Sloan Digital 
Sky Survey (Blanton \etal\ 2003), with normalization 
$\phi_* = (1.49 \pm 0.04) \times 10^{-2} \; h^3 \; {\rm Mpc}^{-3}$ and faint-end slope $\alpha = -1.05 \pm 0.01$, to find:
\begin{eqnarray}
   R_0 &=& \left[ \frac { {d {\cal N} } {dz} } { \pi (c/H_0) \, \phi_* \,  S(z) (\Delta L/L^*) } \right]^{1/2}  \nonumber \\
       & \approx & (110 \pm 10 \, {\rm kpc}) \, h^{-1} \left[  S(z) (\Delta L / L^*) \right] ^{-1/2}  \; .
\end{eqnarray}
The cosmological factor $S(z) \approx 1.57$ at the median absorber redshift ($z = 0.39$). The fractional luminosity 
bandwidth ($\Delta L / L^*$) depends on the minimum luminosity and ranges from 0.56 (integrating down to $0.5 L^*$) 
to 1.83 (integrating down to $0.1 L^*$).  Both scaling factors enter as the square root.  Taking $h \approx 0.7$ and 
recognizing the uncertainty in galactic parameters, we see that both LLS and pLLS have the correct absorber frequencies,
 $d {\cal N} / dz$, to be associated with extended halos of luminous ($0.3-3.0 L^*$) galaxies.   In the COS-Halos Survey, 
the CGM of star-forming galaxies has been detected in \OVI\ absorption (Tumlinson \etal\ 2011) out to distances of 
100-150~kpc.   Tumlinson \etal\ (2013) detected strong \HI\ absorption averaging $\sim1$~\AA\ in \Lya\ equivalent 
width out to 150 kpc, with 100\% covering fraction for star-forming galaxies.  Radial extents of 100-200~kpc are also 
consistent with the region of gravitational influence of $10^{12}~M_{\odot}$ galaxies, estimated from a proper treatment 
of the virial radius (Shull 2014),
\begin{eqnarray}
   R_{\rm vir} (M_h, z_a) &=& (206~{\rm kpc}) \, h_{70}^{-2/3} \, M_{12}^{1/3}  \left( 1+z_a \right)^{-1}  \nonumber \\
       & \times & \left[  \frac {\Omega_m (z_{\rm a})  \, \Delta_{\rm vir} (z_{\rm a})}  {200} \right] ^{-1/3}   \;  .
\end{eqnarray}
This expression differs from often-used formulae by the scaling with overdensity $\Delta_{\rm vir}$ and by the factor 
$(1+z_{\rm a})^{-1}$.  This factor reflects the fact that most galaxies underwent virialization in the past, when the IGM
background density was higher by a factor of $(1+z_a)^3$.   Typical  ``half-mass assembly" redshifts are 
$z_a \approx 0.8-1.2$.  After initial assembly, their proper size changes gradually because of continued mass 
infall into the halo.  From the statistics of current pLLS survey, we are unable to ascertain the \HI\ column density at 
which ``strong \HI\ absorbers" are less directly associated with galaxy halos.  Based on the steep fall-off of the 
observed differential distribution, $f(N_{\rm HI}, z) \propto N_{\rm HI}^{-1.65}$, it likely occurs below  
$N_{\rm HI} = 10^{15}$~cm$^{-2}$.   This prediction is consistent with observations (Figure 5 in Stocke \etal\ 2013) 
which find that virtually no absorbers with $\log N_{\rm HI} \leq 14.5$ are within one virial radius of a galaxy
of any luminosity.  The issue of  \HI\ ``covering factor" is difficult to constrain in our sample, since we have not undertaken 
a program to identify the associated galaxies.  Many of them are at $z > 0.3$ and detecting them would be difficult.  

The degree of pLLS/LLS evolution out to $z \approx 2$ is important for assessing the far-UV opacity to intermediate 
redshift AGN (Section 3.5).  The scarcity of strong absorbers ($\log N_{\rm HI} > 16.5$) results in large fluctuations 
about these mean optical depths.   Far-UV absorption from pLLS could decrease $F_{\lambda}$ and flatten the 
spectral energy distribution in the far-UV.  Selection bias may affect the intermediate-redshift quasars ($1.0 < z < 2.2$) 
used to construct rest-frame LyC composite spectra (Telfer \etal\ 2002; Shull \etal\ 2012b; Stevans \etal\ 2014; 
Tilton \etal\ 2016).   These AGN need to have detectable fluxes in the \GALEX\ far-UV channel covering 
1344--1786~\AA\ (Morrissey \etal\ 2005).  Thus, the observed  AGN sight lines generally avoided encountering LLS 
at $z_{\rm LLS} \approx 0.5-1.0$ with $\log N_{\rm HI} > 17.5$ that block the far-UV absorption.   How strongly this 
selection bias affects the intrinsic AGN composite spectrum is not yet understood.
 
 We would benefit from larger FUV/NUV surveys of pLLS/LLS absorbers toward QSOs with $z_{\rm AGN} = 1-2$, 
 with attention to biases introduced by target selection.  This project will be difficult, since FUV-bright QSOs are 
 needed to extend the absorber distribution, $f(N,z)$,  to higher redshifts, and targeting via the presence or absence 
 of strong \MgII\ absorbers will bias the LLS survey.  The current COS survey provides reliable data at low redshift 
 ($z < 0.5$), but connecting line frequency, $d {\cal N} / dz$, continuously to values at $z > 1.5$ (O'Meara \etal\ 2013)
  is critical for IGM radiative transfer models that rely on FUV opacities.


\noindent
We now summarize the main results of our survey: 
\begin{enumerate}     

\item  Over redshifts $0.24 \leq z \leq 0.84$, the average frequency for strong \HI\ absorbers 
    ($\log N_{\rm HI} \geq 15.0$) is $\langle d {\cal N} / dz \rangle  \approx 4.95 \pm 0.39$.  We parameterize
    the bivariate distribution of \HI\ absorbers as $f(N,z) = C_0 N_{\rm HI} ^{-\beta} (1+z)^{\gamma}$, where a 
    maximum likelihood fit over the ranges $0.24 \leq z \leq 0.84$ and $15 \leq \log N_{\rm HI} \leq 20$ gives 
    $\beta \approx 1.48 \pm 0.05$, $\gamma \approx 1.14^{+0.88}_{-0.89}$, and $C_0 = 4\times10^7$ for 
    column densities  $N_{\rm HI}$ in cm$^{-2}$.  This distribution is poorly determined at $\log N_{\rm HI} > 17.5$.  
    The redshift evolution ($\gamma \approx 1.1$) is consistent with absorbers of constant space density 
    and cross section.

 \item Based on seven true LLS with $\log N_{\rm HI} \geq 17.2$ between $0.24 < z < 0.48$, we derive a LLS
    frequency of $(d{\cal N}/dz)_{\rm LLS} = (0.25^{+0.13}_{-0.09}) (1+z)^{\gamma}$, assuming the best-fitting  
    index $\gamma = 1.14$ derived from MCMC simulations of all strong absorbers.  The frequency for pLLS is
    $(d{\cal N}/dz)_{\rm pLLS} = (1.69 \pm 0.33) (1+z)^{\gamma}$.  If these absorbers are associated with halos 
    of luminous ($0.3-3.0 L^*$) galaxies,  the pLLS frequency implies circumgalactic gas cross sections of 
    $100-150$~kpc radial extent.  
    
 \item Over the range ($0.01 < z < 0.48$) of the COS survey of 159 AGN sensitive to DLAs, we found
    4 DLAs over pathlength $\Delta z_{\rm} \approx 48$.  From this, we estimate an absorber frequency,
    $(d{\cal N}/dz)_{\rm DLA} = 0.083^{+0.066}_{-0.040}$.  Although uncertain, this frequency is lower than 
    the previous (COS-studied) value, $0.25^{+0.24}_{-0.14}$ (Meiring \etal\ 2011), which was based on 3 
    DLAs over $\Delta z = 11.94$, but larger than the values $d{\cal N}/dz \approx 0.027-0.036$ 
    (Neeleman \etal\ 2016; Rao \etal\ 2017).  All of these surveys suffer from small-number statistics.
                  
\item Combining the data from low-redshift COS surveys of pLLS and \Lya\ forest absorbers, we estimate the
   \HI\ photoelectric opacity gradient, $d \tau_{\rm eff}/dz$, for mean optical depth in the far-UV continuum
   (1130~\AA).  For indices $\gamma_{\rm pLLS} = 1.14$ and $\gamma_{\rm LF} = 1.24$, with considerable 
   stochasticity, we estimate mean far-UV optical depths $\tau_{\rm eff} (z_{\rm o}, z) = 0.29$, 0.39, and 0.46 for 
   sources at $z = 1.0$, 1.5, and 2.0. 
    
 \item The observed distributions of strong \HI\ absorbers are expected to have considerable variations among 
    sight lines.  Above $N_{\rm HI} = (10^{17}~{\rm cm}^{-2}) N_{17}$, the fitted cumulative frequency is small, 
    $d{\cal N}(>N_{\rm HI}) / dz = (0.577) N_{17}^{-0.48} \, (1+z)^{\gamma}$.  Our estimates of far-UV continuum 
    opacity suggest that surveys of intermediate-redshift AGN may have a selection bias toward sight lines with low 
    optical depths, lacking strong LLS absorbers. 
  
\end{enumerate}  

\acknowledgments
  
This survey was based on observations made with the NASA/ESA Hubble Space Telescope, obtained from the 
data archive at the Space Telescope Science Institute. STScI is operated by the Association of Universities 
for Research in Astronomy, Inc.\ under NASA contract NAS5-26555.
The project originated from individual and survey observations of AGN taken with the Cosmic Origins Spectrograph 
on the {\it Hubble Space Telescope}.  We appreciate helpful discussions with John Stocke, Todd Tripp, and David 
Turnshek.  In early stages, this research was supported by grant HST-GO-13302.01.A  from the Space Telescope 
Science Institute to the University of Colorado Boulder.  More recent work was carried out through academic support 
from the University of Colorado.

 \newpage


 \newpage
 
  
\appendix


\section{Appendix A:  Notes on LLS and pLLS Absorbers }

In this Appendix, we discuss the column-density determinations for 73 strong \HI\ absorbers 
($\log N_{\rm HI} > 15.75$) at $z_{\rm abs} \geq 0.24$.   In an approximate analysis used in our 
construction of an AGN composite spectrum (Stevans \etal\ 2014), the \HI\ column densities of 
these systems were estimated and used to correct the underlying continuum.  These estimates were 
based on automated CoG fitting, which becomes uncertain when the Lyman lines reach saturation (Table 2) 
and in the presence of multiple velocity components.  
In the current survey, we carefully examined each of these systems, combining multi-component CoG fits with 
an analysis of the flux decrement at the Lyman limit.   In fitting the continuum above and below the LL, we 
looked for contamination by broad AGN emission lines using the template shown in Figure 1. The underlying 
continuum lies below these emission lines.  Systems labeled \#1 through \#26 (Table 6) had initial estimates 
$\log N_{\rm HI} \geq 16.5$, for which we expect a clear Lyman limit with optical depth $\tau_{\rm LL} \geq 0.2$.   
Systems \#27 to \#72 had estimated column densities $15.75 \leq \log N_{\rm HI}  \leq 16.49$.   System \#73 is 
a DLA  with $\log N_{\rm HI} = 20.34\pm0.12$ at $z_{\rm abs}= 0.3211$ not tabulated in Stevans \etal\ (2014).  
Some of these systems were discussed in previous papers (Tripp \etal\ 2011; Ribaudo \etal\ 2011b;  Tumlinson 
\etal\ 2011b; Lehner \etal\ 2013;  Fox \etal\ 2013; Tilton \etal\ 2016).   

For the LLS and pLLS absorbers, our strategy of combining CoG with LL flux decrements, when detected, provided 
confirmation of the column density and its uncertainty.  In a number of cases involving multiple velocity components, 
we used the spectral resolution of G130M and G160M gratings to fit multi-component CoGs. When the components
separations were less than 100~\kms, we combined them into a single system.   In a few instances with clearly
separated components ($\Delta v > 150-200$~\kms) we kept separate track of the individual column densities
and split them into separate absorbers, denoted Systems \#24abc, \#47abc, \#59ab, and \#60ab.  
At the COS spectral resolution, the LL flux decrements and optical depths, $\tau_{\rm LL}$, typically arise from all 
components within 100-400~\kms.    With good S/N in the higher-order Lyman lines, we are able to identify
absorption systems with $\Delta v \geq 100$~\kms.   In comparing our column densities with prior work, we 
usually found good agreement  ($\Delta \log N_{\rm HI} \leq 0.1$), but deviations are noted in the narratives
and in the final column of Table 6.  In our statistical analysis (Section 3 and Appendix B) we use our revised 
column densities and error bars.  

\paragraph{\#1: SDSS\,J092554.70$+$400414.1, $z_{\rm AGN} = 0.471139$, $z_{\rm abs} = 0.24770$.}

See Figure 9.  Because the COS observations for this sight line only covered wavelengths $\lambda>1164$~\AA, 
higher-order Lyman lines (above \Lye) and the Lyman edge are not accessible.  However, \Lya\ shows 
a sub-DLA with damping wings that we fit with $\log N_{\rm HI} =19.45\pm0.05$, a value slightly smaller than
that ($\log N = 19.55 \pm 0.15$) quoted by Battisti \etal\ (2012), Meiring \etal\ (2013), and Lehner \etal\ (2013).  
Although our CoG solution for \Lya\ -  \Lye\ gives a lower value ($\log N = 19.33 \pm 0.03$, $b = 37\pm5$ \kms),
we adopt $\log N _{\rm HI} = 19.45 \pm 0.10$ and $b = 35\pm5$~\kms\ because the lower column density from 
the CoG does not reproduce the observed \Lya\ damping wings.
 
\paragraph{\#2: FIRST\,J020930.7$-$043826, $z_{\rm AGN} =  1.13194$, $z_{\rm abs} = 0.39043$.}
 
See Figure 9.  The flux at $\lambda<1270$~\AA\  is consistent with zero ($<1\%$) transmission in the Lyman 
continuum (optical depth $\tau_{\rm LL} > 4.6$ and $\log N_{\rm HI} > 17.86$).   Lines above Ly12 suffer from
crowding and are difficult to measure accurately.  The CoG fit to \Lya\ - Ly12  constrains $\log N_{\rm HI}$ 
to the range 18.1-18.9 with $b = 45 \pm 5$ \kms, in agreement with the observed line profiles.  In their
Appendix A, Tejos \etal\ (2014) quoted $\log N_{\rm HI} = 18.87\pm0.03$ with $b = 34\pm1$~\kms.  
However, this error bar seems far too small, and we adopt $\log N_{\rm HI} = 18.5 \pm 0.4$.  

\paragraph{\#3: SDSS\,J151428.64$+$361957.9, $z_{\rm AGN} =  0.694596$, $z_{\rm abs} = 0.41065$.}

See Figure 9.  Low-quality data ($S/N \sim 6$ per resolution element) limits the accuracy of the Ly decrement 
method.  The continuum has undulations (1300--1340~\AA) arising from AGN emission lines of \NeVIII\ and 
\OIV.  From transmission shortward of the Ly edge, we find $\log N_{\rm HI} = 17.89 \pm 0.06$.  
The Lyman lines show a single component, with a CoG fit to \Lya\ - Ly12 giving 
$\log N_{\rm HI} =18.04\pm0.20$ and $b= 43 \pm 2$~\kms.  Two absorption components are seen in
 \SiIII\ in this system, not apparent in the \HI\ lines.  We adopt the decrement measurement,
 $\log N_{\rm HI}  = 17.89\pm0.06$, and find that $b=45\pm5$~\kms\  reproduces the observed
 line profiles.
 
\paragraph{\#4: SDSS\,J113327.78$+$032719.1, $z_{\rm AGN} = 0.525073$, $z_{\rm abs} = 0.23740$.}

See Figure 9.  The initial redshift measurement of this system at $z=0.2466$ (Stevans \etal\ 2014) placed it 
within our first redshift bin ($z_{\rm abs} = 0.24-0.28$).  The higher Lyman lines ($n>11$) lie blueward of the edge 
of the COS spectra, and no Ly decrement measurement is possible.  A more careful examination of the 
data shows two absorbers at redshifts slightly below $z = 0.24$, and thus outside bin 1.  Our CoG 
fits to the red and blue components are inconsistent with the higher-order Lyman line profiles, suggesting 
the presence of additional components.  We find qualitative agreement with the line profiles for \Lya\ - Ly10 
using a three-component model:
Component~1 ($z_1 = 0.2374$) with $\log N_1 = 17.2\pm0.1$, $b_1 = 35\pm5$~\kms; 
Component~2 ($z_2 = 0.2378$) with $\log N_2= 15.8\pm0.1$, $b_2 = 25\pm5$~\kms; 
Component~3 ($z_3 = 0.2383$) with $\log N_3 = 16.5\pm0.1$, $b_3 = 25\pm5$~\kms.  
The velocity separations are $\Delta v_{12} = 97$~\kms\ and $\Delta v_{23} = 121$~\kms.
We adopt the summed total column density $\log N_{\rm HI}  = 17.29\pm0.11$.

\paragraph{\#5: SDSS\,J100902.06$+$071343.8, $z_{\rm AGN} = 0.455631$, 
$z_{\rm abs} = 0.35586$.}
 
See Figure 9.  The initial estimate of this system, $\log N = 17.41 \pm 0.04$  (Stevans \etal\ 2014) underpredicts
 the Ly decrement significantly.  The continuum at $\lambda<1240$~\AA\ is consistent with zero flux 
($\tau_{\rm LL} > 5$, $\log N >17.9$), and the lower-order Lyman lines (\Lya\ - Ly6) are contaminated 
with other absorption.  We adopt a CoG fit to Ly7 - Ly12 with $\log N_{\rm HI} = 18.3 \pm 0.2$ and 
$b = 37\pm2$~\kms, consistent with the decrement and providing a good match to the line profiles.  
This column is similar to the range ($18.0-18.8$) quoted by Tumlinson \etal\ (2011b) and the value 
($\log N_{\rm HI} = 18.4\pm0.2$) in Lehner \etal\ (2013).  This sight line also contains a DLA (Meiring
\etal\ 2011) at $z_{\rm abs} = 0.1140$ with $\log N_{\rm HI} = 20.68 \pm 0.10$ (see Table 8). 

\paragraph{\#6: SDSS\,J100102.55$+$594414.3 (SBS\,0957$+$599), 
$z_{\rm AGN} =  0.746236$, $z_{\rm abs} = 0.30355$.}
 
See Figure 9.  The Stevans \etal\ (2014) CoG measurement of this system ($\log N_{\rm HI} = 17.27\pm0.04$) 
significantly underpredicts the observed Lyman decrement. The continuum at $\lambda<1194$~\AA\ 
is consistent with zero  flux ($\tau _{\rm LL} >5$, $\log N_{\rm HI}  > 17.9$), and significant damping 
wings in the \Lya\ profile imply an even higher column density.   Battisti \etal\ (2012) found 
$\log N_{\rm HI} = 19.32\pm0.10$ by fitting the \Lya\ profile wings.  We use a CoG fit to \Lya\ - Ly11, 
constrained by the requirement that $\log N_{\rm HI} > 18$, to find $\log N_{\rm HI}  = 19.41\pm0.12$
and $b = 45 \pm1$ \kms, consistent with both the line profiles and the Ly decrement. 

\paragraph{\#7: SDSS\,J154553.48$+$093620.5, $z_{\rm AGN} =  0.665$, $z_{\rm abs} = 0.47379$.}
 
See Figure 8.  This absorber exhibits a strong Lyman decrement ($\tau_{\rm LL} > 5$) implying 
$\log N_{\rm HI}  > 17.9$.  The transmitted flux shows no recovery from the LL ($\lambda_{\rm obs} =
1344$~\AA) shortward to the end of the COS/G130M data ($\lambda_{\rm obs} \approx 1155$~\AA),
which requires $\log N_{\rm HI} \geq 18.1$.  A CoG fit to \Lyb\ - Ly12 is gives $\log N = 17.67\pm0.15$,  
inconsistent with the Ly decrement.   The \Lya\ absorption line lies redward of the end of the COS/FUV 
data, and we cannot use its damping wings as a measurement of column density.  The redward wing 
of \Lyb\ limits $\log N < 19.0$, but the CoG suggests $\log N_{\rm HI} < 18.5$.  
The dominant absorption in metal lines  (\CII, \SiII, and \OVI) occurs at $z_1 = 0.4738$, with additional 
absorption 150~\kms\ redward ($z_2 = 0.4743$) seen in \CII\ and \OVI.  However, this second component 
cannot explain the redward wing of \Lyb.   We adopt a range $18.1 < \log N_{\rm HI}  \leq 18.5$, treated 
statistically as $\log N_{\rm HI} = 18.3 \pm 0.2$, and find that $b = 25-35$ \kms\ matches the line profiles.
 
\paragraph{\#8: SDSS\,J091029.75$+$101413.6, $z_{\rm AGN} =  0.463194$, $z_{\rm abs} = 0.41924$.}
 
This AGN sight line also contains system \#14 (see below).  The low-S/N G130M data makes line 
measurements somewhat uncertain.  The UV continuum is contaminated by prominent AGN 
emission lines:  \OI\ (1355-1370~\AA\ observed), \CIII\ $\lambda977$ (1430--1450~\AA\ observed), and 
\OVI\ $\lambda1035$ (1490--1520~\AA\ observed).  
Fitting a continuum through line-free regions at 1390-1410~\AA, 1460-1480~\AA, and 1490-1520~\AA, 
we find a Lyman decrement with $\tau_{\rm LL} = 0.49^{+0.05}_{-0.11}$ ($\log N = 16.89^{+0.04}_{-0.11}$).  
A CoG  fit gives $\log N_{\rm HI} = 16.51\pm0.09$ with $b = 38 \pm 8$~\kms, inconsistent with 
the Ly decrement.  We adopt the decrement value, with its asymmetric error bars,
$\log N_{\rm HI} = 16.89^{+0.04}_{-0.11}$.

\paragraph{\#9: SBS\,1108$+$560, $z_{\rm AGN} =  0.766619$, $z_{\rm abs} = 0.46334$.}

See Figure 5. The CoG solution to this system ($\log N \sim 17.1$) significantly underpredicts the flux
in the Lyman continuum.  The spectrum has high S/N, allowing us to estimate the Ly decrement to great 
precision.  The transmitted continuum shortward of the Lyman edge lies {\it above} the interstellar 
absorption line of \CII\ $\lambda 1334.53$; we also see \SiII\ absorption lines at 1190~\AA\ and 1193~\AA.  
The transmitted flux in the Lyman continuum is $1.2\pm0.2$\%, implying optical depth 
$\tau_{\rm LL}  = 4.42\pm0.15$ and our adopted value, $\log N_{\rm HI} =17.85\pm0.02$.  With this column density, 
we find that $b = 18\pm2$ \kms\ provides a good match to the relatively narrow Lyman line profiles.  
A second, blue component visible in \Lye\ - Ly13 is fitted with a CoG having $\log N_{\rm HI} = 15.9\pm0.05$ 
and $b = 20\pm5$~\kms.  The combination of these two components, the weaker of which makes little 
contribution to the Ly decrement, fits the line profiles well.

\paragraph{\#10: PG\,1206$+$459, $z_{\rm AGN} =  1.16254$, $z_{\rm abs} = 0.92772$.}
 
See Figure 9.  This high-redshift absorption system is observable by COS only in the higher Lyman lines 
(Ly6 and above).  However, the high-quality of the data and the Ly decrement allow accurate measurements
for this pLLS (Tripp \etal\ 2011).  Two components appear in the $n \ge 9$ Lyman lines.  The optical
depth,  $\tau_{\rm LL} = 0.86\pm0.02$, implies a total column density $\log N_{\rm tot} =17.15\pm0.05$, 
but the allocation to the two components remains uncertain.  Two components fit the profiles well
($\Delta v = 156$~\kms):  
Component~1 (at $z_1 = 0.9270$) with $\log N_1=16.95\pm0.05$, $b_1 = 28\pm5$~\kms; 
Component~2 (at $z_2 = 0.9280$) with $\log N_2 = 16.3\pm0.1$, $b_2 = 25\pm5$~\kms.
Their column densities sum to $\log N_{\rm tot} = 17.04\pm0.06$ and account for the Ly decrement:  
In their study of this absorber, Tripp \etal\ (2011) fitted three clusters of 9 velocity components, 
which they labeled Group~A ($\log N_A = 15.74$), Group~B ($\log N_B = 16.99$), and Group~C 
($\log N_A = 15.98$).  Their summed column density ($\log N_{\rm tot} = 17.05 \pm 0.11$) is 0.05 dex 
less than our adopted value, $\log N_{\rm HI} = 17.10 \pm 0.10$, which is intermediate between our
CoG and LL measurements.

\paragraph{\#11: PG\,1630$+$377, $z_{\rm AGN} =  1.47607$, $z_{\rm abs} = 0.27395$.}
 
See Figure 9.  Our measurement of the Lyman decrement ($\tau_{\rm LL} = 0.69\pm0.04$) requires total 
column density $\log N=17.04\pm0.05$, comparable to previously reported values of $17.06\pm0.05$ 
(Ribaudo etal\ 2011b), $16.98\pm0.05$ (Lehner \etal\ 2013), and $16.92\pm0.04$ (Stevans \etal\ 2014).  
Higher-order lines (Ly7 - Ly13) show two components ($\Delta v = 75$~\kms):  a strong, blue component 
($z_1 = 0.27395$, $\log N_1 = 16.95\pm0.05$, $b_1 = 28\pm5$~\kms) and a weaker, red component 
($z_2 = 0.27427$, $\log N_2= 16.3\pm0.1$, $b_2 = 25\pm5$~\kms). 
We adopt $\log N_{\rm tot} = 17.04\pm0.05$.  
 
\paragraph{\#12: SDSS\,J100535.24$+$013445.7, $z_{\rm AGN} =  1.0809$, $z_{\rm abs} = 0.83711$.}

The Ly decrement from this absorber is determined by $N_{\rm HI}$ of both system \#12 and \#52 
($z = 0.83938$) located 370~\kms\ to the red.  The LL optical depth was fitted to 
$\tau_{\rm LL} = 0.414 \pm 0.03$ ($\log N_{\rm tot} = 16.82 \pm 0.04$).  System \#12 consists of
two closely separated ($\Delta v = 77$~\kms) velocity components.  CoG fits to the Lyman lines, 
where separable in Ly7 - Ly13, give:  
Component~1 ($z_1 = 0.83690$) with $\log N_1 = 16.53\pm0.07$, $b_1= 25\pm3$~\kms;  and
Component~2 ($z_2 = 0.83737$) with $\log N_2 = 16.31\pm0.09$, $b_2 = 20\pm3$~\kms\
summing to our adopted value $\log N_{\rm tot} = 16.73 \pm 0.10$ for system \#12.  
Including the (CoG-fitted) $\log N_{\rm HI} = 16.09 \pm 0.05$ for system \#52, we find 
$\log N_{\rm tot} = 16.82 \pm 0.10$, consistent with the Ly decrement.  Although system \#12 is 
blended with system \#52, we report them separately because they are easily separable 
($\Delta v \approx 390$~\kms) in their Lyman line absorption.  
 
\paragraph{\#13: SDSS\,J143511.53$+$360437.2, $z_{\rm AGN} =  0.428593$, $z_{\rm abs} = 0.37297$.}
 
The AGN continuum appears to have  broad \OI\ emission features (1310--1335~\AA\ observed frame).
The observed decrement ($\tau_{\rm LL} = 0.42\pm0.05$) implies $\log N_{\rm HI} \approx 16.82\pm0.05$.
The CoG solution to \Lya\ - Ly13, omitting \Lyg, \Lyd, and Ly11 which are blended, gives a consistent 
(adopted) solution, $\log N_{\rm HI} =16.84\pm0.06$, $b=19\pm1$~\kms.  This sight line also contains a 
sub-DLA at $z_{\rm abs} = 0.2026$ with $\log N_{\rm HI} = 19.80 \pm 0.10$ (see Table 7).  
 
\paragraph{\#14: SDSS\,J091029.75$+$101413.6, $z_{\rm AGN} =  0.463194$, $z_{\rm abs}= 0.26340$.}
 
See also system \#8 above.  Low-quality data ($S/N \sim 2$) and an undulating AGN continuum in the 
region of the Lyman limit make decrement measurements unreliable.  However, the uncertain values, 
$\log N_{\rm HI} = 16.86\pm0.5$ (Stevans \etal\ 2014) and $16.58^{+0.04}_{-0.07}$ (Prochaska \etal\ 2017) 
are not supported, as we detect no Ly decrement ($\tau_{\rm LL} < 0.1$).   Our CoG fit to unblended lines 
of \Lyb, \Lyg,  \Lye, and Ly6 - Ly10 gives $\log N_{\rm HI} = 16.21 \pm 0.07$, $b=21\pm2$~\kms, consistent 
with the lack of a measurable Ly decrement 
 
\paragraph{\#15: SDSS\,J161916.54$+$334238.4, $z_{\rm AGN} =  0.470946$, $z_{\rm abs} = 0.47088$.}

This absorber is offset by just 146~\kms\ blueward of the AGN systemic velocity.    The LL lies on the red 
wing of a damped \Lya\ absorber profile at 1332.7~\AA\ ($z_{\rm DLA} = 0.0963$)
with $\log N_{\rm HI} = 20.55\pm0.10$ (see Table 7). However, there is sufficiently recovered continuum in 
the blue wing of the DLA (1310--1320~\AA) to measure a LL decrement with $\tau_{\rm LL} =0.43 \pm 0.05$ 
and $\log N_{\rm HI} =16.83\pm0.05$.  This solution fits the stronger Lyman lines (\Lya\ - Ly8) but increasingly 
overpredicts the higher-order absorption profiles ($n\ge9$).  We adopt $\log N_{\rm HI} = 16.83 \pm 0.10$ 
and $b=32\pm5$~\kms, noting unresolved structure in the system.

\paragraph{\#16: SDSS\,J155048.29$+$400144.9, $z_{\rm AGN} =  0.496843$, $z_{\rm abs} = 0.31257$.}

This sight line has two strong absorbers (\#16 and \#17).  The redshift of this absorber places the LL
in the complicated region ($1200-1220$~\AA) of the Galactic DLA and geocoronal \NI\ and \HI\ emission 
lines.  The AGN continuum may be contaminated by broad \OIV\ $\lambda 788$ emission (observed between 
1170--1185~\AA).  The observed Ly decrement (1190--1197~\AA) is uncertain ($\tau_{\rm LL} = 0.13\pm0.03$) 
implying $\log N_{\rm HI} = 16.31\pm0.10$, a factor of 2 lower than that ($16.62\pm0.06$) found by 
Stevans \etal\ (2014) from a CoG fit.  Additional absorption is present in some of 
the higher-order Lyman lines ($n>8$) which is not accounted for by this lower-$N$ solution.
This may have driven the previous CoG solution to a higher column density. 
We adopt $\log N_{\rm HI}  = 16.3\pm0.1$ and $b=40\pm5$~\kms.
 
\paragraph{\#17: SDSS\,J155048.29$+$400144.9, $z_{\rm AGN} =  0.496843$, $z_{\rm abs} = 0.49200$.}
 
The Lyman decrement ($\tau_{\rm LL}  = 0.25\pm0.03$, $\log N_{\rm HI} = 16.60\pm0.05$)
is consistent with the Stevans \etal\ (2014) value, $\log N_{\rm HI} = 16.57\pm0.05$.  We adopt 
$\log N_{\rm HI}  = 16.60\pm0.05$ with $b = 25\pm5$~\kms.
 
\paragraph{\#18: PKS\,0552$-$640, $z_{\rm AGN} =  0.680$, $z_{\rm abs} = 0.34513$.}
 
See Figure 6.  This system lies on the red edge of a damped \Lya\ profile (the Galactic DLA).  
However, the Lyman edge lies $\sim10$~\AA\ redward of the line center where the continuum 
has recovered from DLA absorption.  The decrement ($\tau_{\rm LL} =0.64\pm0.04$) requires 
$\log N_{\rm HI} =17.01\pm0.03$.  The data are of very high quality ($S/N\sim25$), and we are 
able to measure a single, narrow absorber in 24 lines (\Lya\ to Ly$\omega$).   The highest-order 
lines ($n\ge16$) are blended and not useful in CoG analysis as noted in Table 2.  A fit to 
\Lyb\ - Ly15 gives $\log N_{\rm HI} = 17.01\pm0.05$, $b=16\pm2$~\kms.  This column density 
is above previous values, $16.90\pm0.08$ (Lehner \etal\ 2013) and $16.71\pm0.03$ 
(Stevans \etal\ 2014).  

\paragraph{\#19: SDSS\,J124511.25$+$335610.1, $z_{\rm AGN} =  0.711698$, $z_{\rm abs} = 0.5567$.}
 
This AGN sight line has three strong absorbers (systems \#19, \#20, and \#46) which we fit simultaneously
to set the continuum.  The decrement for this system gives $\tau_{\rm LL} =0.12 \pm 0.02$, although the 
local continua may be contaminated by AGN emission lines.  The LL-inferred column density,
$\log N_{\rm HI} =16.28\pm0.07$, is much smaller than the CoG fit, $\log N_{\rm HI} = 16.63 \pm 0.09$,
but additional absorption appears in the higher ($n>8$) Lyman lines.  
We adopt $\log N_{\rm HI} = 16.28\pm0.10$ based on the Ly decrement, with an increased error.  
 
\paragraph{\#20: SDSS\,J124511.25$+$335610.1, $z_{\rm AGN} =  0.711698$, $z_{\rm abs} = 0.68927$.}

As noted for system \#19, we use a multi-component fit to set the continuum.  The Lyman decrement 
($\tau_{\rm LL}  = 0.37\pm0.03$) implies $\log N_{\rm HI}  = 16.77\pm0.03$, somewhat larger than 
the value ($16.68\pm0.2$) from Stevans \etal\ (2014).   The local continua may be contaminated by 
AGN emission lines.   A two-component profile is seen in the $n>6$ lines.   
Our CoG fits two closely separated ($\Delta z = 75$~\kms) individual components in Ly6 - Ly11 gives:  
Component~1 ($z_1 = 0.68894$) with $\log N_1=16.57 \pm 0.10$, $b_1 = 22\pm3$~\kms, and 
Component~2 ($z_2 = 0.68936$) with $\log N_2 = 16.10\pm0.1$, $b_2 = 35\pm5$~\kms, summing 
to our adopted column density, $\log N_{\rm tot} = 16.70\pm0.11$, similar to the observed decrement.

\paragraph{\#21: SDSS\,J100102.55$+$594414.3 (SBS\,0957$+$599), $z_{\rm AGN} =  0.746236$, 
    $z_{\rm abs} = 0.41595$.}

The AGN continuum is contaminated by the prominent emission lines of \NeVIII\ ($\lambda 770, 780$)
and \OIV\ ($\lambda 788$) observed at 1320--1380~\AA.  Shortward of the Lyman edge, the flux
level is contaminated by the prominent 700~\AA\ lines of \OIII\ and \NIII\ (1220--1250~\AA\ observed frame).
The Ly decrement is somewhat uncertain because of this line emission and continuum placement.   
Choosing continua redward and blueward of the LL to avoid these emission lines, our best fit gives 
$\tau_{\rm LL} = 0.29\pm0.04$ and $\log N_{\rm HI} = 16.66\pm0.06$.
Two closely separated ($\Delta v = 95$~\kms) components are apparent in \Lyd\  - Ly12  
(geocoronal airglow is blended with Ly8 - Ly10).  The redder component is the stronger and 
dominates the LL decrement.  We adopt 
$z_2 = 0.41560$, $\log N_1 = 15.9\pm0.05$,   $b_1 = 24\pm5$~\kms, and
$z_2 = 0.41605$, $\log N_2 = 16.45\pm0.05$, $b_2 = 35\pm5$~\kms, summing to 
$\log N_{\rm tot} = 16.56 \pm 0.05$.  Because of CoG uncertainties, we adopt the Ly decrement value
with an increased error, $\log N_{\rm HI} = 16.66\pm0.08$.

\paragraph{\#22: HB89\,0107$-$025-NED05, $z_{\rm AGN} =  0.956$, $z_{\rm abs} = 0.39909$.}
 
The continuum is well defined above the edge, despite a few AGN emission features (the 700~\AA\
feature from \OIII\ and \NIII\ observed near 1370~\AA).  Our fit to the Lyman decrement 
($\tau_{\rm LL} = 0.50\pm0.02$) implies $\log N_{\rm HI} =16.89\pm0.03$.  A CoG fit to the system
(\Lyg\ - Ly15) gives $\log N_{\rm HI} =16.77\pm0.06$ and $b=25\pm2$~\kms.  We adopt the Ly decrement 
solution with a wider the error range, $\log N_{\rm HI} = 16.89 \pm 0.06$. 
 
\paragraph{\#23: SDSS\,J084349.49$+$411741.6, $z_{\rm AGN} =  0.989986$, $z_{\rm abs} = 0.53255$.}

The LL optical depth, $\tau_{\rm LL} = 0.45\pm0.02$, implies $\log N_{\rm HI} = 16.85\pm0.02$.  However,
the decrement is determined by column densities of both system \#23 and \#51 ($z = 0.53356$), located 
198~\kms\ to the red.   A 2-component CoG to system \#23 reproduces the line profiles well with 
$\log N_{\rm HI} \approx 16.67 \pm 0.08$.  Including $\log N_{\rm HI} = 16.11 \pm 0.05$ for system \#51
gives a total column density $\log N_{\rm tot} = 16.78 \pm 0.07$.  We adopt the Ly decrement value,
but widen the error range to $\log N_{\rm HI} = 16.85\pm0.04$.

\paragraph{\#24: SDSS\,J100535.24$+$013445.7, $z_{\rm AGN} =  1.0809$, $z_{\rm abs} = 0.41865$.}

Three velocity components (denoted \#24abc) contribute to the Ly decrement, with optical depth  
as $\tau_{\rm LL} = 0.58 \pm 0.04$ ($\log N_{\rm HI} = 16.96 \pm 0.03$).    Because the component 
splittings are easily separable, $\Delta v_{12} = 216$~\kms\ and $\Delta v_{23} = 243$~\kms, we
treat these as distinct absorbers, where we adopt:
Component~1 ($z_1 = 0.41755$) with $\log N_1= 14.88 \pm 0.06$, $b_1 = 25\pm2$~\kms; 
Component~2 ($z_2 = 0.41851$) with $\log N_2 = 16.69 \pm 0.04$, $b_2 = 30\pm2$~\kms;
Component~3 ($z_3 = 0.41966$) with $\log N_3 = 15.91 \pm 0.08$, $b_3 = 22\pm2$~\kms, 
summing to $\log N_{\rm tot} = 16.76\pm0.05$.  The LL optical depth is reasonably consistent 
with this 3-component CoG sum.  

\paragraph{\#25: PG\,1522$+$101, $z_{\rm AGN} =  1.32785$, $z_{\rm abs} = 0.72865$.}
 
The decrement ($\tau_{\rm LL} = 0.26\pm0.03$) implies $\log N_{\rm HI} = 16.62\pm0.05$, 
comparable to values ($16.66\pm0.05$ and $16.60\pm0.04$) reported by Lehner \etal\ (2013) 
and Stevans \etal\ (2014).  Combined with our CoG fit to the higher Lyman lines, we adopt 
$\log N_{\rm HI} = 16.62 \pm 0.05$ and $b = 29\pm5$~\kms.  
 
\paragraph{\#26: Q0232$-$042, $z_{\rm AGN} =  1.43737$, $z_{\rm abs} = 0.73888$.}

Our fit to the Ly decrement implies $\log N_{\rm HI} =16.77\pm0.02$, slightly stronger than 
the Stevans \etal\ (2014) CoG solution of $16.64\pm0.08$ and the value $16.72\pm0.03$ 
from Lehner \etal\ (2013).  Including our CoG fit to the higher Lyman lines, we adopt 
$\log N_{\rm HI} = 16.76\pm0.05$ and $b=30\pm5$~\kms.  
 
\paragraph{\#27: PG\,1522$+$101, $z_{\rm AGN} =  1.32785$,  $z_{\rm abs} = 0.51839$.}

Also known as PHL~1377.  The AGN continuum appears contaminated by broad emission 
lines (\OIV\ $\lambda 608$ and \OV\ $\lambda 630$).  A weak Ly decrement is visible 
($\log N_{\rm HI} =16.1-16.3$) with uncertain depth owing to AGN line emission.   We performed 
CoG fits to two closely separated components ($\Delta v = 99$~\kms):  
Component~1 ($z_1 = 0.5180$, $\log N_1 = 15.64\pm0.04$, $b_1 = 15\pm1$~\kms) and 
Component~2 ($z_2 = 0.5185$, $\log N_2 = 16.17\pm0.04$, $b_2 = 17\pm1$~\kms) summing 
to $\log N_{\rm tot} = 16.28\pm0.05$.  This column density is somewhat larger than that
($16.22\pm0.02$) quoted by Lehner \etal\ (2013) but comparable to that ($16.32\pm0.2$) in
Stevans \etal\ (2014).  We adopt our summed 2-component CoG solution, 
$\log N_{\rm tot} = 16.28\pm0.05$, with the error based on combining the CoG with the 
Ly decrement constraint. 

\paragraph{\#28: PG\,1338$+$416, $z_{\rm AGN} =  1.21422$, $z_{\rm abs} = 0.68610$.}

This AGN sight line has three strong absorbers in our survey (\#28, \#29, and \#53).   
The continuum is likely contaminated by AGN broad emission feature of \OIII\ and \NIII\
near 700~\AA\ (observed near 1550~\AA).  The fitted Ly decrement  
($\tau_{\rm LL} = 0.188\pm0.030$) implies $\log N_{\rm HI} =16.47\pm0.07$ with the 
error bar arising from the uncertain placement of the continuum shortward of the LL.  
From CoG fitting, we adopt $\log N_{\rm HI} = 16.49\pm0.05$ with $b = 20\pm5$~\kms, 
comparable to previous values of $16.45\pm0.05$ (Lehner \etal\ 2013), $16.49\pm0.04$
(Stevans \etal\ 2014), and $16.49\pm0.04$ (Tilton \etal\ 2016).  

\paragraph{\#29: PG\,1338$+$416, $z_{\rm AGN} =  1.21422$, $z_{\rm abs} = 0.34885$.}

As noted for system \#28, the AGN continuum is likely contaminated by AGN broad emission 
lines, in this case from \OIV\ $\lambda 554$ and \NeV\ $\lambda 570$ observed between 
1240-1260~\AA.  The Ly decrement ($\tau_{\rm LL} = 0.147\pm0.052$) implies 
$\log N_{\rm HI} = 16.37\pm0.19$.  The higher error arises from uncertain placement of the 
continuum shortward of the LL.   We performed CoG fits to two components separated by
$\Delta v = 76$~\kms:  
Component~1 ($z_1 = 0.34855$, $\log N_1 = 16.28\pm0.02$, $b_1 = 32\pm1$~\kms) and 
Component~2 ($z_2 = 0.34889$, $\log N_2 = 15.44\pm0.18$, $b_2 = 18\pm3$~\kms) summing 
to $\log N_{\rm tot} = 16.34\pm0.05$. 
We adopt $\log N_{\rm HI} = 16.34\pm0.10$, comparable to previous values of $16.30\pm0.13$ 
(Lehner \etal\ 2013) and $16.37\pm0.06$ (Stevans \etal\ 2014).

\paragraph{\#30: PKS\,0637$-$752, $z_{\rm AGN} =  0.653$, $z_{\rm abs} = 0.46850$.}

The fit to the Ly decrement implies $\log N_{\rm HI} =16.48\pm0.02$, but the placement of the continuum 
shortward of the LL is somewhat uncertain.  Using CoG fitting of \Lyb\ - Ly17, we find 
$\log N_{\rm HI}  = 16.43\pm0.05$ and $b = 18\pm5$~\kms, a column density similar to the value of
$16.48\pm0.04$ (Lehner \etal\ 2013) but higher than $16.08\pm0.03$ (Stevans \etal\ 2014).  
We adopt the Ly decrement value with a wider error, $\log N_{\rm HI} = 16.48\pm0.05$.  

\paragraph{\#31: SDSS\,J141910.20$+$420746.9, $z_{\rm AGN} =  0.873501$, $z_{\rm abs} = 0.28895$.}

This AGN sight line has four strong absorbers in our survey (\#31, \#32, \#33, and \#50). The AGN continuum is
contaminated by the 700~\AA\ emission lines (\OIII\ and \NIII) observed between 1300-1320~\AA.  The 
Ly decrement is below the detectable level ($\tau_{\rm HI}  < 0.1$,  $\log N_{\rm HI}  <16.20$).  Using
CoG fitting, we adopt $\log N_{\rm HI} = 16.17\pm0.05$ with $b = 25\pm5$~\kms.  This column is the 
same as quoted by Stevans \etal\ (2014) but below that ($16.40\pm0.07$) quoted in Lehner \etal\ (2013), 
which would produce a larger Lyman decrement than observed. 

\paragraph{\#32: SDSS\,J141910.20$+$420746.9, $z_{\rm AGN} =  0.873501$, $z_{\rm abs} = 0.42555$.}

The continuum near the LL (1300~\AA\ observed frame) is not flat, with no obvious decrement
($\tau_{\rm LL}  < 0.15$ and $\log N_{\rm HI} < 16.38$).  This edge is contaminated by the 700~\AA\ broad
emission feature of \OIII\ and \NIII\ (observed between 1300-1315~\AA).  Using CoG fitting, we adopt
 $\log N_{\rm HI} = 16.24 \pm 0.10$ and $b = 23\pm3$~\kms.  This column density is slightly above previous 
 values of $16.17\pm0.06$ Lehner \etal\ 2013) and $16.02\pm0.02$ (Stevans \etal\ 2014).  

\paragraph{\#33: SDSS\,J141910.20$+$420746.9, $z_{\rm AGN} =  0.873501$, $z_{\rm abs} = 0.53460$.}

The continuum is contaminated by AGN emission lines of \NeVIII\ and \OIV\ observed between 1445-1475~\AA.
We see no obvious Ly decrement ($\tau_{\rm LL}  < 0.1$, $\log N_{\rm HI}  < 16.20$).  Using CoG fitting, 
we adopt $\log N_{\rm HI} = 16.15\pm0.10$ and $b =16\pm2$~\kms, a column intermediate between 
values of $16.34(+0.23,-0.12)$ (Lehner \etal\ 2013) and $16.06\pm0.07$ (Stevans \etal\ 2014).   

\paragraph{\#34: SDSS\,J143511.53$+$360437.2, $z_{\rm AGN} =  0.428593$, $z_{\rm abs} = 0.38766$.}

We see a Ly decrement ($\tau_{\rm LL} = 0.10\pm0.03$, $\log N_{\rm HI} = 16.20\pm0.12$) based on a 
small flux decrement between 1258-1265~\AA).   Prochaska \etal\ (2017) quote $\log N_{\rm HI} <  16.65$
from low-resolution (COS/G140L) data.  
Using CoG fitting, we adopt $\log N_{\rm HI} = 16.17\pm0.05$ and $b =32\pm5$~\kms.  This column 
density is similar to values of $16.18\pm0.05$ (Lehner \etal\ 2013) and $16.15\pm0.02$ (Stevans \etal\ 2014).  

\paragraph{\#35: SDSS\,J094331.61$+$053134.4, $z_{\rm AGN} =  0.564336$, $z_{\rm abs} = 0.35455$.}

Because of continuum undulations (1220-1240~\AA), we cannot measure a reliable Ly decrement.  
CoG fitting gives values identical to those of Lehner \etal\ (2013) and Stevans \etal\ (2014).  We adopt 
$\log N_{\rm HI} = 16.12 \pm 0.05$ with $b = 25\pm5$~\kms\ from CoG fitting.

\paragraph{\#36: PG\,1407$+$265, $z_{\rm AGN} =  0.946$, $z_{\rm abs} = 0.68270$.}

Based on high quality data, we measure a Ly decrement, $\tau_{\rm LL} = 0.153 \pm 0.012$, implying 
$\log N_{\rm HI}  = 16.38\pm0.03$, comparable to previous values of $16.38 \pm 0.02$ (Lehner \etal\ 2013) 
and $16.39\pm0.03$ (Stevans \etal\ 2014).  Including CoG fitting, we adopt 
$\log N_{\rm HI} = 16.38 \pm 0.05$ with $b = 32\pm5$~\kms.  

\paragraph{\#37: PG\,1216$+$069, $z_{\rm AGN} =  0.3313$, $z_{\rm abs} = 0.28231$.}

See Figure 7.  Based on high quality data, we measure a Ly decrement, 
$\tau _{\rm LL} = 0.16 \pm 0.02$, implying $\log N_{\rm HI} = 16.40\pm0.06$.  CoG fitting to 
Lye\ - Ly14 gives $\log N_{\rm HI} = 16.41\pm0.03$ and $b = 25\pm2$~\kms, comparable to 
$16.40\pm0.05$ (Lehner \etal\ 2013). 

\paragraph{\#38: SDSS\,J161916.54$+$334238.4, $z_{\rm AGN} =  0.470946$, $z_{\rm abs} = 0.26938$.}

We see no obvious Ly decrement, but the wavelength calibration near the Lyman edge (1150-1165~\AA) 
is uncertain.  Using CoG fitting, we adopt $\log N_{\rm HI} = 16.40\pm0.08$ and $b = 29\pm2$~\kms.  
This column density is similar to previous values $16.48\pm0.05$ (Lehner \etal\ 2013) and 
$16.40\pm0.03$ (Stevans \etal\ 2014).  

\paragraph{\#39: SBS\,1122$+$594, $z_{\rm AGN} =  0.8514$, $z_{\rm abs} = 0.55810$.}

The AGN continuum near the Ly edge is uncertain, owing to broad emission lines of \NeVIII\ 
and \OIV\ observed between 1440-1470~\AA.  Fitting a continuum below those emission features,
we estimate a Ly decrement of $\tau_{\rm LL}  = 0.18\pm0.04$ and $\log N_{\rm HI} = 16.46\pm0.08$.  
We also fit a CoG with three velocity components separated by $\Delta v_{12} = 129$~\kms\ and
$\Delta v_{23} = 58$~\kms:
Component~1 ($z_1 = 0.55748$) with $\log N_1 = 15.79\pm0.05$, $b_1 = 27\pm3$~\kms; 
Component~2 ($z_2 = 0.55815$) with $\log N_2 = 16.13\pm0.05$, $b_2 = 21\pm2$~\kms; 
Component~3 ($z_3 = 0.55845$) with $\log N_3 = 15.94\pm0.05$, $b_3 = 21\pm3$~\kms,
summing to our adopted value, $\log N_{\rm tot}= 16.45\pm0.06$ with $b = 29\pm2$~\kms.  
This column density is comparable to the value of $16.42\pm0.02$ (Stevans \etal\ 2014) but 
above the value $16.24\pm0.02$ of Lehner \etal\ (2013).   The Ly decrement is consistent with
our higher value.

\paragraph{\#40: HE\,0439$-$5254, $z_{\rm AGN} =  1.053$, $z_{\rm abs} = 0.61512$.}

The AGN continuum below the LL is contaminated by the broad 700~\AA\ emission lines
of \OIII\ and \NIII\ observed between 1445-1455~\AA.  The continuum below the Ly edge
($\lambda < 1473$~\AA) is therefore somewhat uncertain.  From the Ly decrement, we 
estimate $\tau_{\rm LL} = 0.15\pm0.03$, implying $\log N_{\rm HI}  = 16.38\pm0.08$. 
Wit fit a CoG with three velocity components, separated by $\Delta v_{12} = 46$~\kms\ 
and $\Delta v_{23} = 89$~\kms:
Component~1 ($z_1 = 0.61495$) with $\log N_1 = 16.2\pm0.1$, $b_1 = 18\pm2$~\kms; 
Component~2 ($z_2 = 0.61520$) with $\log N_2 = 15.8\pm0.1$, $b_2 = 60\pm10$~\kms; 
Component~3 ($z_3 = 0.61568$) with $\log N_3 = 15.59\pm0.15$, $b_3 = 16\pm5$~\kms.
Our adopted summed total, $\log N_{\rm tot}= 16.42\pm0.10$, is higher than previous values
of $16.28\pm0.04$ (Lehner \etal\ 2013) and $16.25\pm0.04$ (Stevans \etal\ 2014).  We 
believe our 3-component fitting is more accurate, with a summed column density consistent 
with the (less certain) Ly decrement.  


\paragraph{\#41:  B0117$-$2837, $z_{\rm AGN} =  0.348858$, $z_{\rm abs} = 0.348330$.}

The AGN continuum is contaminated by broad emission features of \OI\ (observed at 
1265-1275~\AA) and \CIII\ (observed at 1310-1320~\AA) and by an absorption dip 
(1290-1300~\AA).  The flux shortward of the LL is complicated by the redward damping wing of 
Galactic \Lya\ absorption (1227-1232~\AA).  The Ly decrement is uncertain, 
$\tau_{\rm LL} = 0.16-0.20$ or $\log N_{\rm HI} = 16.40-16.50$, and we place more weight on 
CoG fitting.    We fit two closely separated ($\Delta v = 73$~\kms) absorbers at
$z_1 = 0.34833$ ($\log N_1 = 15.66 \pm 0.03$) and 
$z_2 = 0.34866$ ($\log N_2 = 16.00 \pm 0.04$), summing to $\log N_{\rm tot} = 16.14\pm0.04$, 
well below the estimated Ly decrement.  Because of the continuum uncertainty with the Ly 
decrement, we adopt the summed CoG value, $\log N_{\rm tot} = 16.16 \pm 0.10$ with an 
expanded error bar owing to contaminating AGN line emission. 


\paragraph{\#42:  Ton~576 (SDSS\;J111754.31$+$263416.6), $z_{\rm AGN} =  0.420466$, 
$z_{\rm abs} = 0.35194$.}

This AGN has a well-defined continuum longward of the LL ($\lambda > 1232.6$~\AA), but the 
flux recovery shortward of the Ly edge is complicated by damped Galactic \Lya\ absorption.  
Our fit to the redward wing of the Galactic DLA suggests $\tau_{\rm LL} = 0.14\pm0.03$ 
($\log N_{\rm HI} = 16.35\pm0.08$) for a flat continuum, but a downward-sloping continuum 
to shorter wavelengths gives $\log N_{\rm HI} = 16.22\pm0.08$.  The CoG from Stevans \etal\ 
(2014) gave $16.14\pm0.02$.  Our new two-component ($\Delta v = 69$~\kms) CoG fit to 
\Lyd\ - Ly9 gives values:
$z_1 = 0.35194$, $\log N_1 = 15.89\pm0.09$ and 
$z_2 =  0.35225$, $\log N_2 = 15.70\pm0.06$, 
summing to $\log N_{\rm tot} = 16.11 \pm 0.10$.  We adopt the summed CoG value, with a
slightly larger error bar, $\log N_{\rm HI} = 16.11\pm0.12$, reflecting the Ly decrement.


\paragraph{\#43:  SDSS\,J080908.13$+$461925.6, $z_{\rm AGN} =  0.656338$, $z_{\rm abs} = 0.61917$.}

The continuum is not well defined, owing to likely contamination by \OI\ emission (1560-1570~\AA\
observed frame).  A weak Ly decrement suggests $\tau_{\rm LL} = 0.089 \pm 0.015$ or 
$\log N_{\rm HI} = 16.15\pm 0.07$.  However, the continuum below the LL is uncertain.  We adopt the
value from CoG-fitting, $\log N_{\rm HI} = 16.18 \pm 0.07$ with $b = 36\pm2$ \kms.  


\paragraph{\#44:  3C\,57,  $z_{\rm AGN} =  0.670527$, $z_{\rm abs} = 0.32332$.}

Three velocity components are evident a4 $z_1 = 0.32257$, $z_2 = 0.32303$, and $z_3 = 0.32342$,
with separations ($\Delta v_{12} =  104$~\kms\  and $\Delta v_{23} = 88$~\kms.   The Ly decrement falls 
in the Galactic DLA and is unobservable.  Our new 3-component CoG fit finds 
$\log N_1 = 15.61\pm0.03$, 
$\log N_2 = 15.45 \pm 0.03$, and 
$\log N_3 = 16.12 \pm 0.09$, summing to $\log N_{\rm tot} = 16.30 \pm 0.12$ and similar to the value 
$16.29 \pm 0.01$ (Stevans \etal\ 2014).   We adopt a column density, 
$\log N_{\rm HI} = 16.30 \pm 0.05$, consistent with both CoG and Ly decrement.


\paragraph{\#45:  SDSS\,J113457.62$+$255527.9, $z_{\rm AGN} =  0.710078$, $z_{\rm abs} = 0.43233$.}

A new CoG fit gives $\log N_{\rm HI} = 16.47\pm0.06$, whereas Stevans \etal\ (2014) quoted $16.40\pm0.03$.
The continuum is fairly well defined, but portions may be contaminated by broad emission lines of \NeVIII\ and 
\OIV\ observed at 1320--1350~\AA.   The Ly decrement implies $\tau_{\rm LL} = 0.19 \pm 0.020$ and 
$\log N_{\rm HI} = 16.48\pm0.05$.  With the CoG information, we adopt  
$\log N_{\rm HI} = 16.48\pm0.06$. 


\paragraph{\#46:  SDSS\,J124511.25$+$335610.1, $z_{\rm AGN} =  0.711698$, $z_{\rm abs} = 0.71297$.}

There are three strong absorbers in this sight line, including systems \#19 and \#20.  Absorbers \#20
and \#46 have a complex flux recovery between 1540-1560~\AA, as the envelope of higher Lyman 
lines from system \#46 ($z = 0.71297$) merges with the LL of system \#20 ($z = 0.68918$).   Our fit to 
the Ly decrement gives  $\log N_{\rm HI} = 16.52\pm0.05$.  


\paragraph{\#47:  SDSS\,J143726.14$+$504558.8, $z_{\rm AGN} =  0.783319$, $z_{\rm abs} = 0.77248$.}

The data quality is poor, and the continuum has likely contamination from AGN emission lines of \OI\ 
observed at 1670-1690~\AA\ and \CIII\ $\lambda977$ observed at 1740~\AA.  No Ly decrement is 
evident to a limit $\log N_{\rm HI} < 16.6$.   Because the component splittings are easily separable, 
$\Delta v_{12} = 390$~\kms\ and $\Delta v_{23} = 230$~\kms, we treat these as distinct absorbers, 
denoted \#47abc, where
$z_1 = 0.76890$ ($\log N_1 = 16.15 \pm 0.07$), 
$z_2 = 0.77120$ ($\log N_2 = 15.4 \pm 0.2$), and 
$z_3 = 0.77255$ ($\log N_3 = 16.26 \pm 0.10$).  
The middle component is poorly constrained, with a CoG based on only \Lyg, \Lyd, and \Lye.  
The summed CoG fit gives $\log N_{\rm tot}  = 16.26 \pm 0.11$ with $b = 40\pm6$~\kms.  


\paragraph{\#48:  SDSS\,J234500.43$-$005936.0, $z_{\rm AGN} =  0.789429$, $z_{\rm abs} = 0.253900$.}

Because no data were taken at $\lambda < 1164$~\AA, we cannot measure the Ly decrement (at 1143~\AA).  
We rely on a 2-component CoG solution, which gives $\log N_1 = 15.85\pm0.10$ and $b_1 = 25\pm3$~\kms\ 
(blue component with \Lye\ and \Lyz) and $\log N_2 = 15.77\pm0.16$, $b_1 = 41\pm12$~\kms\ 
(red component with \Lyb\ to \Lyz).  These column densities sum to $\log N_{\rm tot} = 16.11\pm0.13$.


\paragraph{\#49:  SDSS\,J101622.60$+$470643.3, $z_{\rm AGN} =  0.821527$, $z_{\rm abs} = 0.72766$.}

No clear Ly decrement is seen ($\log N_{\rm HI} < 16.34$).  The continuum is uncertain longward of the 
LL because of AGN emission lines. Our CoG fit uses \Lyb\ to Ly10 to find $\log N_{\rm HI} = 16.16\pm0.09$ 
with $b = 19\pm2$ \kms.  


\paragraph{\#50:  SDSS\,J141910.20$+$420746.9, $z_{\rm AGN} =  0.873501$, $z_{\rm abs} = 0.84523$.}

A weak Ly decrement may be present with $\log N_{\rm HI} = 16.23 \pm 0.15$, with an uncertain continuum
redward of the edge owing to \OI\ emission lines observed at 1770-1790~\AA.  We fit a CoG to \Lyd\ up to 
Ly10, with our adopted value $\log N_{\rm HI} = 16.23 \pm 0.05$ and $b = 29 \pm 2$~\kms.  


\paragraph{\#51:  SDSS\,J084349.49$+$411741.6, $z_{\rm AGN} =  0.989986$, $z_{\rm abs} = 0.53556$.}

The source has a well-defined continuum, with a Ly decrement suggesting $\tau_{\rm LL} = 0.41\pm0.02$ or
$\log N_{\rm HI} = 16.81 \pm 0.02$ for a flat continuum.   This decrement includes both systems \#23 and \#51.
However, system \#23 at $z = 0.53255$ is easily separable in the Lyman lines, lying 178~\kms\ blueward.  
These absorbers are visible in blended Lyman lines (\Lyg, \Lyd, \Lye) at redshifts
$z_1 = 0.53255$ for system \#23 ($\log N_1 = 16.67\pm 0.05$) and 
$z_2 = 0.5355$ for system \#51 ($\log N_2 = 16.11 \pm 0.05$).  
Their column densities sum to $\log N_{\rm HI} = 16.78\pm0.05$, consistent with the decrement.  
In our statistics, we treat systems \#23 and \#51 as distinct absorbers, since they are easily separable 
in the Lyman absorption lines.


\paragraph{\#52:  SDSS\,J100535.24$+$013445.7, $z_{\rm AGN} =  1.0809$, $z_{\rm abs} = 0.83938$.}

The source has a well-defined continuum, with LL optical depth $\tau_{\rm LL} = 0.41 \pm 0.03$ 
($\log N_{\rm tot} = 16.82 \pm 0.04$) produced by absorption from system \#12 at $z_1 = 0.83711$ 
and system \#52 at $z_2 = 0.83938$.  These two systems are easily distinguished, separated by 
$\Delta v \approx 390$~\kms\ in the Lyman lines (Ly$\gamma$, Ly$\delta$, Ly$\epsilon$)   Our CoG 
gives $\log N_{\rm HI} = 16.09 \pm 0.05$ for system \#52.  The combined column densities of \#12 and 
\#52 are consistent with the Ly decrement.


\paragraph{\#53:  PG\,1338$+$416, $z_{\rm AGN} =  1.21422$, $z_{\rm abs} = 0.62075$.}  

This AGN sight line has three absorbers with Lyman edges near 1537~\AA\ (system \#28), 
1478~\AA\ (system \#53), and 1225~\AA\ (system \#29).  The continuum between 1540-1580~\AA\ is 
contaminated by broad emission lines of \OII\ and \OIII\ (rest-frame 833-834~\AA).  For this system,
the weak Ly decrement at 1478~\AA\ is poorly determined.   CoG fitting gives 
$\log N_{\rm HI} = 16.17\pm0.06$ for \#53, consistent with that found by Tilton \etal\ (2016).


\paragraph{\#54:  Q\,0232$-$042, $z_{\rm AGN} =  1.43747$, $z_{\rm abs} = 0.322450$.}  

This AGN is also known as PHL\,1377 (see also system \#26).  No Lyman decrement is evident.  
Our CoG fit gives $\log N_{\rm HI} = 16.14\pm0.04$ with $b = 34\pm2$~\kms.


\paragraph{\#55:  PG\,0003$+$158, $z_{\rm AGN} = 0.4509$, $z_{\rm abs} = 0.3478$.}  

The data have high S/N, and a Lyman series is evident up to Ly~12 with higher lines
intruding on the red wing of the Galactic DLA.  We use the excellent CoG fit with
$\log N_{\rm HI} = 16.10 \pm 0.03$ with $b = 17\pm1$~\kms, consistent with a weak
LL flux decrement.  


\paragraph{\#56: SDSS\,J161916.54$+$334238.4, $z_{\rm AGN} =  0.470946$, $z_{\rm abs} = 0.44231$.}  

This AGN also contains system \#15 (Lyman edge at 1341~\AA) and a DLA at 1333~\AA\ 
($z_{\rm abs} = 0.0963$ --see Table 7).  This DLA blocks the LL of system \#56 
as well as lines of Ly7 - Ly9.  A good CoG fit using \Lya\ to Ly6 plus Ly10 and Ly11 yields 
$\log N_{\rm HI} = 15.90 \pm 0.06$ with $b = 26\pm2$~\kms.  


\paragraph{\#57:  PG\,0637$-$752, $z_{\rm AGN} = 0.653 $, $z_{\rm abs} = 0.24326$.}  

The LL at 1133.5~\AA\ falls just below the COS/G130M data range, and blueward continuum
is not visible.  However, a CoG fit to \Lyg, \Lye\, Ly8, and Ly10 gives 
$\log N_{\rm HI} = 15.81\pm0.05$ with $b = 22 \pm 2$~\kms.  


\paragraph{\#58: PKS\,0552$-$640, $z_{\rm AGN} = 0.68 $, $z_{\rm abs} =0.446$.}  

A CoG fit to \Lyb\ through Ly14 gives a very good fit with $\log N_{\rm HI} = 15.96\pm0.02$ 
and $b = 28 \pm 1$~\kms.  This column density is consistent with a weak LL decrement
visible in very good data. 


\paragraph{\#59: SDSS\,J124511.25$+$335610.1, $z_{\rm AGN} = 0.717$, $z_{\rm abs} = 0.63215$.}  

The data quality is poor, with no obvious Ly decrement.  A double-component structure is seen 
in \Lyb\ through Ly8, easily separable as components denoted \#59ab.  A two-component CoG 
gives a reasonable fit, with 
$z_1 = 0.63190$ ($\log N_1 = 15.83 \pm 0.10$, $b_1 = 17\pm1$~\kms) and 
$z_2 = 0.63245$  ($\log N_2 = 15.80 \pm 0.06$, $b_2 = 36\pm2$~\kms) summing to 
$\log N_{\rm HI} = 16.12 \pm 0.10$. 


\paragraph{\#60:  SBS\,1108$+$560, $z_{\rm AGN} = 0.766619$, $z_{\rm abs} = 0.28646$.}  

This sight line has a very strong LLS (system \#9) to the red of absorber \#60.  In the low-flux region 
blueward of \#9 we see two strong velocity components (separation $\Delta v = 233$~\kms) in \Lya\ 
through Ly9, with considerable blending from other absorption.  Because the components are easily 
separable, we treat these as distinct absorbers, denoted \#60ab.
Individual CoG fits to these features give:
$z_1 = 0.2855$ ($\log N_1 = 16.12 \pm 0.09$, $b_1 = 34 \pm 2$~\kms) and 
$z_2 = 0.2865$ ($\log N_2 = 16.14 \pm 0.07$, $b_2 = 57 \pm 3$~\kms) summing to 
$\log N_{\rm HI} = 16.43 \pm 0.08$.  The implied $\tau_{\rm LL} = 0.17$ is difficult to confirm, 
given the poor data at 1173~\AA\ and the likely presence of AGN broad emission lines of 
\OIII\ and  \NIII\  (1150-1170~\AA\ observed frame).  


\paragraph{\#61: SDSS\,J143726.14$+$504555.8, $z_{\rm AGN} = 0.783319$, $z_{\rm abs} = 0.250650$.}  

The LL at 1140~\AA\ is barely within the COS/G130M data range.  The data are quite noisy,
and several Lyman lines are blocked (\Lyg) or contaminated by other absorption (Ly7).  A CoG fit 
with $\log N_{\rm HI} = 16.27 \pm 0.08$ and $b = 24 \pm 2$~\kms\  implies a Ly decrement with 
$\tau_{\rm LL} = 0.12$ that is hard to confirm. However, this column density over-predicts the line 
profiles of Ly8, Ly9, and Ly10.   We widen the error and adopt $\log N_{\rm HI} = 16.27 \pm 0.12$.  


\paragraph{\#62: SDSS\,J234500.43$-$005936.0, $z_{\rm AGN} = 0.789429$, $z_{\rm abs} = 0.54818$.}  

The data quality is quite good, with no LL decrement visible at 1411.6~\AA.  We see hints of two 
velocity components ($\Delta v \approx 40$~\kms) with extra absorption in asymmetric red wings 
of \Lye\ through Ly10.  A two-component CoG gives $\log N_1 = 15.63\pm0,07$ and 
$\log N_2 = 16.10\pm0.17$, summing to $16.23$.  We adopt a single-component CoG to \Lyb\ through 
Ly12 which gives $\log N_{\rm HI} = 16.16 \pm 0.06$ and $b = 26 \pm 2$~\kms.   


\paragraph{\#63: SDSS\,J101622.60$+$470643.3, $z_{\rm AGN} = 0.821527$, $z_{\rm abs} = 0.66475$.}  

This sight line includes system \#49 ($z_{\rm abs} = 0.727$).  The continuum longward of the LL 
(1520-1540~\AA) may be contaminated by AGN emission lines (\OII\ and \OIII\ at 833-834~\AA\ 
rest-frame).  A CoG fit to \Lyb\ through Ly9 gives $\log N_{\rm HI} = 15.99 \pm 0.03$ with
$b = 28 \pm 1$~\kms.


\paragraph{\#64: SBS\,1122$+$594, $z_{\rm AGN} =  0.8514$, $z_{\rm abs} = 0.67835$.}  

This sight line also includes system \#39 ($z_{\rm abs} = 0.5581$).  The data quality is high, with 
a well-constrained CoG fit:  $\log N_{\rm HI} = 16.06 \pm 0.04$ and $b = 19 \pm 1$~\kms.   The 
observed weak Ly decrement is consistent with this column density.  


\paragraph{\#65: SDSS\,J141910.20$+$420746.0, $z_{\rm AGN} =  0.873501$, $z_{\rm abs} = 0.52221$.}  

This sight line also includes systems \#31, \#32, \#33, \#50, and \#66.  Systems \#31 and \#32 
are at much lower redshifts, and system \#50 is at a much higher redshift ($z = 0.84523$).  
System \#33 has a somewhat higher redshift ($z = 0.53460$) but widely separated 
($\Delta v = 2440$~\kms). The data quality is not good, and no Ly decrement is apparent.  
A CoG fit to \Lyb, \Lyg, \Lyd, \Lye, and Ly7 gives $\log N_{\rm HI} = 15.87 \pm 0.07$ with 
$b = 37 \pm 2$~\kms.  


\paragraph{\#66: SDSS\,J141910.20$+$420746.0, $z_{\rm AGN} =  0.873501$, $z_{\rm abs} = 0.80463$.}  

This sight line also includes systems \#31, \#32, \#33, \#50, and \#66.  Systems \#31 and \#32 
are at much lower redshifts, and system \#50 is at somewhat higher redshift ($z = 0.84523$),
although it produces no visible Ly decrement at the location of system \#66 (1645.4~\AA).  
The \Lyg\ line is strong, and lines of \Lyd\ to \Lyz\ show two velocity components
($\Delta v = 66$~\kms) which we fit with CoGs to find:
$z_1 = 0.8044$ ($\log N_1 = 15.71 \pm 0.22$, $b_1 = 16 \pm 3$~\kms) and 
$z_2 = 0.8047$ ($\log N_2 = 15.75 \pm 0.06$, $b_2 = 19 \pm 2$~\kms) summing to 
$\log N_{\rm HI} = 16.03 \pm 0.15$.     


\paragraph{\#67: FBQS\,J0751$+$2919, $z_{\rm AGN} =  0.915$, $z_{\rm abs} = 0.82902$.}  

The data are of very high quality data with a weak Ly decrement.  A CoG fit to \Lyd\ through 
Ly12 gives $\log N_{\rm HI} = 16.06 \pm 0.02$ with $b = 34 \pm 2$~\kms.


\paragraph{\#68: PG\,1407$+$265, $z_{\rm AGN} =  0.946$, $z_{\rm abs} = 0.59964$.}  

The data are of very high quality data, but no Ly decrement is visible.  A CoG fit to \Lyg\ 
through Ly12 gives $\log N_{\rm HI} = 16.08 \pm 0.04$ with $b = 14 \pm 1$~\kms.


\paragraph{\#69: LBQS\,0107$-$0235, $z_{\rm AGN} =  0.957039$, $z_{\rm abs} = 0.536$.}  

The data are of very high quality data, but no Ly decrement is visible.  Two broad, 
well-separated absorption components ($\Delta v = 142$~\kms) are seen in \Lyb\ through 
\Lye, with the stronger (redder) component visible up to Ly10.   A two-component CoG fit finds:
$z_1 = 0.53572$ ($\log N_1 = 15.42 \pm 0.05$, $b_1 = 40 \pm 9$~\kms) and 
$z_2 = 0.53645$ ($\log N_2 = 15.66 \pm 0.05$, $b_2 = 40 \pm 7$~\kms) summing to 
$\log N_{\rm HI} = 15.86 \pm 0.05$.     


\paragraph{\#70:  LBQS\,0107$-$0235,  $z_{\rm AGN} =  0.957039$, $z_{\rm abs} = 0.87636$.}  

A very weak Ly decrement may be present, consistent with $\log N_{\rm HI}  < 16.1$.  
A CoG fit to \Lyd\ through Ly12 gives $\log N_{\rm HI} = 15.96 \pm 0.04$ with 
$b = 30 \pm 5$~\kms.  


\paragraph{\#71: PG\,1522$+$101, $z_{\rm AGN} =  1.32785$, $z_{\rm abs} = 0.67518$.}  

This sight line also includes systems \#25 ($z = 0.72865$), which produces a Ly decrement
at 1576~\AA\ redward of system \#71 and \#27 ($z = 0.51839$).   System \#71 consists of two 
absorption components, separated by $\Delta v \approx 72$~\kms, which we fit with:
$z_1 = 0.6748$ ($\log N_1 = 15.39 \pm 0.05$, $b_1 = 27 \pm 2$~\kms) and 
$z_2 = 0.6752$ ($\log N_2 = 15.71 \pm 0.04$, $b_2 = 42 \pm 4$~\kms) summing to 
$\log N_{\rm HI} = 15.88 \pm 0.05$.     


\paragraph{\#72: PG\,1630$+$377, $z_{\rm AGN} =  1.47607$, $z_{\rm abs} = 0.91449$.}  

Owing to its high redshift (for this survey), the COS spectra show absorption in \Lye\ through 
Ly10, but no Ly decrement is visible.  There appear to be two velocity components 
($\Delta v =  66$~\kms) with a stronger blue component ($z_1 = 0.9143$, $\log N_{\rm HI} = 15.67\pm0.04$, 
but a poorly constrained doppler parameter, $b = 99 \pm 49$~\kms).  A weaker red component at 
$z_2 = 0.9148$ is only reliably detected in \Lye\ through Ly7, with a poorly constrained column density,
$\log N_{\rm HI} = 15.4 \pm 0.2$.  The total system has $\log N_{\rm HI} = 15.86 \pm0.10$,


\paragraph{\#73:   SDSS\,161649.42$+$415416.3, $z_{\rm AGN} =  0.440417$, $z_{\rm abs} = 0.3211$.}  

This DLA did not appear in the Stevans \etal\ (2014) list, but we found it in our new examination
through its Lyman edge at 1204.5~\AA.  Meiring \etal\ (2011) and Battisti \etal\ (2012) quote 
$\log N_{\rm HI} = 20.60 \pm 0.20$ from fitting the \Lya\ damping wings.   Our CoG fit to \Lyb\ and 
\Lyd\ - Ly8 yields a somewhat smaller column density, $\log N_{\rm HI} = 20.34 \pm 0.12$ with 
$b = 49\pm2$ \kms.  Stevans \etal\ (2014) listed no other strong \HI\ absorbers in this sightline with 
$\log N_{\rm HI} > 15.5$.    


 \newpage

\section{Appendix~B:  Maximum-Likelihood Fitting of the pLLS Distribution }

Following the convention in studies of quasar absorption lines (Weymann \etal\ 1998;  Kim \etal\ 2002, among many others), we 
express the column density distribution as separable power laws in column density, $N$, and redshift, $z$,
\begin{equation}
   \label{eqn:fdist}
   f(N,z) \equiv  \frac {d^2 {\cal N}_{\rm abs} } {dz \; d N}  = C_0 N^{-\beta} \, (1+z)^{\gamma} \; .
\end{equation}
As presented in Tables 3-4, the pLLS and LLS absorbers are allocated to bins in $z$ and $\log N$, with the survey sensitivity expressed 
through the effective redshift, $\Delta z_{\rm eff}$, covered by QSOs in our sample.  Many studies (e.g., Janknecht \etal\ 2006; O'Meara 
\etal\ 2013) including our previous work (e.g., Danforth \etal\ 2016, and references therein) have employed variants of least-squares
fitting of $f(N,z)$ to binned histograms in order to determine $\beta$ and $\gamma$. This approach has several benefits, most notably its 
illustrative value in plots or tables and its computational simplicity in the presence of measurement errors and search pathlengths that 
may vary as functions of $N$ and $z$. In the limit of small bins of zero uncertainty, it tends toward the maximum-likelihood results. 
However, in the presence of finite binned data, this least-squares approach does not generally yield the maximum-likelihood estimates 
of the parameters for power-law distributions (Newman 2005; Clauset \etal\ 2009) and it introduces systematic biases in the fit parameters 
and their confidence intervals\footnote{Discussion of these effects are given in published papers (Goldstein \etal\ 2004; Newman 2005; 
Clauset \etal\ 2009 and subsequent on-line revisions, found in https://arxiv.org/abs/cond-mat/0412004 and http://arxiv.org/abs/0706.1062.}.
Because of the prevalence of binned fits in the literature, it is worth explaining in detail the maximum-likelihood approach to obtaining 
these parameters, as implemented in the present study.

From Equation~\ref{eqn:fdist}, the likelihood function for a dataset of $\mathcal{N}$ absorbers given the parameters $\beta$ and $\gamma$ is
\begin{equation}
   \label{eqn:like}
\begin{aligned}
   \mathcal{L}\left( \vec{P}\left({N}, {z}\right) | \beta, \gamma\right)  = \prod_{i=1}^{\mathcal{N}}  & \left( \int_{\rm out}P_i(N, z)dNdz \; + \right. \\
             & \;\; \left. C_0 \int_{z_{\rm min}}^{z_{\rm max}}\int_{N_{\rm min}}^{N_{\rm max}} P_i(N, z) {N}^{-\beta} \, 
             (1+{z})^{\gamma}\, w\left({N}, {z}\right)dNdz\right)\; .
\end{aligned}
\end{equation}
Here, $P_i(N, z)$ is the normalized probability density distribution for measurements of $N$ and $z$ for absorber $i$, and $w\left(N, z\right)$ is
a weight function that accounts for the surveyed pathlength in $z$. The first term in Equation~\ref{eqn:like} represents the total probability that 
absorber $i$ is outside the range in $z$ or $N$ over which we wish to fit the free parameters.

The normalized distribution $P_i(N, z)$ characterizes the data-derived uncertainties and correlations in the measurements, which will depend
on factors such as data quality and line-fitting techniques, and it introduces the need to evaluate the double-integral in Equation~\ref{eqn:like}.
In moderate resolution spectra, such as the COS G130M and G160M data used in this study, the uncertainty in redshift determination is quite
small, and we neglect it.   We assume that $P_i(N, z) \approx P_i(N)\delta(z-z_i)$ where $z_i$ is the measured redshift of the absorber, 
although this assumption is not valid for all datasets used in IGM studies. The column density measurements, on the other hand, can be 
subject to substantial and widely varying uncertainty that can potentially affect the derived parameters. Following Stevans \etal\ (2014), we 
assume that $P(N)$ is a log-normal distribution defined by the measured column density parameters from Table 2 in Stevans \etal\ (2014) 
with modifications listed in Table 6 of the present paper.   The one exception is the absorber at $z=0.47379$ in the SDSSJ154553.48+093620.5 
sightline, for which we have only upper and lower limits ($17.9<\log N<19.0$) and treat as a uniform distribution over that interval.

The function $w\left(N, z\right)$  gives the number of observed targets over which an absorber of column density $N$ and redshift $z$ could 
have been observed. Put another way, the integral of this function is the aforementioned effective pathlength over a given 
redshift and column density interval, $\Delta z_{\rm eff}=\int\int w\left(N, z\right) dNdz$. In general, such a function depends on properties of
 the observations, including but not limited to the observed wavelengths, the redshifts of the background AGNs, and the wavelength-dependent 
 signal-to-noise ratios of the spectra. Often, this function must be evaluated numerically, but in some cases simplifying assumptions can be made. 
 Because the present study is concerned only with relatively strong \HI\ absorbers in high signal-to-noise, well-resolved data, we assume that 
 $w$ is a function only of $z$ and can be expressed as
\begin{equation}
   \label{eqn:zweight}
   w\left(N, z\right)\approx w\left(z\right)\approx \sum_{j=1}^{\mathcal{N}_{\rm AGN}}\left[ H\left( z - z_{\rm min,j}\right)
            - H\left( z - z_{\rm max,j}\right)\right]\; ,
\end{equation}
where $H(x)$ is the Heaviside step function and $z_{\rm min,j}$ and $z_{\rm max,j}$ are the minimum and maximum redshifts at which
an absorber could be detected in sightline $j$, respectively.

Note that if all of the measured column densities and redshifts had no measurement errors and the pathlength was independent of $N$ and
$z$, the likelihood function would remain a pure power law. If upper integration limits are further allowed to go to infinity, it becomes
straightforward to derive analytical formulae for a maximum-likelihood power-law exponent and its uncertainty, as shown in Newman (2005),
Clauset \etal\ (2009), and references therein. Several authors have used these formulae to calculate $\beta$ and/or $\gamma$
(e.g., Tytler 1987a,b; Storrie-Lombardi \etal\ 1994; Stengler-Larrea \etal\ 1995; Songalia \& Cowie 2010; Ribaudo \etal\ 2011a; Rudie \etal\ 2013).
However, typical IGM surveys have a completeness function that varies strongly with $N$ and $z$, as well as large uncertainties in $N$ that
span a wide range of magnitudes depending on $z$ and their location on the CoG. Thus, these surveys may not sufficiently approximate the
assumptions of such analytical formulae. In general, the approach of using these formulae introduces a systematic bias in the derived power-law
exponents.  Although the simplifications chosen for the present study maintain the separability of the likelihood function and thus the
independence of the two fit parameters, this is not necessarily true for all possible forms of $w\left(N, z\right)$ and $P(N, z)$, which could
introduce correlations between $\beta$ and $\gamma$. In such a case, each term in Equation~\ref{eqn:like} must be explicitly evaluated as
a two-dimensional integral at each step in the optimization process. These calculations are not  computationally prohibitive, even if
they must be performed numerically in the absence of an analytical solution.

The normalization $C_0$  is fixed by $\beta$ and $\gamma$ under the requirement that each observed absorber's existence has unity
probability within the dataset. It can be evaluated as
\begin{equation}
   \label{eqn:c0}
   C_0=\mathcal{N^\prime}\left[\int_{z_{\rm min}}^{z_{\rm max}}\int_{N_{\rm min}}^{N_{\rm max}} N^{-\beta} \,
         (1+z)^{\gamma}\, w \left(N, z \right) dN dz \right]^{-1}\; ,
\end{equation}
where the limits of integration are the search ranges of the survey and the single power-law approximation remains valid.  In our standard 
fits, we set $(N_{\rm min},N_{\rm max} )= (10^{15}~{\rm cm}^{-2}, 10^{20}~ {\rm cm}^{-2})$ and $(z_{\rm min},z_{\rm max}) = (0.24, 0.84)$, 
as discussed in Section~3.1. In Equation~\ref{eqn:c0}, we introduced the variable $\mathcal{N}^\prime$ (in contrast to the unprimed 
$\mathcal{N}$). This distinction is necessary to maintain definitional consistency of $C_0$ while accounting for the effects of the finite limits
of integration in the second term of Equation~\ref{eqn:like}, which may allow a fraction of $P(N, z)$ to fall outside the region of integration, 
into the first term of  Equation~\ref{eqn:like}. Therefore, $\mathcal{N}^\prime$ is the non-integer number of (fractional) absorbers contributing 
to the likelihood function,
\begin{equation}
   \label{eqn:nprime}
   \mathcal{N}^\prime = \sum_{i=1}^{\mathcal{N}} \mathcal{N}^\prime_i =
         \sum_{i=1}^{\mathcal{N}}\int_{z_{\rm min}}^{z_{\rm max}}\int_{N_{\rm min}}^{N_{\rm max}} P_i(N, z) dNdz\;,
\end{equation}
where $\mathcal{N}^\prime_i$ is the individual fractional contribution of an absorber $i$.

Using each of our assumptions with Equation~\ref{eqn:like} and taking the logarithm for computational convenience, we obtain the 
log-likelihood function that we use for optimization of $\beta$ and $\gamma$ in the present study:
\begin{equation}
   \label{eqn:lnlike}
   \begin{aligned}
   \ln\mathcal{L}\left( \vec{P}\left({N}\right), \vec{z}\; | \beta, \gamma\right) =  \sum_{i=1}^{\mathcal{N}} \ln 
   &\left(\int_{-\infty}^{N_{\rm min}}P_i(N)dN +\int_{N_{\rm max}}^{\infty}P_i(N)dN\;+\right. \\
& \;\;\left. C_0(1+{z_i})^\gamma w\left({z_i}\right) \int_ {N_{\rm min}}^{N_{\rm max}}  P_i(N) {N}^{-\beta}dN\right)\; .
\end{aligned}
\end{equation}
For our dataset, the second term within the logarithm in Equation~\ref{eqn:lnlike} is negligibly small and therefore not evaluated.  
Because this likelihood function contains multiple numerical integrals that must be evaluated for each absorber, it can be computationally
expensive to evaluate for some choices of $P(N)$ and $(N_{\rm min},N_{\rm max})$. For this reason, we optimize the likelihood function
and determine $\beta$ and $\gamma$ by sampling the posterior probability distribution with version 2.1.0 of \texttt{emcee}
(Foreman-Mackey \etal\ 2013), which implements the affine-invariant Markov-chain Monte Carlo (MCMC) ensemble sampler from
Goodman \& Weare (2010). We adopt uniform priors over $-5<\gamma<10$ and $0<\beta<10$. We initialize 250 walkers randomly
 over the domain, allowing each to take 250 steps, the first 75 of which are discarded as a burn-in period. 
 
 This MCMC procedure yields median values of $\beta = 1.48 \pm 0.05$ and $\gamma = 1.14^{+0.88}_{-0.89}$, and 
 $C_0 = 2.07 \times 10^7$ (for $N $ in cm$^{-2}$).  The error bars indicate the 1$\sigma$ quantiles around the median in the 
 highly-Gaussian, marginalized posterior probability distributions.  Figures~12 and 13 show the fits and differential distributions.



\begin{deluxetable}{cclccc}
\tablecaption{ \bf Detectability Ranges for  \HI\ Lyman Absorption\tablenotemark{a}  }  
\tablecolumns{6}
\tablewidth{0pt}
\tablehead{  \colhead{Feature} & \colhead{$\lambda_0$} & \colhead{$f$} & \colhead{$\log N_{\rm HI}$ } &
\colhead{$(z_{\rm min}-z_{\rm max})$}  &  \colhead{$(z_{\rm min}-z_{\rm max})$}  \\   
                     &  (\AA)   &    &   $N_{\rm HI}$ (cm$^{-2}$)  &    (G130M)   &   (G160M)    }
\startdata

\Lya\      & 1215.67   & 0.4164     & 12.96 &  0.002 -- 0.200  &  0.152 -- 0.477  \\
\Lyb\      & 1025.72   & 0.07912   & 13.83 &  0.107 -- 0.422  &  0.365 -- 0.750  \\
\Lyg\      &   972.54   & 0.02900   & 14.31 &  0.167 -- 0.500  &  0.440 -- 0.846  \\
\Lyd\      &   949.74   & 0.01394   & 14.65 &  0.195 -- 0.536  &  0.474 -- 0.890  \\
\Lye\      &   937.80   & 0.007804 & 14.92 &  0.210 -- 0.556  &  0.493 -- 0.914  \\
\Lyz\      &   930.75   & 0.004816 & 15.13 &  0.219 -- 0.568  &  0.504 -- 0.929  \\
\Lyeta\   &  926.23    & 0.003813 & 15.32 & 0.225 -- 0.575   &  0.512 -- 0.938 \\
\Lytheta &  923.15    & 0.002216 & 15.48 & 0.229 -- 0.581   &  0.517 -- 0.944 \\ 
LL          &   911.75   &                 & 16.20 &  0.245 -- 0.600  &  0.536 -- 0.969  \\

\enddata  

\tablenotetext{a} {Detectability ranges in redshift ($z_{\rm min}$ and $z_{\rm max}$) for
the first eight Lyman lines (and Lyman limit) of H\,I.  Columns 2 and 3 show the absorption 
oscillator strengths ($f$) and wavelengths ($\lambda$).  Column 3  gives the H\,I column 
density detectable in Lyman-series absorption at 50 m\AA\ equivalent width,
$N_{\rm HI} = (5.65 \times 10^{18}~{\rm cm}^{-2}) [ f \, \lambda({\rm \AA})]^{-2}$, 
or in a 10\% flux decrement at the Lyman edge.  Last two columns show redshift coverage 
for HST/COS observations in gratings G130M (1134--1459~\AA) and G160M (1400--1795~\AA). 
Because we usually detect an absorber pattern from \Lya\ through \Lye\, our survey 
should be complete for $\log N_{\rm HI} \geq 15$.  However, most LL decrements would be 
undetected at $\log N_{\rm HI} < 16.2$.  }

\end{deluxetable}



\begin{deluxetable}{llccc}
\tablecaption{\bf Line Overlap Parameters (Higher Lyman Lines)\tablenotemark{a}  }  
\tablecolumns{5}
\tablewidth{0pt}
\tablehead{  \colhead{Lyman ($n$)}  &   \colhead{$\lambda_0$ (\AA) }  
     & \colhead{$\Delta \lambda _{n,n+1}$}  &  \colhead{$f_n$}  & \colhead{$\tau_0 ^{(n)}$}  \\   
            &          &      & (in~$10^{-4})$    &   ($\times N_{17} /b_{25}$)    }
\startdata

Ly12    & 917.1805   & 0.949 \AA  &  7.231  &  3.97  \\
Ly13    & 916.4291   & 0.751 \AA  &  5.777  &  3.17  \\
Ly14    & 915.8238   & 0.605 \AA  &  4.689  &  2.57  \\
Ly15    & 915.3289   & 0.495 \AA  &  3.858  &  2.12   \\
Ly16    & 914.9192   & 0.410 \AA  &  3.212  &  1.76   \\ 
Ly17    & 914.5762   & 0.343 \AA  &  2.703  &  1.48   \\
Ly18    & 914.2861   & 0.290 \AA  &  2.297  &  1.26   \\
Ly19    & 914.0385   & 0.248 \AA  &  1.968  &  1.08   \\
Ly20    & 913.8256   & 0.213 \AA  &  1.699  &  0.930  \\
Ly21    & 913.6411   & 0.185 \AA  &  1.477  &  0.808  \\
Ly22    & 913.4803   & 0.161 \AA  &  1.293  &  0.707 \\
Ly23    & 913.3391   & 0.141 \AA  &  1.137  &  0.622 \\
Ly24    & 913.2146   & 0.125 \AA  &  1.006  &  0.550

\enddata

\tablenotetext{a}{Wavelength separations ($\Delta \lambda_{n,n+1}$) between Lyman 
transitions, $\lambda_n$ and $\lambda_{n+1}$, and optical depths, $\tau_0^{(n)}$.  Here,
Ly\,$n$ denotes transition from $(n+1)p \rightarrow 1s$, and line-center optical depth
$\tau_0^{(n)} = (0.5474) \, N_{17} \, b_{25}^{-1} \,  f_{-4} \, \lambda_{914}$, with
$N_{\rm HI} = (10^{17}~{\rm cm}^{-2})N_{17}$ and doppler parameter scaled to $b = 25$~\kms.  
Oscillator strengths $f_n$ (in units of $10^{-4}$) and wavelengths ($\lambda_0$ scaled to 
914~\AA) are from Morton (2003). }

\end{deluxetable}



\begin{deluxetable} {ccccccccccccccccccc}
\tabletypesize{\scriptsize}
\tablecaption{\bf  Absorber Distribution in Redshift and Column Density\tablenotemark{a} }
\tablecolumns{18}
\tablewidth{0pt}
\tablehead{  \colhead {$z$-bin}  & \colhead {1} & \colhead{2} & \colhead {3} &  \colhead{4} & \colhead {5} &
    \colhead{6} &  \colhead {7} &  \colhead{8} & \colhead {9} &  \colhead{10} & \colhead {11} & \colhead{12} &
     \colhead{13} &  \colhead{14} & \colhead{15} &\colhead {$N_{\rm tot}$} &  \colhead {Redshifts}   }

\startdata
$(i = 1)$  & 2 & 0 & 4 & 5 & 5  & 3  & 2 &       1 & 2 & 2 & 0 & 0 &      1 & 0 & 1 &  [28]  & $0.24 - 0.28$  \\
$(i = 2)$  & 0 & 5 & 1 & 5 & 2  & 4  & 1 &       0 & 3 & 2 & 0 & 0 &      0 & 0 & 1 &  [24]  & $0.28 - 0.32$  \\
$(i = 3)$  & 1 & 1 & 1 & 1 & 2  & 4  & 3 &       0 & 5 & 2 & 0 & 0 &      1 & 0 & 2 &  [23]  & $0.32 - 0.36$  \\
$(i = 4)$  & 0 & 1 & 3 & 5 & 5  & 0  & 1 &       0 & 1 & 0 & 0 & 2 &      0 & 0 & 1 &  [19]  & $0.36 - 0.40$  \\
$(i = 5)$  & 0 & 0 & 3 & 8 & 3  & 1  & 6 &       1 & 1 & 1 & 2 & 1 &      0 & 1 & 0 &  [28]  & $0.40 - 0.44$  \\
$(i = 6)$  & 0 & 1 & 1 & 0 & 1  & 4  & 2 &       2 & 0 & 1 & 0 & 1 &      0 & 1 & 1 &  [15]  & $0.44 - 0.48$  \\
$(i = 7)$  & 0 & 0 & 1 & 0 & 3  & 0  & 3 &       0 & 0 & 2 & 1 & 0 &      0 & 0 & 0 &  [10]   & $0.48 - 0.52$  \\
$(i = 8)$  & 1 & 0 & 0 & 0 & 2  & 1  & 2 &       2 & 3 & 2 & 0 & 1 &      0 & 0 & 0 &  [14]  & $0.52 - 0.56$  \\
$(i = 9)$  & 0 & 0 & 1 & 0 & 3  & 2  & 3 &       0 & 1 & 0 & 0 & 0 &      0 & 0 & 0 &  [11]  & $0.56 - 0.60$  \\
$(i = 10)$ & 0 & 0 & 1& 1 & 1  & 2  & 3 &       2 & 1 & 1 & 1 & 0 &      0 & 0 & 0 &  [12]  & $0.60 - 0.64$  \\
$(i = 11)$ & 0 & 0 & 0 & 0 & 2 & 0  & 1 &       2 & 1 & 0 & 0 & 0 &      0 & 0 & 0 &   [6]   & $0.64 - 0.68$  \\
$(i = 12)$ & 0 & 0 & 0 & 0 & 0 & 2  & 3 &       0 & 0 & 2 & 2 & 0 &      0 & 0 & 0 &   [9]   & $0.68 - 0.72$  \\
$(i = 13)$ & 0 & 0 & 0 & 0 & 0 & 2  & 0 &       0 & 1 & 0 & 1 & 1 &      0 & 0 & 0 &   [5]   & $0.72 - 0.76$  \\
$(i = 14)$ & 0 & 0 & 0 & 1 & 0 & 0  & 1 &       0 & 0 & 0 & 0 & 0 &      0 & 0 & 0 &   [2]   & $0.76 - 0.80$  \\
$(i = 15)$ & 0 & 0 & 0 & 0 & 2 & 0  & 3 &       0 & 3 & 0 & 1 & 0 &      0 & 0 & 0 &   [9]   & $0.80 - 0.84$  \\
                &    &    &    &    &    &     &    &          &    &    &    &    &         &    &    &          &                        \\
Total        & 4 & 8 &16& 26&31&25&34&     10 & 22&15& 8 & 6 &      2 & 2 & 6 & [215] & $0.24 - 0.84$ \\
\enddata

\tablenotetext{a}{Array $F(i,j)$ shows the number of H\,I absorbers per bin, with total numbers 
summed across rows and columns. In total, this table lists 211 absorbers:  
158 ($\log N_{\rm HI} \geq 15.0$), 60 ($\log N_{\rm HI} \geq 16.0$), and 
10 ($\log N_{\rm HI} \geq 17.0$).  Labels ($i,j)$ denote redshift bins ($i=1-15$) of width
$\Delta z = 0.04$ from $0.24 \leq z \leq 0.84$ and column-density bins ($j=1-15$) of width 
$\Delta \log N_{\rm HI}= 0.25$ except for bins 13, 14, 15.  Column density ranges of bins are:  
(1) $\log N_{\rm HI} =$ 14.00-14.25, (2) 14.25-14.50, (3) 14.50-14.75, (4) 14.75-15.00, 
(5) 15.00-15.25, (6) 15.25-15.50, (7) 15.50-15.75, (8) 15.75-16.00, (9) 16.00-16.25, 
(10) 16.25-16.50, (11) 16.50-16.75, (12) 16.75-17.00, (13) 17.00-17.50, (14) 17.50-18.00,
(15) $\log N_{\rm HI} \geq18.00$.  Five absorbers (Systems \#4, 10, 50, 70, 72 in Appendix~A) 
lay at redshifts outside the range of this table.  Although they do not appear in the numbers above, 
they are included in the MCMC analysis of Appendix B.   \\ }

\end{deluxetable}



\begin{deluxetable}{cccccl}
\tabletypesize{\scriptsize}
\tablecaption{\bf Redshift Distribution of \HI\  Absorbers }  
\tablecolumns{6}
\tablewidth{0pt}
\tablehead{  \colhead{Bin ($i$)}  &   \colhead {Redshift  Range}  & \colhead {$(N_{\rm AGN})$\tablenotemark{a} }
    &  \colhead{ $(N_{\rm abs})$\tablenotemark{b} }  &  \colhead { $(\Delta z_{\rm eff})^{(i)}$\tablenotemark{a}  } &  
     \colhead{$(d{\cal N}/dz)$\tablenotemark{b} }   }
                                                                                     
\startdata
   1   &   0.24 -- 0.28   &   102     & 17 (28)   &   3.83      &   $4.44(+1.36,-1.07)$       \\
   2   &   0.28 -- 0.32   &     96     & 13 (24)   &   3.74      &   $3.48(+1.26,-0.95)$       \\
   3   &   0.32 -- 0.36   &     91     & 19 (22)   &   3.39      &   $5.60(+1.60,-1.27)$       \\
   4   &   0.36 -- 0.40   &     81     & 10 (19)   &   3.08      &   $3.25(+1.39,-1.01)$       \\
   5   &   0.40 -- 0.44   &     73     & 17 (28)   &   2.80      &   $6.07(+1.86,-1.46)$       \\
   6   &   0.44 -- 0.48   &     66     & 13 (15)   &   2.40      &   $5.42(+1.96,-1.48)$       \\ 
   7   &   0.48 -- 0.52   &     54     &   9 (10)   &   2.10      &   $4.29(+1.96,-1.46)$       \\
   8   &   0.52 -- 0.56   &     51     & 14 (15)   &   1.98      &   $7.07(+2.44,-1.87)$       \\
   9   &   0.56 -- 0.60   &     49     & 10 (11)   &   1.83      &   $4.92(+2.25,-1.61)$        \\
  10  &   0.60 -- 0.64   &     43     & 10 (12)   &   1.67      &   $5.99(+2.56,-1.86)$       \\
  11  &   0.64 -- 0.68   &     39     &   5 (6)     &   1.40      &   $3.57(+2.42,-1.54)$       \\
  12  &   0.68 -- 0.72   &     31     &   9 (9)     &   1.17      &  $7.69(+3.51,-2.52)$        \\
  13  &   0.72 -- 0.76   &     24     &  5 (5)      &   0.963    &  $5.19(+3.51,-2.24)$        \\
  14  &   0.76 -- 0.80   &     23     &  1 (2)      &   0.849    &  $1.18(+2.71,-0.97)$        \\    
  15  &   0.80 -- 0.84   &    19      &  9 (9)      &   0.738    &  $12.2(+5.6,-4.0)$            \\ 
\enddata

\tablenotetext{a} {Number ($N_{\rm AGN}$) of AGN sight lines providing full or partial 
spectral coverage of the H\,I  Lyman edge (912~\AA) over redshift bins ($i = 1-15$). 
The effective redshift, $\Delta z_{\rm eff}$,  is based on $N_{\rm AGN}$ subtracting 
partial coverage of bins and portions of spectrum blocked by strong absorbers. } 

\tablenotetext{b} {Distribution ($N_{\rm abs}$) of 158 H\,I absorbers with 
$\log N_{\rm HI} \geq 15$ for the 15 redshift bins (from $0.24 \leq z \leq 0.84$).  
Values in parentheses are number of absorbers with $\log N_{\rm HI} \geq 14$ 
(211 in all).  Last column shows the frequency of absorbers per unit redshift, 
$d{\cal N} / dz \equiv N_{\rm abs} / \Delta z_{\rm eff}$, for $\log N_{\rm HI} \geq 15$.  
We compute ($1\sigma$) error bars for one-sided Poisson statistics (Gehrels 1986) 
on $N_{\rm abs}$.  Over total pathlength $\Delta z_{\rm eff} = 31.94$, 
the average absorber frequency is $d {\cal N}/ dz  \approx 4.95 \pm 0.39$.  Full 
statistical analysis (MCMC) of the bivariate distribution, 
$f(N,z) \equiv  (d^2 {\cal N}_{\rm abs} / dz  \,dN)$, is given in Appendix B.}

\end{deluxetable}



\begin{deluxetable}{llllll}
\tabletypesize{\scriptsize}
\tablecaption{\bf Strong  Absorbers and Wavelength Blockage\tablenotemark{a} }  
\tablecolumns{6}
\tablewidth{0pt}
\tablehead{  \colhead { QSO\tablenotemark{b} }  &   \colhead {$z_{\rm abs}$}   &  \colhead {$\log N_{\rm HI}$ }
           &       \colhead  {$z_{\rm QSO}$ } &   \colhead {Affected} &  \colhead  {Wavelength Blockage} }
                                                                                     
\startdata
  1  &  0.24770    & $19.45\pm0.10$  &  0.47114     &  Bin 1         & opaque below 1137.6~\AA\  \\
  2  &  0.39043    & $18.1\pm0.1$      &  1.131         &  Bins 1-4    & opaque below 1267.7~\AA\  \\
  3  &  0.41065    & $17.80\pm 0.05$ &  0.694596   &  Bins 1-5    & opaque below 1286.2~\AA\    \\
  4  &  0.23740    & $17.29\pm0.11$  &   0.5251      &  \dots         & opaque below 1128.2~\AA\   \\
  5  &  0.35586    & $18.3\pm0.2$      &   0.4556      & Bins 1-3     & opaque below 1236.2~\AA\    \\
  6  &  0.30360    & $19.41\pm0.12$  &   0.7462      &  Bins 1-2    & opaque below 1188.6~\AA\    \\
  7  &  0.47379    & $18.3\pm0.2$      &   0.665        & Bins 1-6     & opaque below 1343.7~\AA\      \\
  8  &  0.41924    & $16.80\pm0.05$  &   0.4632      &  Bins 1-5    & minor influence below 1294.0~\AA\  \\
  9  &  0.46334    & $17.85\pm0.02$  &  0.7666       &  Bins 1-6    & opaque below 1334.2~\AA\       \\
 10 &  0.92772    & $17.15\pm0.05$  &  1.1625       &  Bins 1-14  & minor influence below 1757.6~\AA\  \\
\enddata  

\tablenotetext{a} {Potential spectrum blockage by 10 strong absorbers listed with redshifts $z_{\rm abs}$
and fitted column densities, $\log N_{\rm HI}$.  Absorber column densities and redshifts are revised from 
values in Stevans \etal\ (2014), based on analysis of the Lyman limit flux decrement  and new fits to CoG 
for high-order Lyman lines.  See Appendix B for details.  In 8 cases, strong LyC absorption produces 
flux decrements at wavelengths $\lambda \leq (911.753~{\rm \AA})(1+z_{\rm abs})$ listed in 
comments.  Potentially affected bins are noted.  However, our survey can still find pLLS by detecting 
higher Lyman lines longward of the Lyman edge. }
  
\tablenotetext{b} {QSO targets: (1) SDSS$\,$J092554.70+400414.1; (2) FIRST$\,$J020930.7-043826;
(3) SDSS$\,$J151428.64+361957.9; (4) SDSS$\,$J113327.78+032719.1; (5) SDSS$\,$J100902.06+071343.8;
(6) SDSS$\,$J100102.55+594414.3; (7) SDSS$\,$J154553.48+093620.5; (8) SDSS$\,$J091029.75+101413.6;
(9) SBS$\,$1108+560;  (10) PG$\,$1206+459.} 

\end{deluxetable}



\begin{deluxetable}{lllllll}
\tabletypesize{\scriptsize}
\tablecaption{\bf Strong  Absorbers (LLS and pLLS)\tablenotemark{a} }  
\tablecolumns{7}
\tablewidth{0pt}
\tablehead{ \colhead {No.}  & \colhead {QSO~Name}  &  \colhead {$z_{\rm abs}$ } &  \colhead {$z_{\rm AGN}$}  &  
   \colhead {$\log N_{\rm HI}$} & \colhead {$\log N_{\rm HI}$} & \colhead {$\log N_{\rm HI}$}    \\
       &     &    &   &   \colhead{Lehner$^+$\,13}   & \colhead{Stevans$^+$\,14}  &  \colhead{This~Paper}   }
 \startdata 
1  &  J092554.70+400414.1  &  0.2477      & 0.471139   &    $19.55\pm0.15$    &  $19.26\pm0.06$   & $19.45\pm0.10$   \\
2  &  J020930.7-043826        &  0.39035    & 1.131         &                                  &  $18.00\pm0.2$     & $18.5\pm0.4$       \\
3  &  J151428.64+361957.9  &  0.41065    &  0.694596  &                                  &  $17.93\pm 0.2$    & $17.89\pm0.06$    \\                      
4  &  J113327.78+032719.1  &  0.23756    & 0.525073   &                                  &  $17.53\pm0.10$    & $17.29\pm0.11$   \\                    
5  &  J100902.06+071343.8  &  0.35586    & 0.455631   &  $18.40\pm0.20$      &  $17.41\pm0.04$   & $18.3\pm0.2$      \\
6  &  J100102.55+594414.3  &  0.30360    & 0.746236   &                                  &  $17.27\pm0.04$   & $19.41\pm0.12$   \\
7  &  J154553.48+093620.5  &  0.47379    & 0.665         &                                  &  $17.25\pm0.2$     & $18.3\pm0.2$  \\
8  &  J091029.75+101413.6  &  0.41924    & 0.463194   &                                  &  $17.14\pm0.8$     & $16.89^{+0.04}_{-0.11}$  \\
9  & SBS$\,$1108+560          &  0.46334    & 0.76619     &                                  &  $17.06\pm0.1$     & $17.85\pm0.02$ \\
10 & PG$\,$1206+459           &  0.92772    & 1.16254     &  $17.00\pm0.10$      & $17.03\pm0.08$    & $17.10\pm0.10$   \\                          
11 & PG$\,$1630+377            & 0.27395    & 1.47607      & $16.98\pm0.05$       & $16.92\pm0.04$    & $17.04\pm0.05$   \\
12 & J100535.24+013445.7  & 0.41853     & 1.0809        &                                  &  $16.37\pm0.02$   & $16.73\pm0.10$    \\
13 & J143511.53+360437.2   & 0.37292    &  0.428593   &  $16.65\pm0.07$      & $16.72\pm0.06$    & $16.84\pm0.06$    \\
14 & J091029.75+101413.6   & 0.2634      &  0.463914   &                                  & $16.86\pm0.5$      & $16.21\pm0.07$    \\
15 & J161916.54+334238.4   & 0.47091    &  0.470946   &                                  & $16.84\pm0.1$      & $16.83\pm0.10$     \\
16 & J155048.29+400144.9   & 0.31257    &  0.496843   &                                  & $16.62\pm0.06$    & $16.3\pm0.1$        \\
17 & J155048.29+400144.9   & 0.4919      &  0.496843   &                                  & $16.57\pm0.02$    & $16.60\pm0.05$  \\
18 & PKS$\,$0552-640          & 0.34513    &  0.680          &  $16.90\pm0.08$     & $16.71\pm0.03$    & $17.01\pm0.05$   \\
19 & J124511.25+335610.1   & 0.55670    &  0.717          &                                 & $16.50\pm0.2$      & $16.28\pm0.10$    \\
20 & J124511.25+335610.1   & 0.68927    &  0.717          &                                 & $16.68\pm0.2$      & $16.70\pm0.11$    \\
21 & J100102.55+594414.3   & 0.41595    &  0.746236   &                                 & $16.61\pm0.02$    & $16.66\pm0.08$   \\ 
22 & HB89$-$0107$-$25       & 0.39913    &  0.956          &                                & $16.59\pm0.02$    & $16.89\pm0.06$     \\
23 & J084349.49+411741.6   & 0.53255    & 0.989986     &                                & $16.67\pm0.05$    & $16.85\pm0.04$     \\           
24 & J100535.24+013445.7   & 0.83708    & 1.0809         &                                & $16.81\pm0.01$    & $16.96\pm0.08$     \\
25 & PG$\,$1522+101            & 0.72865   &  1.32785       &  $16.66\pm0.05$    & $16.60\pm0.09$    & $16.62\pm0.05$   \\
26 & Q0232-042                     & 0.73888   &  1.43737       &   $16.72\pm0.03$    & $16.64\pm0.08$    & $16.76\pm0.05$    \\
27 & PG$\,$1522+101            & 0.51839   &  1.32785       &  $16.22\pm0.02$    & $16.32\pm0.2$     & $16.28\pm0.05$   \\
28 & PG$\,$1338+416            & 0.68610   &  1.21422       & $16.45\pm0.05$     & $16.49\pm 0.04$  & $16.49\pm0.05$  \\
29 & PG$\,$1338+416            & 0.34885   &   1.2142        & $16.30\pm 0.13$    & $16.37\pm0.06$   & $16.34\pm0.10$   \\
30 & PKS$\,$0637-752           & 0.46850   &   0.653          & $16.48\pm0.04$     & $16.08\pm0.03$   & $16.48\pm0.05$   \\
31 & J141910.20+420746.9    & 0.28895   &   0.873501    & $16.40\pm0.07$     & $16.17\pm0.03$   & $16.17\pm0.05$  \\
32 & J141910.20+420746.9    & 0.42555   &   0.873501    & $16.17\pm0.06$     & $16.02\pm0.02$   & $16.24\pm0.10$  \\
33 & J141910.20+420746.9    & 0.53460   &   0.873501    & $16.34\pm0.17$     & $16.06\pm0.07$   & $16.15\pm0.10$  \\
34 & J143511.53+360437.2    & 0.38766   &  0.428593     & $16.18\pm0.05$     & $16.15\pm0.02$    & $16.17\pm0.05$  \\
35 & J094331.61+053131.4    & 0.35455   &  0.564336     & $16.11\pm0.09$     & $16.12\pm0.09$    & $16.12\pm0.05$   \\
36 & PG\,1407+265                 & 0.68270   &  0.946           & $16.38\pm0.02$     & $16.39\pm0.03$    & $16.38\pm0.05$   \\
37 & PG\,1216+069                 & 0.28231   & 0.3313          & $16.40\pm0.05$     & $16.29\pm0.01$    & $16.41\pm0.03$   \\
38 & J161916.54+334238.4    & 0.26938   & 0.470496      & $16.48\pm0.05$     & $16.40\pm0.03$    & $16.40\pm0.08$   \\
39 & SBS$\,$1122+594           & 0.55810   & 0.852            & $16.24\pm0.03$     & $16.42\pm0.02$    & $16.45\pm0.06$   \\
40 & HE$\,$0439-5254            & 0.61512   & 1.053            & $16.28\pm0.04$     & $16.25\pm0.04$    & $16.42\pm0.10$   \\
73 & J161649.42+415416.3    & 0.3211      & 0.440417      &    $\dots$                &  $\dots$                 & $20.34\pm0.12$    \\                                          
\enddata  

\tablenotetext{a}{Strong absorbers estimated (Stevans \etal\ 2014) with $\log N_{\rm HI} \geq 16.5$ (systems \#1-26) 
and $\log N_{\rm HI} = 16.0-16.5$ (systems \#27-40 ).  System \#73 is a new DLA.   Some were reported previously
(Battisti \etal\ 2012; Lehner \etal\ 2013; Fox \etal\ 2013).  Columns 3 and 4 list  redshifts of absorber ($z_{\rm abs}$) and AGN 
($z_{\rm AGN}$).  Our column densities (this paper) are from new CoG fits to Lyman-series approaching LL.  
Appendix A discusses CoGs and Ly decrements. }

\end{deluxetable}



\begin{deluxetable}{llcllc}
\tabletypesize{\scriptsize}
\tablecaption{\bf Summary of LLS Frequency fitting\tablenotemark{a}  }  
\tablecolumns{6}
\tablewidth{0pt}
\tablehead{  \colhead{Survey} &  \colhead{Instrument}  &   \colhead{Redshift Range} &
      \colhead{$N_0$} & \colhead{$\gamma$ } &  \colhead{$N_{\rm LLS}$} } 
        
\startdata

This Paper                             & COS         & 0.24 -- 0.48 & $0.25^{+0.13}_{-0.09}$ &  $1.14 \pm 0.89$  &  8      \\
Storrie-Lombardi~\etal\ 1994 & FOS         & 0.40 -- 4.69 & $0.27^{+0.20}_{-0.13}$ &  $1.55 \pm 0.45$  &  7     \\
Stengler-Larrea~\etal\  1995 & FOS          & 0.40 -- 4.69 & $0.25^{+0.17}_{-0.10}$ &  $1.50 \pm 0.39$  &  7     \\
Ribaudo~\etal\ 2011a            &FOS/STIS  & 0.25--2.59   & $0.28 \pm 0.05$            &  $1.19 \pm 0.56$  &  17   \\

\enddata  

\tablenotetext{a} {Redshift evolution of low-redshift LLS ($\log N_{\rm HI} \geq 17.2$) studied with UV spectra
and fitted to $(d{\cal N} / dz) = N_0 (1+z)^{\gamma}$.  We include the current COS survey, two \HST\ low-resolution 
surveys with FOS/G140L (Storrie-Lombardi \etal\ 1994; Stengler-Larrea \etal\ 1995) and a low-resolution survey with 
FOS/G140L and STIS/G140L/G230L (Ribaudo \etal\ 2011a).  The last column gives the number $N_{\rm LLS}$ 
of low redshift ($z < 0.84$ LLS in the COS survey, $z < 1.04$ in the FOS Key Project, and $z < 0.84$  in
Ribaudo \etal\ (2011a). These surveys contain many more pLLS, which are used in the fits. }

\end{deluxetable}



\begin{deluxetable}{lllll}
\tabletypesize{\scriptsize}
\tablecaption{\bf List of Strong (DLA and sub-DLA) Absorbers\tablenotemark{a}  }  
\tablecolumns{5}
\tablewidth{0pt}
\tablehead{  \colhead{\rm AGN}           & \colhead{$z_{\rm abs}$ }  & \colhead{$z_{\rm AGN}$ } &
    \colhead { $\log N_{\rm HI}~({\rm cm}^{-2})$\tablenotemark{b}  }  &   {\rm Comments}    }
\startdata
PG\,1216+069               &  0.00635     &   0.331        & $19.32\pm0.03$       &  Sightline to System \#37    \\
J155304.92+354826.6  &  0.0830       &   0.721814  & $19.55 \pm 0.15$     &  Two other strong absorbers ($\log N =15.18, 15.43$)  \\        
J161916.54+334238.4  &  0.0963       &  0.470946   & $20.55 \pm 0.10$     &  Sightline to Systems \#15 and \#56 (Appendix A)   \\       
J100902.06+071343.8  &   0.1140      &  0.455631   & $20.68 \pm 0.10$      &  Sightline to System \#5 (Appendix A)   \\
J092837.98+602521.0  &   0.1538      &  0.29545     & $19.35 \pm 0.15$      &  No other strong absorbers with $\log N > 15.5$  \\    
B0120-28                      &   0.185        &  0.436018    & $20.50 \pm 0.10$     &  No other strong absorbers with $\log N > 15.5$  \\
J143511.53+360437.2  &   0.2026      &  0.428593    & $19.80 \pm 0.10$     &  Sightline to System \#13  (Appendix A) \\   
J1342-0053                   &  0.22711     &  0.326          & $18.5 \pm 0.5$         &  Prochaska \etal\ (2017)     \\
J092554.70+400414.1  &  0.2477       &   0.471139   & $19.45 \pm 0.15$     &  System \#1  (Appendix A)  \\
J100102.55+594414.3  &   0.30360    &   0.746236   & $19.41 \pm 0.12$     &  System \#6 (Appendix A)  \\
J161649.42+415416.3  &   0.3211      &   0.440417   & $20.34 \pm 0.12$     &  System \#73 (Appendix A)  \\         
J0020930.7-043826      &  0.39043     &   1.131         & $18.5 \pm 0.4$         &  System \#2 (Appendix A)    \\         

\enddata  

\tablenotetext{a} {Four low-redshift Damped \Lya\ (DLA) and six sub-DLA systems identified in our survey,
ordered by absorber redshift ($z_{\rm abs}$).   The first eight absorbers lie below the minimum redshift 
($z_{\rm abs} = 0.24$) of our survey and are not included in our statistical analysis.
DLA systems have $\log N_{\rm HI} > 20.3$ and sub-DLAs have $19.0< \log N_{\rm HI} <  20.3$.  
Two other absorbers have column densities $\log N_{\rm HI} = 18.5\pm0.5$) near the sub-DLA range.}

\tablenotetext{b} {~Column densities for several systems in this list were listed in other papers
(Meiring \etal\ 2011; Muzahid \etal\ 2015;  Tejos \etal\ 2014) with column densities determined by
fitting damping wings of \Lya\ profile.   For other absorbers (systems \#1, 2, 6, 7, 73 in Appendix A), 
we use our own fits to  \Lya\ damping wings, the LL decrement, and CoG for higher Lyman lines.  }

\end{deluxetable}


 \end{document}